\documentclass[draft,envcountsame]{svjour3}

\journalname{Higher-Order and Symbolic Computation}

\smartqed

\usepackage{mathptmx}      
\usepackage{latexsym}
\usepackage{amssymb}
\usepackage{amsfonts}
\usepackage{amsmath}
\usepackage{amsopn}
\usepackage{alltt}
\usepackage{array}
\usepackage{stmaryrd}
\usepackage{mathpartir}

\usepackage{ifthen}
\usepackage{moreverb}
\usepackage[all]{xy}
\usepackage{url}
\usepackage{xspace}
\usepackage{xlsyntax}
\usepackage{macros}

\begin{document}

\title{Compilation of extended recursion \\
       in call-by-value functional languages
\thanks{Work partially supported by EPSRC grant GR/R 41545/01.
        This article is a revised and much extended version of
        \cite{Hirscho03b}.}
}
\author{Tom Hirschowitz \and
        Xavier Leroy \and
        J. B. Wells}

\institute{
  Tom Hirschowitz
  \at Department of Mathematics, University of Savoie,
      Campus Scientifique, 73376 Le Bourget-du-Lac, France.
  \email{tom.hirschowitz@univ-savoie.fr}
\and
  Xavier Leroy
  \at INRIA Paris-Rocquencourt,
      Domaine de Voluceau, B.P. 105, 78153 Le Chesnay, France.
  \email{xavier.leroy@inria.fr}
\and
  J. B. Wells
  \at School of Mathematical and Computer Sciences, Heriot-Watt University,
      Riccarton, Edinburgh, EH14 4AS, Great Britain.
  \email{jbw@macs.hw.ac.uk}}

\date{}

\titlerunning{Compilation of extended recursion}

\maketitle

\begin{abstract}
  This paper formalizes and proves correct a compilation scheme for
  mutually-recursive definitions in call-by-value functional
  languages.  This scheme supports a wider range of recursive
  definitions than previous methods.  We formalize our technique as a
  translation scheme to a lambda-calculus featuring in-place update of
  memory blocks, and prove the translation to be correct.
\end{abstract}

\CompileMatrices
\CompilePrefix{xymatrix-}

\section{Introduction}

\subsection{The need for extended recursion}

Functional languages usually feature mutually recursive definition of
values, for example via the \texttt{letrec} construct in Scheme,
\texttt{let rec} in Caml, \texttt{val rec} and \texttt{fun} in
Standard ML, or recursive equations in Haskell.  
Beyond syntax, functional languages differ also in the kind of
expressions they support as right-hand sides of mutually recursive
definitions.  For instance, Haskell \cite{Haskell} allows arbitrary
expressions as right-hand sides of recursive definitions, while
Standard $\ml$ \cite{Milner97} only allows syntactic
$\lambda$-abstractions, and $\caml$ \cite{OCaml} allows both
$\lambda$-abstractions and limited forms of constructor applications.


The range of allowed right-hand sides crucially depends on the
evaluation strategy of the language.  Call-by-name or lazy languages
such as Haskell naturally implement arbitrary recursive definitions:
the on-demand unwinding of the recursive definition performed by lazy
evaluation correctly reaches the fixed point when it exists, or diverges
when the recursive definition is ill-founded, as in $x \eql x + 1$. 
For call-by-value languages, ill-founded definitions are more
problematic: during the evaluation of $x \eql x + 1$, the right-hand side
$x + 1$ must be evaluated while the value of $x$ is still unknown. 
There is no strict call-by-value strategy that allows this.  Thus,
such ill-founded definitions must be rejected, statically or dynamically.

The simplest way to rule out ill-founded definitions and ensure
call-by-value evaluability is to syntactically restrict the right-hand
sides of recursive definitions to be function abstractions, as $\ml$
does.  Such a restriction also enables efficient compilation of the
recursive definitions, for instance using the compilation scheme
described by Appel \cite{Appel92}.  
While generally acceptable for direct programming in ML,
this restriction can be problematic when we wish to encode higher-level
constructs such as objects, classes, recursive modules and mixin modules.  
For instance, Boudol \cite{Boudol04} uses definitions
of the shape $x \eql c~x$ (where $c$ is a variable) in his
recursive record semantics of objects. Similarly,
Hirschowitz and Leroy \cite{Hirscho05} use mutually-dependent sets of such definitions
for representing mixin modules. Putting these works into practice
requires the definition of an efficient, call-by-value intermediate
language supporting such non-standard recursive definitions.  This
definition is the topic of the present article.  

\subsection{From backpatching to immediate in-place update}
\label{subsection-backpatching-ipu}
\paragraph{Backpatching of reference cells}


\begin{figure}

\begin{framed}
1. Initialization:
\expandafter\ifx\csname xgraph\endcsname\relax
   \csname newbox\expandafter\endcsname\csname xgraph\endcsname
\fi
\ifx\graphtemp\undefined
  \csname newdimen\endcsname\graphtemp
\fi
\expandafter\setbox\csname xgraph\endcsname
 =\vtop{\vskip 0pt\hbox{%
    \special{pn 8}%
    \special{pa 0 133}%
    \special{pa 200 133}%
    \special{pa 200 0}%
    \special{pa 0 0}%
    \special{pa 0 133}%
    \special{fp}%
    \special{pa 1033 133}%
    \special{pa 1233 133}%
    \special{pa 1233 0}%
    \special{pa 1033 0}%
    \special{pa 1033 133}%
    \special{fp}%
    \graphtemp=.5ex
    \advance\graphtemp by 0.067in
    \rlap{\kern 0.000in\lower\graphtemp\hbox to 0pt{\hss $x_1~$}}%
    \graphtemp=.5ex
    \advance\graphtemp by 0.067in
    \rlap{\kern 1.033in\lower\graphtemp\hbox to 0pt{\hss $x_2~$}}%
    \special{pa 517 533}%
    \special{pa 717 533}%
    \special{pa 717 400}%
    \special{pa 517 400}%
    \special{pa 517 533}%
    \special{fp}%
    \graphtemp=.5ex
    \advance\graphtemp by 0.467in
    \rlap{\kern 0.617in\lower\graphtemp\hbox to 0pt{\hss $\bullet$\hss}}%
    \special{pa 544 324}%
    \special{pa 615 399}%
    \special{fp}%
    \special{pa 517 366}%
    \special{pa 615 399}%
    \special{fp}%
    \special{pa 100 67}%
    \special{pa 615 399}%
    \special{fp}%
    \special{pa 716 366}%
    \special{pa 619 399}%
    \special{fp}%
    \special{pa 689 324}%
    \special{pa 619 399}%
    \special{fp}%
    \special{pa 1133 67}%
    \special{pa 619 399}%
    \special{fp}%
    \hbox{\vrule depth0.533in width0pt height 0pt}%
    \kern 1.233in
  }%
}%
\showgraph

2. Computation:
\expandafter\ifx\csname xgraph\endcsname\relax
   \csname newbox\expandafter\endcsname\csname xgraph\endcsname
\fi
\ifx\graphtemp\undefined
  \csname newdimen\endcsname\graphtemp
\fi
\expandafter\setbox\csname xgraph\endcsname
 =\vtop{\vskip 0pt\hbox{%
    \special{pn 8}%
    \special{pa 0 733}%
    \special{pa 400 733}%
    \special{fp}%
    \special{pa 400 733}%
    \special{pa 200 333}%
    \special{fp}%
    \special{pa 200 333}%
    \special{pa 0 733}%
    \special{fp}%
    \special{pa 667 733}%
    \special{pa 1067 733}%
    \special{fp}%
    \special{pa 1067 733}%
    \special{pa 867 333}%
    \special{fp}%
    \special{pa 867 333}%
    \special{pa 667 733}%
    \special{fp}%
    \special{pa 1333 733}%
    \special{pa 1733 733}%
    \special{fp}%
    \special{pa 1733 733}%
    \special{pa 1533 333}%
    \special{fp}%
    \special{pa 1533 333}%
    \special{pa 1333 733}%
    \special{fp}%
    \special{pa 333 133}%
    \special{pa 533 133}%
    \special{pa 533 0}%
    \special{pa 333 0}%
    \special{pa 333 133}%
    \special{fp}%
    \special{pa 533 133}%
    \special{pa 733 133}%
    \special{pa 733 0}%
    \special{pa 533 0}%
    \special{pa 533 133}%
    \special{fp}%
    \special{pa 1433 133}%
    \special{pa 1633 133}%
    \special{pa 1633 0}%
    \special{pa 1433 0}%
    \special{pa 1433 133}%
    \special{fp}%
    \graphtemp=.5ex
    \advance\graphtemp by 0.067in
    \rlap{\kern 0.333in\lower\graphtemp\hbox to 0pt{\hss $v_1~$}}%
    \graphtemp=.5ex
    \advance\graphtemp by 0.067in
    \rlap{\kern 1.433in\lower\graphtemp\hbox to 0pt{\hss $v_2~$}}%
    \special{pa 433 1133}%
    \special{pa 633 1133}%
    \special{pa 633 1000}%
    \special{pa 433 1000}%
    \special{pa 433 1133}%
    \special{fp}%
    \special{pa 1433 1133}%
    \special{pa 1633 1133}%
    \special{pa 1633 1000}%
    \special{pa 1433 1000}%
    \special{pa 1433 1133}%
    \special{fp}%
    \graphtemp=.5ex
    \advance\graphtemp by 1.067in
    \rlap{\kern 0.433in\lower\graphtemp\hbox to 0pt{\hss $x_1~$}}%
    \graphtemp=.5ex
    \advance\graphtemp by 1.067in
    \rlap{\kern 1.433in\lower\graphtemp\hbox to 0pt{\hss $x_2~$}}%
    \special{pa 933 1533}%
    \special{pa 1133 1533}%
    \special{pa 1133 1400}%
    \special{pa 933 1400}%
    \special{pa 933 1533}%
    \special{fp}%
    \graphtemp=.5ex
    \advance\graphtemp by 1.467in
    \rlap{\kern 1.033in\lower\graphtemp\hbox to 0pt{\hss $\bullet$\hss}}%
    \special{pa 962 1322}%
    \special{pa 1031 1399}%
    \special{fp}%
    \special{pa 934 1364}%
    \special{pa 1031 1399}%
    \special{fp}%
    \special{pa 533 1067}%
    \special{pa 1031 1399}%
    \special{fp}%
    \special{pa 1132 1364}%
    \special{pa 1035 1399}%
    \special{fp}%
    \special{pa 1105 1322}%
    \special{pa 1035 1399}%
    \special{fp}%
    \special{pa 1533 1067}%
    \special{pa 1035 1399}%
    \special{fp}%
    \special{pa 286 273}%
    \special{pa 202 332}%
    \special{fp}%
    \special{pa 249 240}%
    \special{pa 202 332}%
    \special{fp}%
    \special{pa 433 67}%
    \special{pa 202 332}%
    \special{fp}%
    \special{pa 818 240}%
    \special{pa 865 332}%
    \special{fp}%
    \special{pa 780 273}%
    \special{pa 865 332}%
    \special{fp}%
    \special{pa 633 67}%
    \special{pa 865 332}%
    \special{fp}%
    \special{pa 1558 231}%
    \special{pa 1533 331}%
    \special{fp}%
    \special{pa 1508 231}%
    \special{pa 1533 331}%
    \special{fp}%
    \special{pa 1533 67}%
    \special{pa 1533 331}%
    \special{fp}%
    \special{pa 469 917}%
    \special{pa 532 999}%
    \special{fp}%
    \special{pa 438 956}%
    \special{pa 532 999}%
    \special{fp}%
    \special{pa 200 733}%
    \special{pa 532 999}%
    \special{fp}%
    \special{pa 623 945}%
    \special{pa 535 998}%
    \special{fp}%
    \special{pa 588 910}%
    \special{pa 535 998}%
    \special{fp}%
    \special{pa 800 733}%
    \special{pa 535 998}%
    \special{fp}%
    \special{pa 1450 936}%
    \special{pa 1531 999}%
    \special{fp}%
    \special{pa 1430 981}%
    \special{pa 1531 999}%
    \special{fp}%
    \special{pa 933 733}%
    \special{pa 1531 999}%
    \special{fp}%
    \special{pa 1558 898}%
    \special{pa 1533 998}%
    \special{fp}%
    \special{pa 1508 898}%
    \special{pa 1533 998}%
    \special{fp}%
    \special{pa 1533 733}%
    \special{pa 1533 998}%
    \special{fp}%
    \hbox{\vrule depth1.533in width0pt height 0pt}%
    \kern 1.733in
  }%
}%
\showgraph

3. Reference update:
\expandafter\ifx\csname xgraph\endcsname\relax
   \csname newbox\expandafter\endcsname\csname xgraph\endcsname
\fi
\ifx\graphtemp\undefined
  \csname newdimen\endcsname\graphtemp
\fi
\expandafter\setbox\csname xgraph\endcsname
 =\vtop{\vskip 0pt\hbox{%
    \special{pn 8}%
    \special{pa 67 850}%
    \special{pa 467 850}%
    \special{fp}%
    \special{pa 467 850}%
    \special{pa 267 450}%
    \special{fp}%
    \special{pa 267 450}%
    \special{pa 67 850}%
    \special{fp}%
    \special{pa 733 850}%
    \special{pa 1133 850}%
    \special{fp}%
    \special{pa 1133 850}%
    \special{pa 933 450}%
    \special{fp}%
    \special{pa 933 450}%
    \special{pa 733 850}%
    \special{fp}%
    \special{pa 1400 850}%
    \special{pa 1800 850}%
    \special{fp}%
    \special{pa 1800 850}%
    \special{pa 1600 450}%
    \special{fp}%
    \special{pa 1600 450}%
    \special{pa 1400 850}%
    \special{fp}%
    \special{pa 400 250}%
    \special{pa 600 250}%
    \special{pa 600 117}%
    \special{pa 400 117}%
    \special{pa 400 250}%
    \special{fp}%
    \special{pa 600 250}%
    \special{pa 800 250}%
    \special{pa 800 117}%
    \special{pa 600 117}%
    \special{pa 600 250}%
    \special{fp}%
    \special{pa 1500 250}%
    \special{pa 1700 250}%
    \special{pa 1700 117}%
    \special{pa 1500 117}%
    \special{pa 1500 250}%
    \special{fp}%
    \graphtemp=.5ex
    \advance\graphtemp by 0.183in
    \rlap{\kern 0.400in\lower\graphtemp\hbox to 0pt{\hss $v_1~$}}%
    \graphtemp=.5ex
    \advance\graphtemp by 0.183in
    \rlap{\kern 1.500in\lower\graphtemp\hbox to 0pt{\hss $v_2~$}}%
    \special{pa 500 1250}%
    \special{pa 700 1250}%
    \special{pa 700 1117}%
    \special{pa 500 1117}%
    \special{pa 500 1250}%
    \special{fp}%
    \special{pa 1500 1250}%
    \special{pa 1700 1250}%
    \special{pa 1700 1117}%
    \special{pa 1500 1117}%
    \special{pa 1500 1250}%
    \special{fp}%
    \graphtemp=.5ex
    \advance\graphtemp by 1.183in
    \rlap{\kern 0.500in\lower\graphtemp\hbox to 0pt{\hss $x_1~$}}%
    \graphtemp=.5ex
    \advance\graphtemp by 1.183in
    \rlap{\kern 1.500in\lower\graphtemp\hbox to 0pt{\hss $x_2~$}}%
    \special{pa 545 27}%
    \special{pa 598 115}%
    \special{fp}%
    \special{pa 510 62}%
    \special{pa 598 115}%
    \special{fp}%
    \special{pa 600 1183}%
    \special{pa 467 1317}%
    \special{pa -66 1317}%
    \special{pa -66 -16}%
    \special{pa 467 -16}%
    \special{pa 598 115}%
    \special{sp}%
    \special{pa 1668 36}%
    \special{pa 1601 115}%
    \special{fp}%
    \special{pa 1623 14}%
    \special{pa 1601 115}%
    \special{fp}%
    \special{pa 1600 1183}%
    \special{pa 1733 1317}%
    \special{pa 2200 1317}%
    \special{pa 2200 -16}%
    \special{pa 1667 -16}%
    \special{pa 1601 115}%
    \special{sp}%
    \special{pa 353 389}%
    \special{pa 268 448}%
    \special{fp}%
    \special{pa 315 357}%
    \special{pa 268 448}%
    \special{fp}%
    \special{pa 500 183}%
    \special{pa 268 448}%
    \special{fp}%
    \special{pa 885 357}%
    \special{pa 932 448}%
    \special{fp}%
    \special{pa 847 389}%
    \special{pa 932 448}%
    \special{fp}%
    \special{pa 700 183}%
    \special{pa 932 448}%
    \special{fp}%
    \special{pa 1625 348}%
    \special{pa 1600 448}%
    \special{fp}%
    \special{pa 1575 348}%
    \special{pa 1600 448}%
    \special{fp}%
    \special{pa 1600 183}%
    \special{pa 1600 448}%
    \special{fp}%
    \special{pa 536 1033}%
    \special{pa 598 1115}%
    \special{fp}%
    \special{pa 504 1072}%
    \special{pa 598 1115}%
    \special{fp}%
    \special{pa 267 850}%
    \special{pa 598 1115}%
    \special{fp}%
    \special{pa 690 1062}%
    \special{pa 602 1115}%
    \special{fp}%
    \special{pa 655 1027}%
    \special{pa 602 1115}%
    \special{fp}%
    \special{pa 867 850}%
    \special{pa 602 1115}%
    \special{fp}%
    \special{pa 1517 1052}%
    \special{pa 1598 1116}%
    \special{fp}%
    \special{pa 1496 1098}%
    \special{pa 1598 1116}%
    \special{fp}%
    \special{pa 1000 850}%
    \special{pa 1598 1116}%
    \special{fp}%
    \special{pa 1625 1014}%
    \special{pa 1600 1114}%
    \special{fp}%
    \special{pa 1575 1014}%
    \special{pa 1600 1114}%
    \special{fp}%
    \special{pa 1600 850}%
    \special{pa 1600 1114}%
    \special{fp}%
    \hbox{\vrule depth1.300in width0pt height 0pt}%
    \kern 2.142in
  }%
}%
\showgraph
\end{framed}
\caption{The backpatching scheme}
\label{figure-backpatching}
\end{figure}

A famous example of a call-by-value language that does not statically
restrict the right-hand sides of recursive definitions is Scheme
\cite{Kelsey98}.  The operational semantics of the
\texttt{letrec} construct of Scheme is known as the
\intro{backpatching} semantics\footnote{The \intro{immediate in-place
    update} compilation scheme studied in this paper also uses a kind
  of backpatching, but we only use ``backpatching'' to refer
  to the schemes described in this section, i.e., to
  abbreviate ``backpatching of reference cells''.}.  It is illustrated
in Figure~\ref{figure-backpatching}.  Consider two mutually-dependent
definitions $x_1 \eql \e_1$ and $x_2 \eql \e_2$.  First, a reference
cell is assigned to each recursive variable, and initialized to some
dummy value $\bdummy$ (represented by $\bullet$ in
Figure~\ref{figure-backpatching}). Then, the right-hand sides are
evaluated, building data structures that possibly include the
reference cells, to obtain some values $\val_1$ and $\val_2$. Until
this point, any attempt to dereference the cells is a run-time error.
Finally, the reference cells are updated with $\val_1$ and $\val_2$,
and the definitions can be considered fully evaluated.

The backpatching scheme leaves some flexibility as to when the
reference cells bound to recursively-defined variables are
dereferenced.  In Scheme, every occurrence of these variables that is
evaluated in the lexical scope of the \texttt{letrec} binding causes
an immediate dereference.  Boudol and Zimmer \cite{Boudol02} propose a compilation
scheme for a call-by-value $\lambda$-calculus with unrestricted
mutually recursive definitions where
the dereferencing is further delayed because arguments to
functions are passed by reference rather than by value.  The
difference is best illustrated on the definition $x \eql (\fun y .
\fun z. \ifte{z \eql 0}{1}{y~(z - 1)})~x$.  In Scheme, it compiles
down to the following intermediate code (written in ML-style notation)
$$\begin{array}{l}
\letseq x \eql \bref~\bdummy ~\inlet \\
x \meql (\fun y . \fun z
. \ifte{z \eql 0}{1}{y~(z - 1)})~!x
\end{array}$$
and therefore fails at run-time because the reference $x$ is accessed
at a time when it still contains $\bdummy$.  In Boudol and Zimmer's
compilation scheme, the $y$ parameter is passed by reference,
resulting in the following compiled code:
$$\begin{array}{l}
\letseq x \eql \bref~\bdummy ~\inlet \\
x \meql (\fun y . \fun z
. \ifte{z \eql 0}{1}{!y~(z - 1)})~x
\end{array}$$
Here, $x$ is passed as a function argument without being dereferenced,
therefore ensuring that the recursive definition evaluates correctly. 
The downside is that the recursive call to $y$ has now to be
preceded by a dereferencing of $y$. 

In summary, the backpatching semantics featured in Scheme enables a
wider range of recursive definitions to be evaluated under a
call-by-value regime than the syntactic restriction of ML.  This range
is even wider in Boudol and Zimmer's variant \cite{Boudol02}.  In both cases,
a drawback of this approach is that, in general, recursive calls to a
recursively-defined function must go through one additional
indirection.  For well-founded definitions, this indirection
seems superfluous, since no further update of the reference cells is
needed.
Scheme compilers optimize this indirection away in some cases, typically when
the right-hand sides are syntactic functions; but removing it
in all cases requires alternative approaches, which we now
describe.

\paragraph{In-place update}
\label{subsubsection-original-scheme}

The \intro{in-place update} scheme~\cite{Cousineau87} is a variant of
the backpatching implementation of recursive definitions that
avoids the additional indirection just mentioned.  It is used in
the $\caml$ compilers~\cite{OCaml}. 


\begin{figure}

\begin{framed}
1. Pre-allocation:
\expandafter\ifx\csname xgraph\endcsname\relax
   \csname newbox\expandafter\endcsname\csname xgraph\endcsname
\fi
\ifx\graphtemp\undefined
  \csname newdimen\endcsname\graphtemp
\fi
\expandafter\setbox\csname xgraph\endcsname
 =\vtop{\vskip 0pt\hbox{%
    \special{pn 8}%
    \special{pa 233 133}%
    \special{pa 433 133}%
    \special{pa 433 0}%
    \special{pa 233 0}%
    \special{pa 233 133}%
    \special{fp}%
    \graphtemp=.5ex
    \advance\graphtemp by 0.067in
    \rlap{\kern 0.333in\lower\graphtemp\hbox to 0pt{\hss $\bullet$\hss}}%
    \special{pa 433 133}%
    \special{pa 633 133}%
    \special{pa 633 0}%
    \special{pa 433 0}%
    \special{pa 433 133}%
    \special{fp}%
    \graphtemp=.5ex
    \advance\graphtemp by 0.067in
    \rlap{\kern 0.533in\lower\graphtemp\hbox to 0pt{\hss $\bullet$\hss}}%
    \special{pa 1333 133}%
    \special{pa 1533 133}%
    \special{pa 1533 0}%
    \special{pa 1333 0}%
    \special{pa 1333 133}%
    \special{fp}%
    \graphtemp=.5ex
    \advance\graphtemp by 0.067in
    \rlap{\kern 1.433in\lower\graphtemp\hbox to 0pt{\hss $\bullet$\hss}}%
    \graphtemp=.5ex
    \advance\graphtemp by 0.067in
    \rlap{\kern 0.233in\lower\graphtemp\hbox to 0pt{\hss $x_1~$}}%
    \graphtemp=.5ex
    \advance\graphtemp by 0.067in
    \rlap{\kern 1.333in\lower\graphtemp\hbox to 0pt{\hss $x_2~$}}%
    \hbox{\vrule depth0.133in width0pt height 0pt}%
    \kern 1.533in
  }%
}%
\showgraph

2. Computation:
\expandafter\ifx\csname xgraph\endcsname\relax
   \csname newbox\expandafter\endcsname\csname xgraph\endcsname
\fi
\ifx\graphtemp\undefined
  \csname newdimen\endcsname\graphtemp
\fi
\expandafter\setbox\csname xgraph\endcsname
 =\vtop{\vskip 0pt\hbox{%
    \special{pn 8}%
    \special{pa 0 733}%
    \special{pa 400 733}%
    \special{fp}%
    \special{pa 400 733}%
    \special{pa 200 333}%
    \special{fp}%
    \special{pa 200 333}%
    \special{pa 0 733}%
    \special{fp}%
    \special{pa 667 733}%
    \special{pa 1067 733}%
    \special{fp}%
    \special{pa 1067 733}%
    \special{pa 867 333}%
    \special{fp}%
    \special{pa 867 333}%
    \special{pa 667 733}%
    \special{fp}%
    \special{pa 1333 733}%
    \special{pa 1733 733}%
    \special{fp}%
    \special{pa 1733 733}%
    \special{pa 1533 333}%
    \special{fp}%
    \special{pa 1533 333}%
    \special{pa 1333 733}%
    \special{fp}%
    \special{pa 333 133}%
    \special{pa 533 133}%
    \special{pa 533 0}%
    \special{pa 333 0}%
    \special{pa 333 133}%
    \special{fp}%
    \special{pa 533 133}%
    \special{pa 733 133}%
    \special{pa 733 0}%
    \special{pa 533 0}%
    \special{pa 533 133}%
    \special{fp}%
    \special{pa 1433 133}%
    \special{pa 1633 133}%
    \special{pa 1633 0}%
    \special{pa 1433 0}%
    \special{pa 1433 133}%
    \special{fp}%
    \graphtemp=.5ex
    \advance\graphtemp by 0.067in
    \rlap{\kern 0.333in\lower\graphtemp\hbox to 0pt{\hss $v_1~$}}%
    \graphtemp=.5ex
    \advance\graphtemp by 0.067in
    \rlap{\kern 1.433in\lower\graphtemp\hbox to 0pt{\hss $v_2~$}}%
    \special{pa 333 1133}%
    \special{pa 533 1133}%
    \special{pa 533 1000}%
    \special{pa 333 1000}%
    \special{pa 333 1133}%
    \special{fp}%
    \graphtemp=.5ex
    \advance\graphtemp by 1.067in
    \rlap{\kern 0.433in\lower\graphtemp\hbox to 0pt{\hss $\bullet$\hss}}%
    \special{pa 533 1133}%
    \special{pa 733 1133}%
    \special{pa 733 1000}%
    \special{pa 533 1000}%
    \special{pa 533 1133}%
    \special{fp}%
    \graphtemp=.5ex
    \advance\graphtemp by 1.067in
    \rlap{\kern 0.633in\lower\graphtemp\hbox to 0pt{\hss $\bullet$\hss}}%
    \special{pa 1433 1133}%
    \special{pa 1633 1133}%
    \special{pa 1633 1000}%
    \special{pa 1433 1000}%
    \special{pa 1433 1133}%
    \special{fp}%
    \graphtemp=.5ex
    \advance\graphtemp by 1.067in
    \rlap{\kern 1.533in\lower\graphtemp\hbox to 0pt{\hss $\bullet$\hss}}%
    \graphtemp=.5ex
    \advance\graphtemp by 1.067in
    \rlap{\kern 0.333in\lower\graphtemp\hbox to 0pt{\hss $x_1~$}}%
    \graphtemp=.5ex
    \advance\graphtemp by 1.067in
    \rlap{\kern 1.433in\lower\graphtemp\hbox to 0pt{\hss $x_2~$}}%
    \special{pa 286 273}%
    \special{pa 202 332}%
    \special{fp}%
    \special{pa 249 240}%
    \special{pa 202 332}%
    \special{fp}%
    \special{pa 433 67}%
    \special{pa 202 332}%
    \special{fp}%
    \special{pa 818 240}%
    \special{pa 865 332}%
    \special{fp}%
    \special{pa 780 273}%
    \special{pa 865 332}%
    \special{fp}%
    \special{pa 633 67}%
    \special{pa 865 332}%
    \special{fp}%
    \special{pa 1558 231}%
    \special{pa 1533 331}%
    \special{fp}%
    \special{pa 1508 231}%
    \special{pa 1533 331}%
    \special{fp}%
    \special{pa 1533 67}%
    \special{pa 1533 331}%
    \special{fp}%
    \special{pa 440 913}%
    \special{pa 498 998}%
    \special{fp}%
    \special{pa 407 951}%
    \special{pa 498 998}%
    \special{fp}%
    \special{pa 200 733}%
    \special{pa 498 998}%
    \special{fp}%
    \special{pa 653 939}%
    \special{pa 568 998}%
    \special{fp}%
    \special{pa 615 907}%
    \special{pa 568 998}%
    \special{fp}%
    \special{pa 800 733}%
    \special{pa 568 998}%
    \special{fp}%
    \special{pa 1418 934}%
    \special{pa 1498 999}%
    \special{fp}%
    \special{pa 1397 979}%
    \special{pa 1498 999}%
    \special{fp}%
    \special{pa 933 733}%
    \special{pa 1498 999}%
    \special{fp}%
    \special{pa 1592 898}%
    \special{pa 1567 998}%
    \special{fp}%
    \special{pa 1542 898}%
    \special{pa 1567 998}%
    \special{fp}%
    \special{pa 1567 733}%
    \special{pa 1567 998}%
    \special{fp}%
    \hbox{\vrule depth1.133in width0pt height 0pt}%
    \kern 1.733in
  }%
}%
\showgraph

3. In-place update:
\expandafter\ifx\csname xgraph\endcsname\relax
   \csname newbox\expandafter\endcsname\csname xgraph\endcsname
\fi
\ifx\graphtemp\undefined
  \csname newdimen\endcsname\graphtemp
\fi
\expandafter\setbox\csname xgraph\endcsname
 =\vtop{\vskip 0pt\hbox{%
    \special{pn 8}%
    \special{pa 125 517}%
    \special{pa 525 517}%
    \special{fp}%
    \special{pa 525 517}%
    \special{pa 325 117}%
    \special{fp}%
    \special{pa 325 117}%
    \special{pa 125 517}%
    \special{fp}%
    \special{pa 792 517}%
    \special{pa 1192 517}%
    \special{fp}%
    \special{pa 1192 517}%
    \special{pa 992 117}%
    \special{fp}%
    \special{pa 992 117}%
    \special{pa 792 517}%
    \special{fp}%
    \special{pa 1458 517}%
    \special{pa 1858 517}%
    \special{fp}%
    \special{pa 1858 517}%
    \special{pa 1658 117}%
    \special{fp}%
    \special{pa 1658 117}%
    \special{pa 1458 517}%
    \special{fp}%
    \special{pa 458 917}%
    \special{pa 658 917}%
    \special{pa 658 783}%
    \special{pa 458 783}%
    \special{pa 458 917}%
    \special{fp}%
    \special{pa 658 917}%
    \special{pa 858 917}%
    \special{pa 858 783}%
    \special{pa 658 783}%
    \special{pa 658 917}%
    \special{fp}%
    \special{pa 1558 917}%
    \special{pa 1758 917}%
    \special{pa 1758 783}%
    \special{pa 1558 783}%
    \special{pa 1558 917}%
    \special{fp}%
    \graphtemp=.5ex
    \advance\graphtemp by 0.850in
    \rlap{\kern 0.458in\lower\graphtemp\hbox to 0pt{\hss $x_1~$}}%
    \graphtemp=.5ex
    \advance\graphtemp by 0.850in
    \rlap{\kern 1.558in\lower\graphtemp\hbox to 0pt{\hss $x_2~$}}%
    \special{pa 324 11}%
    \special{pa 324 114}%
    \special{fp}%
    \special{pa 276 23}%
    \special{pa 324 114}%
    \special{fp}%
    \special{pa 558 850}%
    \special{pa 425 983}%
    \special{pa -41 983}%
    \special{pa -41 -16}%
    \special{pa 292 -16}%
    \special{pa 324 114}%
    \special{sp}%
    \special{pa 1041 23}%
    \special{pa 992 114}%
    \special{fp}%
    \special{pa 992 11}%
    \special{pa 992 114}%
    \special{fp}%
    \special{pa 758 850}%
    \special{pa 892 983}%
    \special{pa 1358 983}%
    \special{pa 1358 -16}%
    \special{pa 1025 -16}%
    \special{pa 992 114}%
    \special{sp}%
    \special{pa 1726 36}%
    \special{pa 1659 115}%
    \special{fp}%
    \special{pa 1682 14}%
    \special{pa 1659 115}%
    \special{fp}%
    \special{pa 1658 850}%
    \special{pa 1792 983}%
    \special{pa 2058 983}%
    \special{pa 2058 -16}%
    \special{pa 1725 -16}%
    \special{pa 1659 115}%
    \special{sp}%
    \special{pa 565 697}%
    \special{pa 623 782}%
    \special{fp}%
    \special{pa 532 734}%
    \special{pa 623 782}%
    \special{fp}%
    \special{pa 325 517}%
    \special{pa 623 782}%
    \special{fp}%
    \special{pa 778 723}%
    \special{pa 693 782}%
    \special{fp}%
    \special{pa 740 690}%
    \special{pa 693 782}%
    \special{fp}%
    \special{pa 925 517}%
    \special{pa 693 782}%
    \special{fp}%
    \special{pa 1543 717}%
    \special{pa 1623 782}%
    \special{fp}%
    \special{pa 1522 762}%
    \special{pa 1623 782}%
    \special{fp}%
    \special{pa 1058 517}%
    \special{pa 1623 782}%
    \special{fp}%
    \special{pa 1717 681}%
    \special{pa 1692 781}%
    \special{fp}%
    \special{pa 1667 681}%
    \special{pa 1692 781}%
    \special{fp}%
    \special{pa 1692 517}%
    \special{pa 1692 781}%
    \special{fp}%
    \hbox{\vrule depth0.967in width0pt height 0pt}%
    \kern 2.025in
  }%
}%
\showgraph
\end{framed}
\caption{The in-place update scheme}
\label{figure-picture-inplace-updating-trick}
\end{figure}

The in-place update scheme
implements mutually recursive definitions that satisfy the following two
conditions.  For a mutually recursive definition $x_1 \eql \e_1 , \ldots, 
x_n \eql \e_n$, first, the value of each definition should be represented
at run-time by a heap allocated block of statically predictable size;
second, for each $i$, the computation of $\e_i$ should not need the
value of any of the definitions $\e_j$, but only their names $x_j$. 
As an example of the second condition, the recursive definition
$f \eql \fun x.(... f ...)$ is accepted, since the computation of
the right-hand side does not need the value of $f$. We say that
it \intro{weakly} depends on $f$. In contrast, the recursive definition
$f \eql (f~0)$ is rejected. We say that the right-hand side
\intro{strongly} depends on $f$. Several techniques to check this condition
have been proposed~\cite{Boudol04,Hirscho05,Hirscho-PhD,Dreyer04}.

The evaluation of a set of mutually recursive definitions with
in-place update consists of three steps.  First, for each definition,
allocate an uninitialized block of the expected size, and bind it to
the recursively-defined identifier.  Those blocks are called
\intro{dummy} blocks, and this step is called the
\intro{pre-allocation} step.  Second, compute the right-hand sides of the
definitions.  Recursively-defined identifiers thus refer to the
corresponding dummy blocks.  Owing to the second condition, no attempt
is made to access the contents of the dummy blocks. This step leads,
for each definition, to a block of the expected size. Third, update
the dummy blocks in place with the contents of the computed blocks.
(Alternatively, the second step could store directly its results
in the dummy blocks. However, this would require a special evaluation
scheme for right-hand sides of recursive definitions whereas, here,
they are evaluated just like any other expression.)

For example, consider a mutually recursive definition $x_1 \eql \e_1,
x_2 \eql \e_2$, where it is statically predictable that the values of
the expressions $\e_1$ and $\e_2$ will be represented at runtime by
heap-allocated blocks of sizes 2 and 1, respectively.  Here is what
the compiled code does, as depicted in
Figure~\ref{figure-picture-inplace-updating-trick}.  First, it
allocates two uninitialized heap blocks, at addresses~$\loc_1$
and~$\loc_2$, of respective sizes~2 and~1.  
Then, it computes $\e_1$, where
$x_1$ and $x_2$ are bound to~$\loc_1$ and~$\loc_2$, respectively. The
result is a heap block of size~2, possibly containing references to~$\loc_1$
and~$\loc_2$. The same process is carried on for~$\e_2$, resulting in
a heap block of size~1.  The third and final step copies the contents
of the two obtained blocks to~$\loc_1$ and~$\loc_2$, respectively, then
garbage-collects the useless blocks.  The result is that the two
initially dummy blocks now contain the proper cyclic data structures,
without the indirection inherent in the backpatching semantics. 

\paragraph{Immediate in-place update}
\label{subsubsection-ipu}
The scheme described above computes all definitions in sequence, and
only then updates the dummy blocks in place.  From the example above,
it seems quite clear that in-place update for a definition could be
done as soon as its value is available.  Such an improvement has been
proposed for the backpatching semantics~\cite{Waddell05}, and we
merely adapt it to our setting here. 
We call this method the \intro{immediate in-place update} scheme
and concentrate on it in the remainder of this paper. 

As long as definitions weakly depend on each other, as happens with
functions for instance, both schemes behave identically.
Nevertheless, in the case where $\e_2$ strongly depends on $x_1$, for
example if $\e_2 \eql \fst{x_1} + 1$, the original scheme can go
wrong.  Indeed, the contents of $\loc_1$ are still undefined when
$\e_2$ is computed. Instead, with immediate in-place update, the value
$\val_1$ is already available when computing $\e_2$.  This trivial
modification to the scheme thus increases the expressive power of
mutually recursive definitions. It allows definitions to de-structure
the values of previous definitions.  Furthermore, it allows some
of the mutually-recursive definitions to have statically unknown sizes,
as shown by the following example.


\begin{figure}
\begin{framed}
1. Pre-allocation:
\expandafter\ifx\csname xgraph\endcsname\relax
   \csname newbox\expandafter\endcsname\csname xgraph\endcsname
\fi
\ifx\graphtemp\undefined
  \csname newdimen\endcsname\graphtemp
\fi
\expandafter\setbox\csname xgraph\endcsname
 =\vtop{\vskip 0pt\hbox{%
    \special{pn 8}%
    \special{pa 267 133}%
    \special{pa 467 133}%
    \special{pa 467 0}%
    \special{pa 267 0}%
    \special{pa 267 133}%
    \special{fp}%
    \graphtemp=.5ex
    \advance\graphtemp by 0.067in
    \rlap{\kern 0.367in\lower\graphtemp\hbox to 0pt{\hss $\bullet$\hss}}%
    \special{pa 467 133}%
    \special{pa 667 133}%
    \special{pa 667 0}%
    \special{pa 467 0}%
    \special{pa 467 133}%
    \special{fp}%
    \graphtemp=.5ex
    \advance\graphtemp by 0.067in
    \rlap{\kern 0.567in\lower\graphtemp\hbox to 0pt{\hss $\bullet$\hss}}%
    \special{pa 2033 133}%
    \special{pa 2233 133}%
    \special{pa 2233 0}%
    \special{pa 2033 0}%
    \special{pa 2033 133}%
    \special{fp}%
    \graphtemp=.5ex
    \advance\graphtemp by 0.067in
    \rlap{\kern 2.133in\lower\graphtemp\hbox to 0pt{\hss $\bullet$\hss}}%
    \graphtemp=.5ex
    \advance\graphtemp by 0.067in
    \rlap{\kern 0.267in\lower\graphtemp\hbox to 0pt{\hss $x_1~$}}%
    \graphtemp=.5ex
    \advance\graphtemp by 0.067in
    \rlap{\kern 2.033in\lower\graphtemp\hbox to 0pt{\hss $x_3~$}}%
    \hbox{\vrule depth0.133in width0pt height 0pt}%
    \kern 2.233in
  }%
}%
\showgraph

2. Computation of $e_1$:
\expandafter\ifx\csname xgraph\endcsname\relax
   \csname newbox\expandafter\endcsname\csname xgraph\endcsname
\fi
\ifx\graphtemp\undefined
  \csname newdimen\endcsname\graphtemp
\fi
\expandafter\setbox\csname xgraph\endcsname
 =\vtop{\vskip 0pt\hbox{%
    \special{pn 8}%
    \special{pa 0 733}%
    \special{pa 400 733}%
    \special{fp}%
    \special{pa 400 733}%
    \special{pa 200 333}%
    \special{fp}%
    \special{pa 200 333}%
    \special{pa 0 733}%
    \special{fp}%
    \special{pa 667 733}%
    \special{pa 1067 733}%
    \special{fp}%
    \special{pa 1067 733}%
    \special{pa 867 333}%
    \special{fp}%
    \special{pa 867 333}%
    \special{pa 667 733}%
    \special{fp}%
    \special{pa 333 133}%
    \special{pa 533 133}%
    \special{pa 533 0}%
    \special{pa 333 0}%
    \special{pa 333 133}%
    \special{fp}%
    \special{pa 533 133}%
    \special{pa 733 133}%
    \special{pa 733 0}%
    \special{pa 533 0}%
    \special{pa 533 133}%
    \special{fp}%
    \graphtemp=.5ex
    \advance\graphtemp by 0.067in
    \rlap{\kern 0.333in\lower\graphtemp\hbox to 0pt{\hss $v_1~$}}%
    \special{pa 333 1133}%
    \special{pa 533 1133}%
    \special{pa 533 1000}%
    \special{pa 333 1000}%
    \special{pa 333 1133}%
    \special{fp}%
    \graphtemp=.5ex
    \advance\graphtemp by 1.067in
    \rlap{\kern 0.433in\lower\graphtemp\hbox to 0pt{\hss $\bullet$\hss}}%
    \special{pa 533 1133}%
    \special{pa 733 1133}%
    \special{pa 733 1000}%
    \special{pa 533 1000}%
    \special{pa 533 1133}%
    \special{fp}%
    \graphtemp=.5ex
    \advance\graphtemp by 1.067in
    \rlap{\kern 0.633in\lower\graphtemp\hbox to 0pt{\hss $\bullet$\hss}}%
    \special{pa 2100 1133}%
    \special{pa 2300 1133}%
    \special{pa 2300 1000}%
    \special{pa 2100 1000}%
    \special{pa 2100 1133}%
    \special{fp}%
    \graphtemp=.5ex
    \advance\graphtemp by 1.067in
    \rlap{\kern 2.200in\lower\graphtemp\hbox to 0pt{\hss $\bullet$\hss}}%
    \graphtemp=.5ex
    \advance\graphtemp by 1.067in
    \rlap{\kern 0.333in\lower\graphtemp\hbox to 0pt{\hss $x_1~$}}%
    \graphtemp=.5ex
    \advance\graphtemp by 1.067in
    \rlap{\kern 2.100in\lower\graphtemp\hbox to 0pt{\hss $x_3~$}}%
    \special{pa 286 273}%
    \special{pa 202 332}%
    \special{fp}%
    \special{pa 249 240}%
    \special{pa 202 332}%
    \special{fp}%
    \special{pa 433 67}%
    \special{pa 202 332}%
    \special{fp}%
    \special{pa 818 240}%
    \special{pa 865 332}%
    \special{fp}%
    \special{pa 780 273}%
    \special{pa 865 332}%
    \special{fp}%
    \special{pa 633 67}%
    \special{pa 865 332}%
    \special{fp}%
    \special{pa 440 913}%
    \special{pa 498 998}%
    \special{fp}%
    \special{pa 407 951}%
    \special{pa 498 998}%
    \special{fp}%
    \special{pa 200 733}%
    \special{pa 498 998}%
    \special{fp}%
    \special{pa 653 939}%
    \special{pa 568 998}%
    \special{fp}%
    \special{pa 615 907}%
    \special{pa 568 998}%
    \special{fp}%
    \special{pa 800 733}%
    \special{pa 568 998}%
    \special{fp}%
    \special{pa 2072 954}%
    \special{pa 2164 1000}%
    \special{fp}%
    \special{pa 2061 1003}%
    \special{pa 2164 1000}%
    \special{fp}%
    \special{pa 933 733}%
    \special{pa 2164 1000}%
    \special{fp}%
    \hbox{\vrule depth1.133in width0pt height 0pt}%
    \kern 2.400in
  }%
}%
\showgraph

3. Update of $x_1$ with $v_1$:
\expandafter\ifx\csname xgraph\endcsname\relax
   \csname newbox\expandafter\endcsname\csname xgraph\endcsname
\fi
\ifx\graphtemp\undefined
  \csname newdimen\endcsname\graphtemp
\fi
\expandafter\setbox\csname xgraph\endcsname
 =\vtop{\vskip 0pt\hbox{%
    \special{pn 8}%
    \special{pa 125 517}%
    \special{pa 525 517}%
    \special{fp}%
    \special{pa 525 517}%
    \special{pa 325 117}%
    \special{fp}%
    \special{pa 325 117}%
    \special{pa 125 517}%
    \special{fp}%
    \special{pa 792 517}%
    \special{pa 1192 517}%
    \special{fp}%
    \special{pa 1192 517}%
    \special{pa 992 117}%
    \special{fp}%
    \special{pa 992 117}%
    \special{pa 792 517}%
    \special{fp}%
    \special{pa 458 917}%
    \special{pa 658 917}%
    \special{pa 658 783}%
    \special{pa 458 783}%
    \special{pa 458 917}%
    \special{fp}%
    \special{pa 658 917}%
    \special{pa 858 917}%
    \special{pa 858 783}%
    \special{pa 658 783}%
    \special{pa 658 917}%
    \special{fp}%
    \special{pa 2225 917}%
    \special{pa 2425 917}%
    \special{pa 2425 783}%
    \special{pa 2225 783}%
    \special{pa 2225 917}%
    \special{fp}%
    \graphtemp=.5ex
    \advance\graphtemp by 0.850in
    \rlap{\kern 2.325in\lower\graphtemp\hbox to 0pt{\hss $\bullet$\hss}}%
    \graphtemp=.5ex
    \advance\graphtemp by 0.850in
    \rlap{\kern 0.458in\lower\graphtemp\hbox to 0pt{\hss $x_1~$}}%
    \graphtemp=.5ex
    \advance\graphtemp by 0.850in
    \rlap{\kern 2.225in\lower\graphtemp\hbox to 0pt{\hss $x_3~$}}%
    \special{pa 324 11}%
    \special{pa 324 114}%
    \special{fp}%
    \special{pa 276 23}%
    \special{pa 324 114}%
    \special{fp}%
    \special{pa 558 850}%
    \special{pa 425 983}%
    \special{pa -41 983}%
    \special{pa -41 -16}%
    \special{pa 292 -16}%
    \special{pa 324 114}%
    \special{sp}%
    \special{pa 1041 23}%
    \special{pa 992 114}%
    \special{fp}%
    \special{pa 992 11}%
    \special{pa 992 114}%
    \special{fp}%
    \special{pa 758 850}%
    \special{pa 892 983}%
    \special{pa 1358 983}%
    \special{pa 1358 -16}%
    \special{pa 1025 -16}%
    \special{pa 992 114}%
    \special{sp}%
    \special{pa 565 697}%
    \special{pa 623 782}%
    \special{fp}%
    \special{pa 532 734}%
    \special{pa 623 782}%
    \special{fp}%
    \special{pa 325 517}%
    \special{pa 623 782}%
    \special{fp}%
    \special{pa 778 723}%
    \special{pa 693 782}%
    \special{fp}%
    \special{pa 740 690}%
    \special{pa 693 782}%
    \special{fp}%
    \special{pa 925 517}%
    \special{pa 693 782}%
    \special{fp}%
    \special{pa 2197 737}%
    \special{pa 2289 783}%
    \special{fp}%
    \special{pa 2186 786}%
    \special{pa 2289 783}%
    \special{fp}%
    \special{pa 1058 517}%
    \special{pa 2289 783}%
    \special{fp}%
    \hbox{\vrule depth0.967in width0pt height 0pt}%
    \kern 2.525in
  }%
}%
\showgraph

4. Computation of $e_2$ and binding of its value to $x_2$:
\expandafter\ifx\csname xgraph\endcsname\relax
   \csname newbox\expandafter\endcsname\csname xgraph\endcsname
\fi
\ifx\graphtemp\undefined
  \csname newdimen\endcsname\graphtemp
\fi
\expandafter\setbox\csname xgraph\endcsname
 =\vtop{\vskip 0pt\hbox{%
    \special{pn 8}%
    \special{pa 125 517}%
    \special{pa 525 517}%
    \special{fp}%
    \special{pa 525 517}%
    \special{pa 325 117}%
    \special{fp}%
    \special{pa 325 117}%
    \special{pa 125 517}%
    \special{fp}%
    \special{pa 792 517}%
    \special{pa 1192 517}%
    \special{fp}%
    \special{pa 1192 517}%
    \special{pa 992 117}%
    \special{fp}%
    \special{pa 992 117}%
    \special{pa 792 517}%
    \special{fp}%
    \special{pa 1458 517}%
    \special{pa 1858 517}%
    \special{fp}%
    \special{pa 1858 517}%
    \special{pa 1658 117}%
    \special{fp}%
    \special{pa 1658 117}%
    \special{pa 1458 517}%
    \special{fp}%
    \special{pa 458 917}%
    \special{pa 658 917}%
    \special{pa 658 783}%
    \special{pa 458 783}%
    \special{pa 458 917}%
    \special{fp}%
    \special{pa 658 917}%
    \special{pa 858 917}%
    \special{pa 858 783}%
    \special{pa 658 783}%
    \special{pa 658 917}%
    \special{fp}%
    \special{pa 2225 917}%
    \special{pa 2425 917}%
    \special{pa 2425 783}%
    \special{pa 2225 783}%
    \special{pa 2225 917}%
    \special{fp}%
    \graphtemp=.5ex
    \advance\graphtemp by 0.850in
    \rlap{\kern 2.325in\lower\graphtemp\hbox to 0pt{\hss $\bullet$\hss}}%
    \graphtemp=.5ex
    \advance\graphtemp by 0.850in
    \rlap{\kern 0.458in\lower\graphtemp\hbox to 0pt{\hss $x_1~$}}%
    \graphtemp=.5ex
    \advance\graphtemp by 0.850in
    \rlap{\kern 2.225in\lower\graphtemp\hbox to 0pt{\hss $x_3~$}}%
    \graphtemp=\baselineskip
    \multiply\graphtemp by -1
    \divide\graphtemp by 2
    \advance\graphtemp by .5ex
    \advance\graphtemp by 0.117in
    \rlap{\kern 1.658in\lower\graphtemp\hbox to 0pt{\hss $x_2$\hss}}%
    \special{pa 324 11}%
    \special{pa 324 114}%
    \special{fp}%
    \special{pa 276 23}%
    \special{pa 324 114}%
    \special{fp}%
    \special{pa 558 850}%
    \special{pa 425 983}%
    \special{pa -41 983}%
    \special{pa -41 -16}%
    \special{pa 292 -16}%
    \special{pa 324 114}%
    \special{sp}%
    \special{pa 1041 23}%
    \special{pa 992 114}%
    \special{fp}%
    \special{pa 992 11}%
    \special{pa 992 114}%
    \special{fp}%
    \special{pa 758 850}%
    \special{pa 892 983}%
    \special{pa 1358 983}%
    \special{pa 1358 -16}%
    \special{pa 1025 -16}%
    \special{pa 992 114}%
    \special{sp}%
    \special{pa 565 697}%
    \special{pa 623 782}%
    \special{fp}%
    \special{pa 532 734}%
    \special{pa 623 782}%
    \special{fp}%
    \special{pa 325 517}%
    \special{pa 623 782}%
    \special{fp}%
    \special{pa 778 723}%
    \special{pa 693 782}%
    \special{fp}%
    \special{pa 740 690}%
    \special{pa 693 782}%
    \special{fp}%
    \special{pa 925 517}%
    \special{pa 693 782}%
    \special{fp}%
    \special{pa 2197 737}%
    \special{pa 2289 783}%
    \special{fp}%
    \special{pa 2186 786}%
    \special{pa 2289 783}%
    \special{fp}%
    \special{pa 1058 517}%
    \special{pa 2289 783}%
    \special{fp}%
    \special{pa 1194 342}%
    \special{pa 1094 317}%
    \special{fp}%
    \special{pa 1194 292}%
    \special{pa 1094 317}%
    \special{fp}%
    \special{pa 1558 317}%
    \special{pa 1094 317}%
    \special{fp}%
    \special{pa 2222 940}%
    \special{pa 2323 918}%
    \special{fp}%
    \special{pa 2245 985}%
    \special{pa 2323 918}%
    \special{fp}%
    \special{pa 1658 517}%
    \special{pa 1658 1050}%
    \special{pa 2058 1050}%
    \special{pa 2323 918}%
    \special{sp}%
    \hbox{\vrule depth1.033in width0pt height 0pt}%
    \kern 2.525in
  }%
}%
\showgraph

5. Computation and update of $e_3$:
\expandafter\ifx\csname xgraph\endcsname\relax
   \csname newbox\expandafter\endcsname\csname xgraph\endcsname
\fi
\ifx\graphtemp\undefined
  \csname newdimen\endcsname\graphtemp
\fi
\expandafter\setbox\csname xgraph\endcsname
 =\vtop{\vskip 0pt\hbox{%
    \special{pn 8}%
    \special{pa 125 517}%
    \special{pa 525 517}%
    \special{fp}%
    \special{pa 525 517}%
    \special{pa 325 117}%
    \special{fp}%
    \special{pa 325 117}%
    \special{pa 125 517}%
    \special{fp}%
    \special{pa 792 517}%
    \special{pa 1192 517}%
    \special{fp}%
    \special{pa 1192 517}%
    \special{pa 992 117}%
    \special{fp}%
    \special{pa 992 117}%
    \special{pa 792 517}%
    \special{fp}%
    \special{pa 1458 517}%
    \special{pa 1858 517}%
    \special{fp}%
    \special{pa 1858 517}%
    \special{pa 1658 117}%
    \special{fp}%
    \special{pa 1658 117}%
    \special{pa 1458 517}%
    \special{fp}%
    \special{pa 2125 517}%
    \special{pa 2525 517}%
    \special{fp}%
    \special{pa 2525 517}%
    \special{pa 2325 117}%
    \special{fp}%
    \special{pa 2325 117}%
    \special{pa 2125 517}%
    \special{fp}%
    \special{pa 458 917}%
    \special{pa 658 917}%
    \special{pa 658 783}%
    \special{pa 458 783}%
    \special{pa 458 917}%
    \special{fp}%
    \special{pa 658 917}%
    \special{pa 858 917}%
    \special{pa 858 783}%
    \special{pa 658 783}%
    \special{pa 658 917}%
    \special{fp}%
    \special{pa 2225 917}%
    \special{pa 2425 917}%
    \special{pa 2425 783}%
    \special{pa 2225 783}%
    \special{pa 2225 917}%
    \special{fp}%
    \graphtemp=.5ex
    \advance\graphtemp by 0.850in
    \rlap{\kern 0.458in\lower\graphtemp\hbox to 0pt{\hss $x_1~$}}%
    \graphtemp=.5ex
    \advance\graphtemp by 0.850in
    \rlap{\kern 2.225in\lower\graphtemp\hbox to 0pt{\hss $x_3~$}}%
    \graphtemp=\baselineskip
    \multiply\graphtemp by -1
    \divide\graphtemp by 2
    \advance\graphtemp by .5ex
    \advance\graphtemp by 0.117in
    \rlap{\kern 1.658in\lower\graphtemp\hbox to 0pt{\hss $x_2$\hss}}%
    \special{pa 324 11}%
    \special{pa 324 114}%
    \special{fp}%
    \special{pa 276 23}%
    \special{pa 324 114}%
    \special{fp}%
    \special{pa 558 850}%
    \special{pa 425 983}%
    \special{pa -41 983}%
    \special{pa -41 -16}%
    \special{pa 292 -16}%
    \special{pa 324 114}%
    \special{sp}%
    \special{pa 1041 23}%
    \special{pa 992 114}%
    \special{fp}%
    \special{pa 992 11}%
    \special{pa 992 114}%
    \special{fp}%
    \special{pa 758 850}%
    \special{pa 892 983}%
    \special{pa 1358 983}%
    \special{pa 1358 -16}%
    \special{pa 1025 -16}%
    \special{pa 992 114}%
    \special{sp}%
    \special{pa 565 697}%
    \special{pa 623 782}%
    \special{fp}%
    \special{pa 532 734}%
    \special{pa 623 782}%
    \special{fp}%
    \special{pa 325 517}%
    \special{pa 623 782}%
    \special{fp}%
    \special{pa 778 723}%
    \special{pa 693 782}%
    \special{fp}%
    \special{pa 740 690}%
    \special{pa 693 782}%
    \special{fp}%
    \special{pa 925 517}%
    \special{pa 693 782}%
    \special{fp}%
    \special{pa 2197 737}%
    \special{pa 2289 783}%
    \special{fp}%
    \special{pa 2186 786}%
    \special{pa 2289 783}%
    \special{fp}%
    \special{pa 1058 517}%
    \special{pa 2289 783}%
    \special{fp}%
    \special{pa 2383 681}%
    \special{pa 2358 781}%
    \special{fp}%
    \special{pa 2333 681}%
    \special{pa 2358 781}%
    \special{fp}%
    \special{pa 2358 517}%
    \special{pa 2358 781}%
    \special{fp}%
    \special{pa 1194 342}%
    \special{pa 1094 317}%
    \special{fp}%
    \special{pa 1194 292}%
    \special{pa 1094 317}%
    \special{fp}%
    \special{pa 1558 317}%
    \special{pa 1094 317}%
    \special{fp}%
    \special{pa 2222 940}%
    \special{pa 2323 918}%
    \special{fp}%
    \special{pa 2245 985}%
    \special{pa 2323 918}%
    \special{fp}%
    \special{pa 1658 517}%
    \special{pa 1658 1050}%
    \special{pa 2058 1050}%
    \special{pa 2323 918}%
    \special{sp}%
    \special{pa 1861 342}%
    \special{pa 1761 317}%
    \special{fp}%
    \special{pa 1861 292}%
    \special{pa 1761 317}%
    \special{fp}%
    \special{pa 2225 317}%
    \special{pa 1761 317}%
    \special{fp}%
    \special{pa 2393 36}%
    \special{pa 2326 115}%
    \special{fp}%
    \special{pa 2348 14}%
    \special{pa 2326 115}%
    \special{fp}%
    \special{pa 2325 850}%
    \special{pa 2458 983}%
    \special{pa 2725 983}%
    \special{pa 2725 -16}%
    \special{pa 2392 -16}%
    \special{pa 2326 115}%
    \special{sp}%
    \hbox{\vrule depth1.033in width0pt height 0pt}%
    \kern 2.692in
  }%
}%
\showgraph

\end{framed}
\caption{The immediate in-place update scheme}
\label{figure-picture-refined-inplace-updating-trick}
\end{figure}

An example of execution is presented in
Figure~\ref{figure-picture-refined-inplace-updating-trick}.  The
definition is $x_1 \eql \e_1, x_2 \eql \e_2, x_3 \eql \e_3$, where $\e_1$ and
$\e_3$ are expected to evaluate to blocks of sizes 2 and 1,
respectively, but where the representation for the value of $\e_2$ is
not statically predictable.  The pre-allocation step allocates
dummy blocks for $x_1$ and $x_3$ only.  The value $\val_1$ of $\e_1$ is
then computed. It can reference $x_1$ and $x_3$, which correspond to
pointers to the dummy blocks, but not $x_2$, which would not make any
sense here. This value is copied to the corresponding dummy block. 
Then, the value $\val_2$ of $\e_2$ is computed.  The computation can
refer to both dummy blocks, and can also strongly depend on $x_1$, but
not on $x_2$.  Finally, the value $\val_3$ of $\e_3$ is computed and
copied to the corresponding dummy block. 

The immediate in-place update scheme implements more definitions than
the original in-place update scheme. In fact, it implements arbitrary
non-recursive definitions, thus allowing to merge the traditionally
distinct constructs $\letseq$ and \Letrec.

\paragraph{Restrictions imposed on the source language}
What are the restrictions put on recursive definitions in the source
language if we are to compile them with the immediate in-place update
scheme? 
We adopt the following sufficient conditions.
First, the values of forward referenced definitions must be
represented by heap-allocated blocks.  Second, the sizes of these
blocks must be known statically.  Third, the contents of these blocks
should not be accessed  before they have been
updated with proper values.  These restrictions are highly dependent
on the data representation strategy implemented by the compiler.  The
second restriction also depends on how expected sizes are computed at
compile-time, which entails a static analysis that is necessarily
conservative.  For instance, Hirschowitz \cite{Hirscho-PhD} derives the
sizes from the static types of the right-hand sides of recursive
definitions, while the $\caml$ compiler proceeds by syntactic
inspection of the shapes of the right-hand sides.  More sophisticated
static analyses, such as 0-CFA \cite{Shivers-88} or enriched type
systems, could also be used.

In this article, we abstract over these compiler-dependent issues as
follows.  We define a source language where each recursive definition is
annotated by the expected size of the representation of the right-hand
side, if known.  These annotations reflect the result of a prior size
analysis of the kind mentioned earlier.  Both our source and target
languages feature a notion of size, which we only assume to be
preserved by the translation (Hypothesis~\ref{hypothesis-size}) and
satisfy a few natural requirements
(Hypotheses~\ref{hypothesis-size-ab} and~\ref{hypothesis-size-alloc}).

\subsection{Summary of contributions}


The contributions of this article are threefold.  First, we introduce
and formalize a call-by-value functional language called $\lambdaab$,
featuring an extended recursion construct that is not restricted to
$\lambda$-abstractions as right-hand sides of recursive definitions,
but also supports recursive definitions of data structures ($x \eql
\texttt{cons}~1~x$) and of fixed points of certain higher-order
functions ($x \eql f~x$).  This recursion construct subsumes both the
standard recursive and non-recursive value binding constructs
$\letseq$ and \Letrec, and is compilable by immediate in-place
update. 

Second, we provide the first formalization of the in-place update
implementation scheme.  It is formalized as a translation from $\lambdaab$
to a target language $\lambdaalloc$ that does not feature recursive
definitions, but instead explicitly manipulates a heap via allocation
and update operations.   This language is designed to closely match 
the \texttt{Lambda} intermediate languages used by the $\caml$
compiler~\cite{OCaml}, attesting that it can be implemented efficiently.



Third, we prove that the evaluation of any $\lambdaab$ expression is
correctly simulated by its translation. This is the first formal
correctness proof for the in-place update scheme.

The remainder of this paper is organized as follows.  In
Section~\ref{section-lambdaab}, we formalize the source language
$\lambdaab$.  Section~\ref{section-lambdaalloc} defines the target
language $\lambdaalloc$.  We define the compilation scheme from
$\lambdaab$ to $\lambdaalloc$ in Section~\ref{section-compilation},
and prove its correctness in Section~\ref{section-correctness}.
Finally, we discuss related work in Section~\ref{section-related-work}
and conclusions and future work in Section~\ref{section-future-work}.

\section{The source language $\lambdaab$}
\label{section-lambdaab}

\subsection{Notations}
\label{subsection-notation}
Given two sets $A$ and $B$, $A \orth B$ means that $A$ and $B$ are
disjoint, $\Parties{A}$ denotes the set of all subsets of $A$, and
$\cardinal{A}$ denotes the cardinal of $A$.

\blurb{A binary relation $R$ is \intro{functional} iff for all $a,b,c$,
$(a,b) \in R$ and $(a,c) \in R$ implies $b \meq c$, and furthermore
for all $a$ there exists a $b$ such that $(a, b) \in R$.  A
\intro{function} is a triple $(R,A,B)$ of two sets $A$ and
$B$, and a functional binary relation $R \subseteq {A \times
  B}$. }

For all sets $A$ and $B$ and functions $f: A \to B$, $\dom{f}$ denotes
the \intro{domain} $A$ of $f$, and $\cod{f}$ denotes its
\intro{codomain} $B$.  Moreover, $\cosieve{f}{C}$ denotes $f$
restricted to $A \setminus C$.  We also write $f \where{a \mapsto b}$
for the unique function $f': (A \cup \ens{a}) \to (B \cup \ens{b})$
such that $f'(a) \meq b$ and for all $a' \in A \setminus \ens{a}$,
$f'(a') \meq f(a')$.  Moreover, for all functions $f_1: A_1 \to B_1$
and $f_2: A_2 \to B_2$, if $A_1 \orth A_2$, then $f_1 \fu f_2$ denotes
the union of $f_1$ and $f_2$ as graphs.

For any syntactic entity ranged over by a meta variable $X$, with
variables ranged over by $x$, the notation $\sbst{x_1 \repl X_1,
  \ldots, x_n \repl X_n}$ (for $n \geq 1$) denotes a
\intro{substitution} function $\sub$ that maps $x_i$ to $X_i$ for $1
\leq i \leq n$, and maps all other variables to themselves.  The
identity substitution is written $\identity$. The application of a
substitution to a syntactic entity with bindings must use standard
techniques to avoid variable capture. The domain of this substitution
is the set of all variables, and its \intro{support} $\supp{\sub}$ is
$\ens{x \alt x \neq \sub(x)}$.  Substitutions are required to have
finite support.  Accordingly, the \intro{cosupport} is defined by
$\cosupp{\sub} \meq \ens{\sub(x) \alt x \in \supp{\sub}}$. For all
substitutions $\sub_1$ and $\sub_2$, if $\supp{\sub_1} \orth
\supp{\sub_2}$, we define their disjoint union $\sub_1 \fu \sub_2$ by
$(\sub_1 \fu \sub_2)(x) \meq \sub_1(x)$ for all $x \in \supp{\sub_1}$,
$(\sub_1 \fu \sub_2)(x) \meq \sub_2(x)$ for all $x \in \supp{\sub_2}$,
and $(\sub_1 \fu \sub_2)(x) \meq x$ for all $x \notin (\supp{\sub_1}
\du \supp{\sub_2})$.  (This overloads the previous notation $f_1 \fu
f_2$ for functions with disjoint domains.)  For all substitutions
$\sub_1$ and $\sub_2$, we write $\sub_1(\sub_2)$ for the unique substitution
of support $\supp{\sub_2}$ such that for all $x \in \supp{\sub_2}$,
$\sub_1(\sub_2)(x) \meq \sub_1(\sub_2(x))$.  It is in general
different from the composition $\sub_1 \rond \sub_2$, since if $x \in
\supp{\sub_1} \setminus \supp{\sub_2}$, $(\sub_1 \rond \sub_2)(x) \meq
\sub_1(x)$, whereas $(\sub_1(\sub_2))(x) \meq x$.

\subsection{Syntax}
\label{subsection-syntax}

\begin{figure}
\begin{framed}
$$\begin{array}{l@{\quad}rcll}
\mbox{Variable:} &x & \in  & \Vars & \\
\mbox{Name:}     &\X & \in & \Namesfun \\[8pt]
\mbox{Expression:}
& \e \in \expr 
  & \bnf & x \alt \fun x.\e \alt \e_1~\e_2 & \textrm{$\lambda$-calculus}\\
&& \alt & \record{\s} \alt \e \rsel \X & \textrm{Record operations} \\
&& \alt & \letrecin{\bb}{\e} & \textrm{Recursive definitions} \\[8pt]
\mbox{Record row:} &\s & \bnf & \emptysequence \alt \X \eql x, \s \\
\mbox{Binding:}    &\bb & \bnf & \emptysequence \alt x \is \e, \bb \\[8pt]
\mbox{Size indication:}
& \is & \bnf & \ex{n} \alt \uu & \textrm{($n$ a natural number)}
\end{array}$$
\end{framed}
\caption{Syntax of $\lambdaab$}
\label{figure-syntax-lambdaab}
\end{figure}

The syntax of $\lambdaab$ is defined in
Figure~\ref{figure-syntax-lambdaab}. 
The meta-variables $\X$ and $x$ range over names and variables,
respectively.  Variables are used in binders, as usual. Names are used
for labeling record fields.
The metavariables for other syntactic entities are in
lowercase, in order to ease the distinction with the metavariables
for syntactic entities of the target language
(Section~\ref{section-lambdaalloc}), which will be in upper case.
The syntax includes the $\lambda$-calculus: variable $x$,
abstraction $\fun x . \e$, and application $\e_1~\e_2$. 
The language also features records, record selection $\e \rsel \X$ and
a binding construct written $\letrec$. By convention, the $\letrec$
construct has lowest precedence, so that for instance
$\letrecin{\bb}{\e_1~\e_2}$ means $\letrecin{\bb}{(\e_1~\e_2)}$.  In
a $\letrecin{\bb}{\e}$ expression, $\e$ is called the \intro{body}.
To simplify the formalization and without loss of expressiveness,
records are restricted to contain only variables, i.e., be of the
shape $\record{\X_1 \eql x_1, \ldots, \X_n \eql x_n}$.
\intro{Bindings} $\bb$ have the shape $x_1 \is_1 \e_1, \ldots, x_n \is_n
\e_n$, where arbitrary expressions are syntactically allowed as the
right-hand sides of definitions, and every definition is annotated
with a \intro{size indication} $\is$.  A size indication can be either
the unknown size indication $\uu$, or a known size indication
$\ex{n}$, where $n$ is a natural number. We write $\emptysequence$ for
the empty binding.

\paragraph{Implicit syntactic constraints} In what follows, we
implicitly restrict ourselves to record rows, bindings and expressions
satisfying the following conditions:
\begin{enumerate}
\item Record rows do not define the same name twice;
\item Bindings do not define the same variable twice;
\item Bindings do not contain \emph{forward references} to definitions
  of unknown size, in the sense made precise next.
\end{enumerate}

\begin{figure}
\begin{framed}
$$\begin{array}{l@{{} \meq {}}l@{\qquad}l@{{} \meq {}}l}
\FV{x} & \ens{x} &
\FV{\fun x . \e} & \FV{\e} \setminus \ens{x} \\
\FV{\e_1\ \e_2} & \FV{\e_1} \cup \FV{\e_2} &
\FV{\record{\s}} & \FV{\s} \\
\FV{\e.\X} & \FV{\e} &
\FV{\letrecin{\bb}{\e}} & (\FV{\bb} \cup \FV{\e}) \setminus \dom{\bb} \\
\FV{\bb} & \Union{(x \is \e) \in \bb}{\ens{x} \cup \FV{\e}} &
\FV{\s} & \ens{\s (X) \alt X \in \dom{\s}}
\end{array}$$
\end{framed}
\caption{Free variables in $\lambdaab$}
\label{figure-scope}
\end{figure}

The \intro{free variables} $\FV{\e}$ of expressions, bindings, and
record rows are defined inductively by the rules in
Figure~\ref{figure-scope}.  In a $\letrec$ binding $\bb \meq (x_1
\is_1 \e_1, \ldots, x_n \is_n \e_n)$, we say that there is a
\intro{forward reference} of $x_i$ to $x_j$ if $i \leq j$ and $x_j \in
\FV{\e_i}$.  Condition~3 requires that for all bindings $\bb$ and
forward reference of $x_i$ to $x_j$ in $\bb$, the size indication
$\is_j$ is $\ex{n}$ for some $n$.  This is consistent with the immediate
in-place update scheme, where no blocks are pre-allocated for
definitions of unknown size, so previous definitions must not refer to
them.

Finally, taking advantage of conditions~1 and~2 above, we implicitly
view record rows as finite functions from names to variables and
bindings as finite functions from variables to expressions, and use
standard notations for domain, codomain, application, etc.
Also, we write $\s_1,\s_2$
for the concatenation of $\s_1$ and $\s_2$, and similarly for
bindings.  Finally, we implicitly view records and bindings as sets of
pairs $(\X,x)$ (resp.~of triples $(x, \is, \e)$), for example to write
$(\X \eql x) \in \s$ (resp.~$(x \is \e) \in \bb$).

\paragraph{Structural equivalence} We consider expressions
equivalent up to $\alpha$-conversion\footnote{The notion
  of structural equivalence could include reordering of record fields, but
  we do not need it, so we just consider $\alpha$-equivalence.}, i.e.,
renaming of bound variables, in functions and $\letrec$ expressions.
In the following, to avoid ambiguity, we call \emph{raw} expressions
not considered up to $\alpha$-conversion.  Let $\meq$ denote equality
of raw expressions and $\aeq$ denote equality modulo $\alpha$
conversion.


%

\subsection{Dynamic semantics}
\label{subsection-semantics-lambdaab}

We now define the dynamic semantics of $\lambdaab$.  
Figure~\ref{figure-values-lambdaab} defines $\lambdaab$ values to be
variables, functions, and records. 

\subsubsection{Overview: sizes and recursive definitions}

We have seen that $\letrec$-bound definitions can
be annotated with natural numbers representing their sizes. The role
of these size indications is to declare in advance the expected sizes
of the memory blocks representing the values of definitions.
Technically, they will be required to match the size of allocated
blocks in the sense of our target calculus.  For definitions that are
not forward-referenced from previous definitions, there is no need for
annotations.  

In $\lambdaab$, during the evaluation of a binding, if the currently
evaluated definition is expected to have size $n$, then it must
evaluate to a non-variable value whose size equals~$n$.  Otherwise,
evaluation gets stuck.

\begin{Hypothesis}[Size in $\lambdaab$]
\label{hypothesis-size-ab}
We assume given a partial function $\Sizeabfun$ from $\lambdaab$ values
to natural numbers, defined exactly on $\values \setminus \Vars$. 
\end{Hypothesis}

An evaluated definition not matching its size indication is considered
an error, in the sense that it prevents further reductions.  Thus,
only \intro{size-respecting} bindings $\bv$, as defined in
Figure~\ref{figure-values-lambdaab}, are considered fully evaluated.

\begin{figure}
\begin{framed}
$$\begin{array}{ll@{\hspace*{12pt}}lcll}
\mbox{Value:} &
&\val \in \values 
& \bnf & x \alt \fun x.\e \alt \record{\sv} \vspace{8pt} \\
\mbox{Answer:} &
&\answer \in \answers
& \bnf & \val \alt \letrecin{\bv}{\val} \vspace{8pt} \\
\mbox{Size-respecting binding:} &
&\bv & \bnf & \emptysequence \\
&&& \alt & x \uu \val, \bv \\
&&& \alt & x \ex{n} \val, \bv & \textrm{where $\Sizeab{\val} \meq n$}
\end{array}$$
\end{framed}
\caption{Values and answers in $\lambdaab$}
\label{figure-values-lambdaab}
\end{figure}

Note that size-respecting bindings define only values.  
The intuition is that, given a
definition $(x \ex{n} \e)$, this forces the topmost block of the value
of $\e$ to be determined by previous definitions.  For instance,
suppose that $\Sizeab{\record{\X \eql x}} \meq n$. Then, the binding
$(y \ex{n} \record{\X \eql x}, z \ex{n} y)$ is not fully evaluated,
but we will see below that it evaluates correctly to $(y \ex{n}
\record{\X \eql x}, z \ex{n} \record{\X \eql x})$.  On the contrary,
the binding $(z \ex{n} y, y \ex{n} \record{\X \eql x})$ is invalid:
$y$ can not be replaced with its value, according to the reduction
relation defined below. (Such a reduction step could not be
implemented by immediate in-place update as depicted in
Figure~\ref{figure-picture-refined-inplace-updating-trick}.)

Besides the non-standard notion of size, the dynamic semantics of
$\lambdaab$ is unusual in its handling of mutually recursive
definitions, which is adapted from the equational theory of
Ariola and Blom \cite{Ariola98}. There is no rule for eliminating $\letrec$:
evaluated bindings remain at top-level in the expression and also in
evaluation \emph{answers}, as defined in
Figure~\ref{figure-values-lambdaab}.  This top-level binding serves as
a kind of heap or recursive evaluation environment.  An answer
$\answer$ is defined to be a value, possibly surrounded by an
evaluated, size-respecting binding. It thus may have the shape
$\letrecin{\bv}{\val}$.

The dynamic semantics of $\letrec$ relies on five fundamental
equations, which resemble the rules used by
Wright and Felleisen \cite{Felleisen92}.  We start with an informal
presentation of these equations using 
\intro{contexts} $\anycont$, i.e., terms with a
hole $\trou$. Context application $\appcontext{\anycont}{\e}$ is
textual, possibly capturing replacement of $\trou$ with $\e$ in
$\anycont$.  The rules rely on additional conditions defined later to
(1) avoid variable captures and (2) enforce the reduction strategy of
the language, but are roughly as follows.
\begin{enumerate}
\item The first equation is \intro{lifting}. It lifts a $\letrec$ node
  up one level in an expression. An expression of the
  shape $\e_1~(\letrecin{\bb}{\e_2})$ is equated with
  $\letrecin{\bb}{(\e_1~\e_2)}$.

\item The second equation is \intro{internal merging}. 
In a  binding, when one of the definitions starts with another
binding, then this binding can be merged with the enclosing one. An
expression of the shape
$\letrecin{\bb_1, x \eql (\letrecin{\bb_2}{\e_1}), \bb_3}{\e_2}$ 
is equated with
$\letrecin{\bb_1, \bb_2, x \eql \e_1, \bb_3}{\e_2}$.  

\item The third equation is \intro{external merging},
which merges two consecutive  bindings.  An expression of the
shape
$\letrecin{\bb_1}{\letrecin{\bb_2}{\e}}$ 
is equated with
$\letrecin{\bb_1, \bb_2}{\e}$. 

\item The fourth equation, \intro{external substitution}, replaces
  variables defined in an enclosing binding with their definitions.
  Given a context $\anycont$, an expression of the shape
  $\letrecin{\bb}{\appcontext{\anycont}{x}}$ is equated with $\letrec
  {\bb} \inlet {\appcontext{\anycont}{\e}}$, if $x \eql \e$ appears in
  $\bb$.

\item The last equation, \intro{internal substitution}, replaces
  variables defined in the same binding with their definitions.  Given
  a context $\anycont$, an expression of the shape
$\letrecin{\bb_1, y \eql \appcontext{\anycont}{x}, \bb_2}{\e_1}$
is equated with
$\letrec {\bb_1, y \eql \appcontext{\anycont}{\e_2}, \bb_2} \inlet {\e_1}$
if $x \eql \e_2$ appears in $\bb_1, y \eql \anycontext{x}, \bb_2$. 
\end{enumerate}

The issue is how to arrange these operations to make the evaluation
deterministic and to ensure that it reaches the answer when it exists.
Our choice can be summarized as follows.  First, bindings that are not
at top-level in the expression must be lifted before their evaluation
can begin. Thus, only the top-level binding can be evaluated.  As soon
as one of its definitions gets evaluated, evaluation can proceed with
the next one, or with the body if there is no definition left. If
evaluation encounters a binding inside the considered expression, then this
binding is lifted up to the top level of the expression, or just
before the top-level binding if there is one. In this case, it is
merged with the latter, internally or externally, according to the
context. External substitution is used to replace a variable in
\emph{dereferencing} position (like $x$ in $x.\X$ or $x~\val$, see
the precise definition of dereferencing contexts below) with its
value, fetched from the top-level binding.  Internal substitution is
used similarly, but inside the top-level binding, and only from left
to right (i.e., when the copied definition comes from the left of the
current evaluation point).

\begin{Remark}[Policy on substitution and call-by-value]
\label{remark-substitution-and-cbv}
The substitution rules only replace one occurrence of a variable at a
time, which has to be in destructive position. This strategy
w.r.t.~substitution, called \intro{destruct-time} by
Sewell et al \cite{Bierman08}, does not contradict the fact that $\lambdaab$ is
call-by-value. Indeed, only values are copied, and any expression
reached by the evaluation is immediately evaluated. The fact that
evaluated definitions are not immediately substituted with their
values in the rest of the expression is rather a matter of
presentation. Notably, this presentation allows $\lambdaab$ to
properly represent recursive data structures, as shown in
Section~\ref{subsection-examples} and Figure~\ref{figure-example-5}.
\end{Remark}

To implement our strategy, we remark that evaluation should not be the
same at top-level and inside an evaluation context.  For example,
consider $\e \aeq ((\letrecin{x \uu \e_0}{x~y})~z)$, where $\e_0$
reduces to $\e_1$.  According to the informal specification above,
before the evaluation of $\e_0$ can start, the binding should first be
lifted to the top level to obtain $\e' \aeq (\letrecin{x \uu
  \e_0}{(x~y~z)})$. So, our reduction relation should not respect the
usual rule saying that for any $\e_0$ and $\e_1$, if $\e_0 \reduct
\e_1$, then $\context{\e_0} \reduct \context{\e_1}$
for any evaluation context $\cont$.
This leads us to define two relations: the
\intro{subreduction relation} $\ccontraction,$ handling reductions
inside expressions, and the \intro{reduction relation} $\reduct$,
handling top-level reductions.  We write $\ccontraction^+$
(resp.~$\ccontraction^*$) for the transitive (resp.~transitive
reflexive) closure of the relation $\ccontraction$, and similarly for
$\reduct$.

\begin{figure}
\begin{framed}
\begin{tabular}{*{2}{p{.45\linewidth}}}
\centering $\begin{array}[t]{rcll}
\multicolumn{4}{l}{\mbox{Lift context:}}\\
\ \ \lcont &\bnf& \e~\trou \alt \trou~\val 
\alt \trou \rsel \X &\\
\multicolumn{3}{l}{\mbox{Nested lift context:}}\\
\ \ \fcont &\bnf& \trou \alt \e~\fcont \alt \fcont~\val 
\alt \fcont \rsel \X \\
\multicolumn{4}{l}{\mbox{Evaluation context:}}\\
\ \ \cont &\bnf& \fcont \\
              & \alt & \letrecin{\bv}{\fcont} \\
              & \alt & \letrecin{\bv, x \is \fcont, \bb}{\e}
\end{array}$ &
\centering
$\begin{array}[t]{rcll}
\multicolumn{3}{l}{\mbox{Binding context:}}\\
\ \ \bcont &\bnf& \bv, x \is \trou, \bb \\
\multicolumn{4}{l}{\mbox{Nested dereferencing context:}}\\
\ \ \ncont &\bnf& \trou~\val
\alt \trou \rsel \X \\
& \alt & 
\e~\ncont \alt \ncont~\val 
\alt \ncont \rsel \X \\
\multicolumn{4}{l}{\mbox{Dereferencing context:}}\\
\ \ \dcont &\bnf& \ncont \\
              & \alt & \letrecin{\bv}{\ncont} \\
              & \alt & \letrecin{\bv, x \is \ncont, \bb}{\e} \\
               & \alt & \letrecin{\bcontof{\ex{n}}}{\e}
\end{array}$
\end{tabular}
\end{framed}
\caption{Evaluation contexts of $\lambdaab$}
\label{figure-evaluation-contexts-lambdaab}
\end{figure}

\begin{figure}
\begin{framed}
Alpha equivalence:
  \begin{mathpar}
%
\inferrule{
\e \aeq \e'
}{
\e\ \fcont \aeq \e'\ \fcont
}
\and
%
\inferrule{
\val \aeq \val'
}{
\fcont\ \val \aeq \fcont\ \val'
}
\and
%
\inferrule{
\bv \aeq \bv'
}{
\letrecin{\bv}{\fcont} \aeq
\letrecin{\bv'}{\fcont}
}
\and
%
\inferrule{
\bv \aeq \bv' \\
\bb \aeq \bb' \\
\e \aeq \e'
}{
(\letrecin{\bv, x \is \fcont, \bb}{\e}) \aeq
(\letrecin{\bv', x \is \fcont, \bb'}{\e'})
}
\and
%
\inferrule{
\fcont \aeq \fcont'
}{
\context{\fcont} \aeq \context{\fcont'}
}
\and
%
\inferrule{
\e \aeq \e'
}{
(\bb_1, x \is \e, \bb_2) \aeq (\bb_1, x \is \e', \bb_2)
}
\and
\end{mathpar}

$\begin{array}{ll@{{} \meq {}}l}
\mbox{Free variables:} &
\FV{\trou} & \emptyset \\
&\FV{\e\ \fcont} & \FV{\e} \cup \FV{\fcont} \\
&\FV{\fcont\ \val} & \FV{\fcont} \cup \FV{\val} \\
&\FV{\fcont.\X} & \FV{\fcont} \\
&\FV{\letrecin{\bv}{\fcont}} & \FV{\bv} \cup \FV{\fcont} \\
&\FV{\letrecin{\bv, x \is \fcont, \bb}{\e}} & \ens{x} \cup \FV{\bv,\bb} \cup \FV{\fcont} \cup \FV{\e} \\[3mm]
\mbox{Captured variables:} &
\Captcont{\letrecin{\bv}{\fcont}} & \dom{\bv} \\
&\Captcont{\letrecin{\bv, x \is \fcont, \bb}{\e}} & \ens{x} \cup \dom{\bv,\bb} \\
&\Captcont{\fcont} & \emptyset 
\end{array}$
\end{framed}
\caption{Structural equivalence of $\lambdaab$ evaluation contexts}
\label{figure-scope-contexts-ab}
\end{figure}

\subsubsection{The subreduction relation} 
First, we define subreduction in
Figure~\ref{figure-evaluation-lambdaab}, using notions defined in
Figures~\ref{figure-evaluation-contexts-lambdaab}
and~\ref{figure-scope-contexts-ab}.  It is first defined on raw
expressions, then lifted to $\alpha$-equivalence classes of
expressions by the usual rule

\begin{mathpar}
\inferrule{
\e_1 \aeq \e'_1 \\
\e'_1 \ccontraction \e'_2 \\
\e'_2 \aeq \e_2
}{
\e_1 \ccontraction \e_2
}
\end{mathpar}

Record projection selects the appropriate field in the record (rule
\RuleSelectab).  The application of a function $\fun x . \e$ to a
value $\val$ reduces to the body of the function where the argument
has been $\letrec$-bound to $x$ (rule \RuleBetaab).  Rule \RuleLiftab
describes how bindings are lifted up to the top of the term.
\intro{Lift contexts} $\lcont$ are defined in
Figure~\ref{figure-evaluation-contexts-lambdaab}.  Rule \RuleLiftab
states that an expression of the shape
\(\lcontext{\letrecin{\bb}{\e}}\) subreduces to
\(\letrecin{\bb}{\lcontext{\e}}\), provided no variable capture
occurs. Alpha-equivalence is defined over contexts as follows: all
variables may be $\alpha$-renamed, except those that have $\trou$ in
their scope. More formally, $\alpha$-equivalence for evaluation
contexts is the smallest equivalence relation over evaluation
contexts respecting the rules in
Figure~\ref{figure-scope-contexts-ab}. In the same figure, we define
the \intro{captured} variables $\Captcont{\cont}$ of an evaluation
context $\cont$, and the free variables of an evaluation context.  We
have $\Captcont{\cont} \subseteq \FV{\cont}$ for all $\cont$.  

\begin{Remark}[Evaluation order]
  Function applications are evaluated from right to left.  This
  nonstandard choice is explained in
  Remark~\ref{remark-non-determinism-and-evaluation-order}, in
  light of the semantics of the target language $\lambdaalloc$.  The
  results of the paper can be adapted to a left-to-right evaluation
  setting with some additional work.
\end{Remark}

\begin{figure}
\begin{framed}
\begin{tabular}{@{}l@{}}
{ \textrm{Subreduction rules ($\ccontraction$)}}\vspace{8pt}\\
\begin{mathpar}
\inferrule{}{\record{\sv} \rsel \X
\ccontraction \sv(\X)}
~~(\RuleSelectab)
\and
\inferrule{x \notin \FV{\val}}{
(\fun x.\e)~\val \ccontraction \letrecin{x \uu \val}{\e}}
~~(\RuleBetaab)
\and
\inferrule{
\dom{\bb} \orth \FV{\lcont}}{
\lcontext{\letrecin{\bb}{\e}}
\ccontraction
\letrecin{\bb}{\lcontext{\e}}}
~~(\RuleLiftab)
\end{mathpar}\vspace{8pt}\\
{ \textrm{Reduction rules ($\reduct$)}}\vspace{8pt}\\
\begin{mathpar}
\inferrule{\e \ccontraction \e'}{
\context{\e} \reduct \context{\e'}}
~~(\RuleContextab)
\and
\inferrule{
\dom{\bb_1} \orth (\ens{x} \cup \FV{\bv, \bb_2} \cup \FV{\e'})
}{
(\letrecin{\bv, x \is (\letrecin{\bb_1}{\e}), \bb_2}{\e'})
~~{} \reduct
(\letrecin{\bv, \bb_1, x \is \e, \bb_2}{\e'})}
~~(\RuleIMab)
\and 
\inferrule{
\dom{\bb} \orth \FV{\bv}}
{(\letrecin{\bv}{\letrecin{\bb}{\e}})
\reduct 
\letrecin{\bv, \bb}{\e}}
~~(\RuleEMab)
\and
\inferrule{}{
\dcontext{x}
\reduct
\dcontext{\dcont(x)}}
~~(\RuleSubstab)
\end{mathpar}
\end{tabular}
\end{framed}
\caption{Dynamic semantics of $\lambdaab$}
\label{figure-evaluation-lambdaab}
\end{figure}

\subsubsection{The reduction relation}
The reduction relation is defined in
Figure~\ref{figure-evaluation-lambdaab}.  It is first defined on
raw expressions, then lifted to $\alpha$-equivalence classes
of expressions by the usual rule
\begin{mathpar}
\inferrule{
\e_1 \aeq \e'_1 \\
\e'_1 \reduct \e'_2 \\
\e'_2 \aeq \e_2
}{
\e_1 \reduct \e_2
}\cdot
\end{mathpar}

Rule \RuleContextab extends the subreduction relation (as a relation
over raw expressions) to any evaluation context. As defined in
Figure~\ref{figure-evaluation-contexts-lambdaab}, we call a
\intro{nested lift context} $\fcont$ a series of lift contexts.
Moreover, we call a \intro{binding context} $\bcontof{\is}$ of size
$\is$ a binding $(\bv, x \is \trou, \bb)$ where the context hole
$\trou$ corresponds to the next definition to be evaluated, and this
definition is annotated by $\is$. An \intro{evaluation context}
$\cont$ is a nested lift context, possibly appearing as the next
definition to evaluate in the top-level binding, or enclosed inside a
fully evaluated top-level binding. 
Our unusual, staged formulation of evaluation contexts enforces the
determinism of the reduction relation w.r.t. bindings: evaluation
never takes place inside or after a binding, except the top-level one.
Other bindings inside the expression first have to be lifted to the
top by rule \RuleLiftab, then be merged with the top-level binding, if
any, by rules \RuleEMab and \RuleIMab (respectively for external and
internal merging). If the top-level binding is of the shape $\bv, x
\is (\letrecin{\bb_1}{\e}), \bb_2$, rule \RuleIMab allows to merge
$\bb_1$ with it, obtaining $\bv, \bb_1, x \is \e, \bb_2$.  When an
inner binding has been lifted to the top level, if there is already a
top-level binding, then the two bindings are merged together by rule
\RuleEMab. This implements the strategy informally described above.

Finally, rule \RuleSubstab describe how the variables defined by the
top-level binding are replaced with their values when needed, i.e.,
when they appear in a \intro{dereferencing context}, as defined in
Figure~\ref{figure-evaluation-contexts-lambdaab}.  Dereferencing
contexts may take two forms.  First, they can be binding contexts of
known size $\letrecin{\bv, x \ex{n} \trou, \bb}{\e}.$ In the immediate
in-place update compilation scheme, any definition of known size
yields an allocation of a dummy block, which has to be updated.  This
is reflected here by requiring that in definitions of the shape $(x
\ex{n} y)$, $y$ be eventually replaced with a non-variable value of
size~$n$.  Dereferencing contexts can also be \intro{nested
  dereferencing contexts}, i.e., function applications $\trou\ \val$
or record field selection $\trou.\X$, wrapped by an evaluation
context, as defined in
Figure~\ref{figure-evaluation-contexts-lambdaab}.  Therefore, in $\lambdaab$,
the value of a variable is copied only when needed for function
application or record selection (or in-place update, implicitly). The
value of a variable $x$ is found in the current evaluation context, as
formalized by the following notion of access in evaluation contexts.

\begin{Definition}
  Define $\begin{array}[t]{lcl}
    \Binding{\fcont} &\aeq& \emptysequence \\
    \Binding{\letrecin{\bv}{\fcont}} &\aeq& \bv \\
    \Binding{\letrecin{\bv, x \is \fcont, \bb}{\e}} &\aeq& \bv, \\[.5em]
  \end{array}$

\noindent  The value $\cont(x)$ of $x$ in $\cont$ is $(\Binding{\cont})(x)$, when the latter is defined.
\end{Definition}


\begin{Lemma}[Determinism of evaluation]
The $\reduct$ relation is deterministic.
\end{Lemma}

\begin{pf}
  We prove the result for raw expressions first, and then extend
  it to $\alpha$-equivalence classes.  First, subreduction is
  obviously deterministic, on raw expressions as well as on
  $\alpha$-equivalence classes. Furthermore, both on raw
  expressions and on $\alpha$-equivalence classes, the reduction rules
  do not overlap, so we only have to prove that each rule is
  deterministic.

  First consider the case of raw expressions.  For all
  evaluation contexts $\cont_1,\cont_2$ and subreduction redexes
  $\e_1$ and $\e_2$, if $\appcontext{\cont_1}{\e_1} \meq
  \appcontext{\cont_2}{\e_2}$, we show that $\cont_1 \meq \cont_2$. This is
  shown in three steps: for lift contexts (by case analysis), nested
  lift contexts (by induction), and evaluation contexts (by case
  analysis).  Hence, rule \RuleContextab is deterministic.  Similarly,
  rule \RuleSubstab is deterministic.

  Consider now the case of $\alpha$-equivalence classes. Let a
  \intro{renaming} $\vren$ be a substitution function (as defined in
  Section~\ref{subsection-notation}) from variables to variables, and
  let $\vren (\e)$ denote capture-avoiding substitution in the usual
  sense. For all evaluation contexts $\cont_1,\cont_2$ and sub
  reduction redexes $\e_1$ and $\e_2$, if $\appcontext{\cont_1}{\e_1}
  \aeq \appcontext{\cont_2}{\e_2}$, then there exists a renaming
  $\vren$, such that $\supp{\vren} \subseteq \Captcont{\cont_1}$ and
  $\vren(\cont_1) \aeq \cont_2$ and $\vren(\e_1) \aeq \e_2$.  This
  entails that rule \RuleContextab is deterministic. We proceed
  similarly for rule \RuleSubstab.
\end{pf}

\begin{Definition}[Faulty $\lambdaab$ expression]
  A \intro{faulty} $\lambdaab$ expression is an expression whose
  reduction gets stuck on an expression that is not an answer.  By
  determinism, a non-faulty expression is an expression whose
  evaluation either does not terminate or reaches an answer.  
\end{Definition}

We now characterize faulty expressions, using the following notion of
\emph{decomposition} of an expression $\e$: a \emph{decomposition} of
an expression $\e$ is a pair $(\cont, \e')$ such that $\e \hts
\context{\e'}$.  (We consider pairs $(\cont, \e)$ modulo renaming of
the captured variables of $\cont$, hence decomposition is well-defined
on $\alpha$-equivalence classes of expressions.)

Let us now define an ordering over decompositions. Decompositions
$(\cont, \e')$ induce occurrences in the abstract syntax tree of
expressions, i.e., paths from its root to the designated occurrence of
$\e'$. This assignment is injective, i.e., these paths characterize
decompositions. However, it is not onto since some paths do not
correspond to any evaluation context.  Given two decompositions
$(\cont, \e)$ and $(\cont', \e')$ of some given $\e_0$, corresponding
to paths $p$ and $p'$, consider their maximal common prefix
$p''$.  We say that $p \lesst p'$ when either:
\begin{itemize}
\item $p'' \meq p$, or
\item $p'$, after $p''$, turns left in an application, i.e., $p''$
  corresponds to a decomposition $(\cont'', (\fcontext{\e}\ \val))$
  and $\cont' \hts \appcontext{\cont''}{\fcont\ \val}$ (the other
  decomposition thus has $\cont \hts
  \appcontext{\cont''}{\fcontext{\e}\ \trou}$), or
\item $p'$, after $p''$, goes further in the top-level binding than
  $p$, i.e., $\cont \hts (\letrecin{\bv, x \is \trou, \bb}{\e_1})$, $\e$
  is a value (of the expected size if needed), and $\cont'$ has shape
  $\letrecin{\bv, x \is \val, \bv', y \is \fcont, \bb'}{\e_1}$ or
  $\letrecin{\bv, \bv'}{\fcont}$.
\end{itemize}
This relation $\lesst$ defines a total ordering on the set of
decompositions of any expression $\e$, which furthermore has a maximal
element -- the decomposition turning left in applications when
possible, and going as far as possible in the top-level binding.
Using this notion, we prove the following characterization.

\begin{Proposition}
\label{proposition-faulty-ab}
For all $\e$, the following are equivalent:
\begin{enumerate}\item 
  $\e$ is faulty;
\item 
  $\e$ reduces to an expression $\dcontext{\val}$ in normal form, such
  that if $\dcont \aeq \letrecin{\bcontof{\ex{n}}}{\e}$ for some $n$,
  $\bcontof{\ex{n}}$, and $\e$, then $\Sizeab{\val}$, if defined, is
  not $n$;
\item 
  $\e$ reduces to an expression $\e_0$ such that:
  \begin{itemize}
  \item $\e_0 \hts \dcontext{x}$, with $\dcont(x)$ undefined,
  \item $\e_0 \hts \letrecin{\bv, x \ex{n} \val, \bb}{\e'}$ with
    $\Sizeab{\val} \nmeq n$ and $\val \notin \Vars$,
  \item $\e_0 \hts \context{\record{\sv}~\val}$,
  \item $\e_0 \hts \context{\record{\sv}.\X}$ with $X \notin \dom{\sv}$,
  \item $\e_0 \hts \context{(\fun x. \e').\X}$.
  \end{itemize}
\end{enumerate}
  Moreover, for all $x$ and $\dcont$, $\dcont(x)$ is undefined if and
  only if
  \begin{itemize}
  \item either $x \notin \Captcont{\dcont}$,
  \item or $\dcont \hts \letrecin{\bv, x' \is \fcont, \bb}{\e'}$, with
    $x \in \ens{x'} \cup \dom{\bb}$.
  \end{itemize}
\end{Proposition}
\begin{pf}
  First, observe that all cases of $(3)$ are faulty, hence $(3)$
  implies $(1)$.  We now show that $(1)$ implies $(3)$.

  Consider an expression with a normal form $\e$ which is not an
  answer.  Consider its maximal decomposition $(\cont, \e')$ w.r.t.\
  the ordering $\lesst$.  The expression $\e$ is an answer exactly
  when $\e'$ is a value and $\cont$ is either empty or of the shape
  $\letrecin{\bv}{\trou}$. We proceed by case analysis on the other
  cases.
  
  If $\e'$ is not a value, then by maximality, it has the shape
  $\letrecin{\bb'}{\e''}$ for some $\bb'$ and $\e''$, and $\cont$ is
  not empty. But then one of rules \RuleIMab, \RuleEMab, and
  \RuleLiftab applies, contradicting the fact that $\e$ is in normal
  form.

  If $\e'$ is a value $\val$, then $\cont$ must have shape
  $\letrecin{\bv}{\fcont}$ or $\letrecin{\bv, x \is \fcont,
    \bb}{\e''}$.

  If $\fcont$ is not empty, then $\cont$ has the shape
  $\appcontext{\cont'}{\lcont}$.  Now, if $\lcont \aeq (\e\ \trou)$,
  the decomposition $(\cont, \e')$ cannot be maximal, since the
  decomposition $(\appcontext{\cont'}{\trou\ \val}, \e)$ is greater.
  Otherwise, if $\lcont \aeq (\trou\ \val')$, then we have
  $\appcontext{\cont'}{\val\ \val'}$ in normal form, hence either
  $\val$ is a variable undefined in $\cont'$, or is a record.
  Otherwise, $\lcont \aeq (\trou.\X)$, hence either $\val$ is a variable
  undefined in $\cont'$, or is a function, or is a record without
  an $\X$ field. All these cases are covered by $(3)$.

  If otherwise $\fcont$ is empty, then $\cont$ must have the 
  shape $\letrecin{\bv, x \is \trou, \bb}{\e''}$. But then, for 
  the decomposition $(\cont, \e')$ to be maximal, we must
  have ${\is} \meq {\ex{n}}$ for some $n$, and either
  \begin{itemize}
  \item $\val$ is a variable undefined in $\cont$ (first case of
    $(3)$), or
  \item $\Sizeab{\val}$ is defined and different from $n$ (second
    case).
  \end{itemize}

  Finally, to show the equivalence with $(2)$, all the cases of $(3)$
  are covered by $(2)$, so $(3)$ implies $(2)$, and the only
  possibility for an expression $\dcontext{\val}$ in normal form to be
  an answer is that $\dcont$ has the shape
  $\letrecin{\bcontof{\ex{n}}}{\e}$ with $\Sizeab{\val} \meq n$, so
  $(2)$ implies $(1)$.
\end{pf}
\begin{Remark}
  In $\lambdaab$, we restrict record values to contain only
  variables.  Actually, we could permit other kinds of values in
  record expressions, but not in record values, because it would break
  the properties of $\lambdaab$ w.r.t.~sharing. In particular, as we
  also mention in Section~\ref{subsection-examples}, the sharing
  properties of $\lambdaab$ make it directly extensible
  with mutable values.  If we allowed non-variable values in record
  values, then this would no longer be the case. 

  To see this, assume that $\lambdaab$ is extended with such record
  values and a ternary operator $\e.\X \shortleftarrow \e'$ for
  mutation of record fields. Then, consider $\e \aeq (\letrecin{x \uu
    \record{\X \eql \record{\Y \eql \val}}}{x.\X.\Y \shortleftarrow \val'})$.
  The evaluation of $\e$ is as follows: first, the record is copied,
  then its $\X$ field is projected, which gives $\letrecin{x \uu
    \record{\X \eql \record{\Y \eql \val}}}{\record{\Y \eql \val}.\Y
    \shortleftarrow \val'}$, which is impossible to rewrite to the
  expected result.

  In addition to this undesirable behavior, enriching $\lambdaab$ with
  non-variable values in record values would force us to considerably
  enrich the equational theory of our target language $\lambdaalloc$. 
  Indeed, $\lambdaalloc$ gives a rather fine-grained account of
  sharing, and we would have to add equations to reason modulo
  sharing. 
\end{Remark}

\subsection{Examples}
\label{subsection-examples}

In this section, we show examples of $\lambdaab$
reduction and give intuitions on important applications of
$\lambdaab$, namely mixin modules and recursive modules.  These
examples demonstrate the expressive power of $\lambdaab$,
compared to the recursion constructs of both ML and Scheme, and
also compared to the conference version of this
paper~\cite{Hirscho03b}.  Other possible applications include
encodings of objects following Boudol~\cite{Boudol04}. However,
$\lambdaab$ would have to be (straightforwardly) extended with mutable
records to support this encoding. 

\subsubsection{Basic examples}

\begin{figure}
\begin{framed}
  \begin{center}
  \begin{tabular}{lp{0.4\textwidth}}
    \multicolumn{1}{c}{\textrm{Expression}} & Comments \\ \hline \\[-0.5em]
    $\letrecin{x \uu \fun y . y}{x~x}$
    & Not a valid answer because, $x~x$ is not a value, 
      and the only possible reduction is by rule \RuleSubstab. \\
    \multicolumn{1}{c}{$\downarrow$} & \\[0.5em]
    $\letrecin{x \uu \fun y . y}{(\fun y . y)~x}$
    & Only the first occurrence of $x$ is substituted.
      We then apply rule \RuleBetaab. \\
    \multicolumn{1}{c}{$\downarrow$} & \\[0.5em]
    $\begin{array}[t]{l}
      \letrec~x \uu \fun y .  y ~\inlet \\
      \letrecin{y \uu x}{y}
    \end{array}$
    & Not yet a valid answer. We apply rule \RuleEMab. \\[0.5em]
    \multicolumn{1}{c}{$\downarrow$} & \\[0.5em]
    $\begin{array}[t]{l}
      \letrecin{x \uu \fun y . y, y \uu x}{y} 
    \end{array}$
    & The binding is now size-respecting, because of the $\uu$.
  \end{tabular}
  \end{center}
\end{framed}
\caption{Substitution and function application}
\label{figure-example-1}
\end{figure}


We start with small examples to give some intuition on the
semantics. First, as noted in
Remark~\ref{remark-substitution-and-cbv}, substitution occurs at
destruct-time in $\lambdaab$, following the terminology of
\cite{Bierman08}. This means that substitution of an occurrence of
a variable is only performed when this occurrence has to be replaced
with a non-variable value in order for the evaluation to continue. 
This is illustrated in Figure~\ref{figure-example-1}, which shows an
example of substitution at function application time.  The first
expression is partitioned into $\dcont \aeq
\letrecin{x \uu \fun y . y}{\trou~x}$ and~$x$. 

\begin{figure}
\begin{framed}
  \begin{center}
  \begin{tabular}{lp{0.4\textwidth}}
    \multicolumn{1}{c}{\textrm{Expression}} & Comments \\ \hline \\[-0.5em]
    $\begin{array}[t]{l}
      \letrec 
      \begin{array}[t]{l} 
        z \uu x~x, \\
        x \ex{n} \fun y . y 
      \end{array} \\
      \inlet~z \\
      \hspace{3em} \diagup\!\!\!\!\!\!\!\!\downarrow
    \end{array}$
    & The forward reference is syntactically correct (even if  $n \neq \Sizeab{\fun y . y}$), but
      the value of $x$ cannot be copied, because it would be from right to left.
      This is consistent with the in-place update compilation 
      scheme sketched in Section~\ref{subsection-backpatching-ipu}.
    \\[0.5em]
    \hline \\[-0.5em]
    $\begin{array}[t]{l}
      \letrec 
      \begin{array}[t]{l} 
        x \ex{n} \fun y . y, \\
        z \uu x~x 
      \end{array} \\
      \inlet~z \\[0.5em]
      \hspace{3em}\downarrow
    \end{array}$
    & The value of $x$ can be copied, but only if the size indication is correct, 
      otherwise
      the first definition is not considered valid. Note that
      the size indication is in fact not necessary here because $x$ is not forward 
      referenced. 
    \\
    $\begin{array}[t]{l}
      \letrec 
      \begin{array}[t]{l} 
        x \ex{n} \fun y . y, \\
        z \uu (\fun y . y)~x 
      \end{array} \\
      \inlet~z \\
      \hspace{3em}\vdots
    \end{array}$
    & 
\end{tabular}
\end{center}
\end{framed}
\caption{Forward references}
\label{figure-example-2}
\end{figure}


Figure~\ref{figure-example-2} illustrates the left-to-right evaluation
of bindings in $\lambdaab$ and the semantics of size indications. In
particular, it emphasizes the fact that if a size indication turns out
to be wrong, then the reduction is stuck.  With respect to
compilation, this models the fact that in the in-place update method,
pre-allocated blocks should not be updated with larger blocks,
otherwise execution might go wrong.  In the second example of
Figure~\ref{figure-example-2}, 
whose evaluation is correct, the first
expression is partitioned into $\dcont \aeq \letrecin{x \ex{n} \fun y
. y, z \uu \trou~x}{z}$ and $x$. 

\begin{figure}
\begin{framed}
  \begin{center}
  \begin{tabular}{lp{0.4\textwidth}}
    \multicolumn{1}{c}{\textrm{Expression}} & Comments \\ \hline \\[-0.5em]
    $\begin{array}[t]{l}
      \letrec 
      \begin{array}[t]{l} 
        y \uu \record{\X \eql \record{}}, \\
        z \uu y
      \end{array} \\
      \inlet~z
    \end{array}$
    & The definition $z \uu y$ respects sizes, so the whole expression
      is an answer. \\
    \\ \hline \\[-0.5em]
    $\begin{array}[t]{l}
      \letrec 
      \begin{array}[t]{l} 
        y \uu \record{\X \eql \record{}}, \\
        z \ex{n} y
      \end{array} \\
      \inlet~z \\
      \hspace{3em}\downarrow \\
    \end{array}$
    & The definition $z \ex{n} y$ does not respect sizes, so the expression
      reduces by rule \RuleSubstab.
    \\ 
    $\begin{array}[t]{l}
      \letrec 
      \begin{array}[t]{l} 
        y \uu \record{\X \eql \record{}}, \\
        z \ex{n} \record{\X \eql \record{}}
      \end{array} \\
      \inlet~z
    \end{array}$
    & We eventually reach an answer.
\end{tabular}
\end{center}
\end{framed}
\caption{Size indications and dereferencing contexts}
\label{figure-example-4}
\end{figure}


Figure~\ref{figure-example-4} shows a subtle point of the semantics. 
Namely, the size indications change the degree of sharing of
definitions, in case they are just variables. From
Figure~\ref{figure-evaluation-contexts-lambdaab}, we remark that a
binding context of the shape $x \ex{n} \trou$ is dereferencing. Therefore, if
it is filled with a variable, this variable has to be substituted
with its value in order for evaluation to continue. 
Figure~\ref{figure-example-4} provides two examples differing only by
one size indication. In the first case, the expression is a valid
answer. In the second case, at the level of compiled code, a block is
pre-allocated for $z$, which will eventually represent its value, so
we must update it: the value of $y$ is copied to this block. At the
source language level, this copying enables $\lambdaab$ to correctly
reflect sharing in the compiled code, and therefore makes it ready for
extension with mutable values. 

\begin{figure}
\begin{framed}
  \begin{center}
    \begin{tabular}{lp{0.35\textwidth}}
      \multicolumn{1}{c}{\textrm{Expression}} & Comments \\ \hline \\[-0.5ex]
      $\begin{array}[t]{l}
        \letrec 
        \begin{array}[t]{l} 
          \id{even} \uu \fun x . \begin{array}[t]{l}(x \eql 0)~\OR \\ (\id{odd}~(x - 1)), \end{array} \\
          \id{odd} \ex{n} \fun x . \begin{array}[t]{l}(x > 0)~\AND \\ (\id{even}~(x - 1)) \end{array}
        \end{array} \\
        \inlet~\id{even}~56 \\
        \hspace{6em}\downarrow
      \end{array}$
      & The forward reference to $\id{odd}$ is syntactically correct, and
        $\id{odd}$ evaluates correctly if $n$ is the right size. 
        We apply rule \RuleSubstab to replace $\id{even}$ with its definition.
      \\
      $\begin{array}[t]{l}
        \letrec 
        \begin{array}[t]{l} 
          \id{even} \uu \ldots, \\
          \id{odd} \ex{n} \ldots 
        \end{array} \\
        \inlet~\begin{array}[t]{l}(\fun x . (x \eql 0) \OR (\id{odd}~(x - 1)))~56 \end{array} \\
        \hspace{6em}\downarrow_{+} \\
      \end{array}$
      & We apply rule \RuleBetaab, followed by rule \RuleEMab.
      \\
      $\begin{array}[t]{l}
        \letrec 
        \begin{array}[t]{l} 
          \id{even} \uu \ldots, \\
          \id{odd} \ex{n} \ldots,  \\
          \id{x_1} \uu 56
        \end{array} \\
        \inlet~\begin{array}[t]{l}(x_1 \eql 0)~\OR~(\id{odd}~(x_1 - 1))\end{array} \\[0.5em]
        \hspace{6em}\downarrow_+ \\
      \end{array}$
      & We then perform the boolean test unsuccessfully, obtaining
        $\id{odd}~(x_1 - 1)$, where we then replace $x_1$ with its
        value and obtain $\id{odd}~55$. We can then replace $\id{odd}$ with
        its value and apply rule \RuleBetaab again, and so on.
      \\
      $\begin{array}[t]{l}
        \letrec 
        \begin{array}[t]{l} 
          \id{even} \uu \ldots, \\
          \id{odd} \ex{n} \ldots, \\
          \id{x_1} \uu 56
        \end{array} \\
        \inlet~\id{odd}~55 \\
        \hspace{6em}\vdots
      \end{array}$
      &
    \end{tabular}
  \end{center}
  \end{framed}
  \caption{Mutual recursion}
  \label{figure-example-3}
\end{figure}


Figure~\ref{figure-example-3} shows an example of mutually
recursive functions, assuming that $\lambdaab$ has been extended
with standard operations on booleans and integers.  Finally, one
may wonder why we do not perform substitution immediately after
evaluation, as usual, but use destruct-time substitution instead. 
The reason is that it better represents the semantics of the
construct we want to define. First, as previously mentioned, sharing
is propertly modeled.  Second,
as shown in Figure~\ref{figure-example-5}, it allows to represent
recursive data structures such as infinite lists.

\begin{figure}
\begin{framed}
  \begin{center}
  \begin{tabular}{lc}
    \multicolumn{1}{c}{\textrm{Expression}} & Comments \\ \hline \\[-0.5em]
    $\begin{array}[t]{l}
      \letrec 
      \begin{array}[t]{l} 
        x \ex{n} \record{\id{Head} \eql 0, \id{Tail} \eql x}
      \end{array} \\
      \inlet~x
    \end{array}$
    & \begin{minipage}[t]{.35\textwidth}
      This is a valid answer, representing an infinite (cyclic) list
      of zeroes.
    \end{minipage}
  \end{tabular}
\end{center}
\end{framed}
\caption{Recursive data structure}
\label{figure-example-5}
\end{figure}


\subsubsection{Mixin modules}
\label{subsubsection-mixin-modules}

We now consider a more elaborate example, namely an encoding of a
simple language of mixin modules, following the approach of \cite{Hirscho05}.
The design of mixin modules in a call-by-value setting raises a number of
issues that fall outside the scope of this paper; see
\cite{Hirscho-PhD} for a discussion.  Our goal here is to
informally explain why $\lambdaab$ is an adequate target language for
compiling mixin modules. Thus, we briefly describe a simple language
of call-by-value mixin modules, for which we sketch a compilation
scheme. 


\paragraph{Mixin modules} Mixin modules are unevaluated modules
with holes.  Mixin modules are to ML-style modules what classes are to
objects in object-oriented languages. The language provides a
\verb/close/ operator to instantiate a complete mixin module into
a module, thus triggering the evaluation of its components (see
below). In order to obtain a complete mixin module, the language
provides modularity operators, such as composition and deletion.  For
instance, one can define the mixin modules \verb|Even| and \verb|Odd| as
follows.
\begin{verbatim}
        mixin Even = import
          odd : int -> int
        export 
          even x = (x = 0) or (odd (x - 1))
        end

        mixin Odd = import
          even : int -> int
        export 
          odd x = (x > 0) and (even (x - 1))
        end
\end{verbatim}

The holes of a mixin module are called its \intro{imports}, and its
defined components are its \intro{exports}.  The contents of mixin
modules are not evaluated until instantiation, as described below. 
One can \intro{compose} \verb|Even| and \verb|Odd| to obtain
\begin{verbatim}
        mixin Nat1_Open = Even + Odd
\end{verbatim}
which is equivalent to
\begin{verbatim}
        mixin Nat1_Open = import
        export 
          even x = (x = 0) or (odd (x - 1))
          odd x = (x > 0) and (even (x - 1))
        end
\end{verbatim}

The name \verb|Nat1_Open| refers to the fact that the definitions of
this mixin module are still late bound and can be overridden.  Then,
this mixin module can be \intro{instantiated} into a proper module by
\begin{verbatim}
        module Nat1 = close Nat1_Open
\end{verbatim}
which is equivalent to
\begin{verbatim}
        module Nat1 = struct
          let rec even x = (x = 0) or (odd (x - 1))
          and odd x = (x > 0) and (even (x - 1))
        end
\end{verbatim}
One can then select components from \verb/Nat1/, and write for instance
\verb/Nat1.even 56/. 

As an example of \intro{overriding}, one can optimize the definition
of \verb|even| in \verb|Nat1_Open| by first removing it from
\verb/Nat1_Open/, and then composing the result with a mixin
module containing the new definition:
\begin{verbatim}
        mixin Nat2_Open = (Nat1_Open - even) +
          import 
          export 
            even x = ((x mod 2) = 0)
          end
\end{verbatim}
which is equivalent to 
\begin{verbatim}
        mixin Nat2_Open = import
        export 
          odd x = (x > 0) and (even (x - 1))
          even x = ((x mod 2) = 0)
        end
\end{verbatim}

The obtained mixin module can then be instantiated into a plain
module, as above. 
Finally, we extend \verb/Nat1_Open/ with a computation using 
the defined functions:
\begin{verbatim}
        mixin Nat_Test_Open = Nat1_Open +
          import
            even : int -> int
          export 
            test = even 56
          end  
\end{verbatim}

The obtained mixin module is equivalent to
\begin{verbatim}
        mixin Nat_Test_Open = import
        export 
          even x = (x = 0) or (odd (x - 1))
          odd x = (x > 0) and (even (x - 1))
          test = even 56
        end
\end{verbatim}

\paragraph{An incorrect encoding in $\lambdaab$} A reasonable idea for
encoding mixin modules in $\lambdaab$ would be to adapt the
standard encoding of objects and classes as recursive
records~\cite{Cardelli84}.  However, this encoding allows to
represent mixin modules, but not to instantiate them.  Consider
for instance \verb/Nat_Test_Open/. It would be translated into
a \intro{generator}, that is, a function over records:
$$\begin{array}{l}
\letrec~\id{Nat\_Test\_Open} \uu \fun \id{self} . \\
\ \ \recordl\begin{array}[t]{l}
  \id{even} \eql \fun x . (x \eql 0)~\OR~(\id{self}.\id{odd}~(x - 1)) \\
  \id{odd} \eql \fun x . (x > 0)~\AND~(\id{self}.\id{even}~(x - 1)) \\
  \id{test} \eql \id{self}.\id{even}~56 \ \recordr 
  \end{array}\\
\inlet\ \ldots
\end{array}$$

Then, the instantiation of \verb/Nat_Test_Open/ would consist
of taking its fixed point, which gives
$$\begin{array}[t]{l}
\letrecin{\id{Nat\_Test} \ex{n} \id{Nat\_Test\_Open}~\id{Nat\_Test}}{\ldots}
\end{array}$$
(assuming $n$ to be the correct size), which gives after substitution
$$\begin{array}[t]{l}
\letrec~\id{Nat\_Test} \eql (\fun \id{self} . \recordl\begin{array}[t]{l}
  \id{even} \eql \fun x . \ldots \id{self}.\id{odd} \ldots \\
  \id{odd} \eql \fun x . \ldots \id{self}.\id{even} \ldots \\
  \id{test} \eql \id{self}.\id{even}~56 \ \recordr) \\
  \ \ \id{Nat\_Test}  \\
  \end{array}\\
\inlet\ \ldots
\end{array}$$
$$\begin{array}[t]{ll}
{} \reduct^+ &
\letrec \begin{array}[t]{l}
  \id{self} \uu \id{Nat\_Test} \\
  \id{Nat\_Test} \ex{n} \recordl\begin{array}[t]{l}
  \id{even} \eql \fun x . \ldots \id{self}.\id{odd} \ldots \\
  \id{odd} \eql \fun x . \ldots \id{self}.\id{even} \ldots \\
  \id{test} \eql \id{self}.\id{even}~56 \ \recordr) \\
  \end{array} 
  \end{array}\\
& \inlet\ \ldots
\end{array}$$
$$\begin{array}[t]{ll}
{} \reduct &
\letrec \begin{array}[t]{l}
  \id{self} \uu \id{Nat\_Test} \\
  \id{Nat\_Test} \ex{n} \recordl\begin{array}[t]{l}
  \id{even} \eql \fun x . \ldots \id{self}.\id{odd} \ldots \\
  \id{odd} \eql \fun x . \ldots \id{self}.\id{even} \ldots \\
  \id{test} \eql \id{Nat\_Test}.\id{even}~56 \ \recordr) \\
  \end{array} 
  \end{array}\\
& \inlet\ \ldots
\end{array}$$
whose evaluation is stuck, because $\id{Nat\_Test}$ is not yet
evaluated and its definition is already requested. So the recursive
record semantics of objects and classes does not directly adapt to
mixin modules. The reason is that the components of a mixin module may
strongly depend on each other, in the sense of
Section~\ref{subsubsection-original-scheme}, while the components of a
class are essentially methods, which only weakly depend on each
other. 

\begin{Remark}[Objects and strong dependencies]
  In Java, initialization of instance and static fields
  by arbitrary expressions can lead to strong dependencies between the
  fields.  However, the semantics of field initialization in Java
  does not guarantee that a fixed point is reached
  \cite[section 8.3.2.3]{Java-spec}.
  Here is an example. 
\begin{verbatim}
        static int f() { return x + 1; }
        static int x = f() * 2;
\end{verbatim}
This code assigns \verb!2! to \verb!x! instead of causing an
error as expected.
\end{Remark}

\paragraph{A correct encoding in $\lambdaab$} We must find another way to
compile mixin modules.  In \cite{Hirscho05}, a mixin module is
translated into a record of functions, whose fields correspond to the
exports of the source mixin module. Each export is abstracted over the
other components upon which it depends, and over a dummy argument,
useful for suspending the computation in the absence of dependencies. 
For instance, the mixin module \verb|Even| has only one export
\verb|even|, which depends on the import \verb|odd|, so it is
represented~by
$$\begin{array}{l}
\letrec~\id{Even} \eql \record{ \id{even} \eql \fun \id{odd} . \fun \_ . \fun x . (x \eql 0)~\OR~(\id{odd}~(x - 1)) }
\end{array}$$
where $\_$ denotes an unused variable. 
Similarly, \verb|Odd| is represented by
$$\begin{array}{l}
\letrec~\id{Odd} \eql \record{ \id{odd} \eql \fun \id{even} . \fun \_ . \fun x . (x > 0)~\AND~(\id{even}~(x - 1)) }
\end{array}$$
The translation of composition merely consists of picking the right
fields in the arguments. For example, composing \verb/Even/ and \verb/Odd/ yields
$$\begin{array}{l}
\letrec~\id{Nat1\_Open} \eql \record{ \id{even} \eql \id{Even}.\id{even}, \id{odd} \eql \id{Odd}.\id{odd}}
\end{array}$$
The composition can be generated even in a separate compilation
setting, where only the types of \verb|Even| and \verb|Odd| are
available. Indeed, it only relies on the names exported by the two
mixin modules, which are mentioned in their types. Deletion is as
easy as composition, since we only have to pick the non deleted fields
of the argument. 

Instantiation is more difficult, because of strong dependencies and
sizes. Consider for example the instantiation of
\verb|Nat_Test_Open|.  Here, 
$\id{even}$ and $\id{odd}$ must be defined before $\id{test}$, which
strongly depends on them.  Thus, we obtain
$$\begin{array}{ll}
\letrec & \begin{array}[t]{l}
        \id{even} \uu \id{Nat\_Test\_Open}.\id{even}~\id{odd}~\record{}, \\
        \id{odd} \ex{n} \id{Nat\_Test\_Open}.\id{odd}~\id{even}~\record{}, \\
        \id{test} \uu \id{Nat\_Test\_Open}.\id{test}~\id{even}~\record{} \\
        \end{array} \\
\inlet & \record{\id{even} \eql \id{even}, \id{odd} \eql \id{odd}, \id{test} \eql \id{test}} 
\end{array}$$
This translation evaluates as expected, provided we can statically
guess the correct size $n$ for the $\id{odd}$ component.  For some
data representation strategies, this size can be computed from the
static type of $\id{odd}$, but not always for other strategies; see
Section~\ref{section-future-work} for a discussion.  

Another difficulty of the translation outlined here is to determine a
correct order in which to evaluate the components of the mixin being
closed.  The approach proposed in \cite{Hirscho05} and refined in
\cite{Hirscho04b} relies on exploiting dependency information added to the
static types of mixin modules.  Another approach, outlined in
\cite{Hirscho-PhD,Hirscho04a}, is to embed dependency information in
the run-time representation of mixin modules, and determine a correct
evaluation order at run-time.   

\subsubsection{Recursive modules}
Another possible application of $\lambdaab$ is for compiling 
recursive modules in extensions of the ML module system
\cite{Harper99,Russo01,OCaml,Dreyer07}.  Recursive structures are
easily encoded in $\lambdaab$.  For example, consider the following
two mutually recursive structures:
\begin{verbatim}
        module Even = struct
          let even x = (x = 0) or (Odd.odd (x - 1))
        end
        and Odd = struct
          let odd x = (x > 0) and (Even.even (x - 1))
        end
\end{verbatim}
Define the syntactic sugar
$\module{\Bb}$, where $\Bb$ is a list of declarations of the shape
$\X_1 \ass x_1 
\is_1 \e_1, \ldots, \X_n \ass x_n \is_n \e_n$, to denote
$\letrecin{x_1 \is_1 \e_1, \ldots, x_n \is_n \e_n}{\record{\X_1 \eql
    x_1, \ldots, \X_n \eql x_n}}$.
Using this notation, the example above can be expressed as
$$\begin{array}[t]{l}
\letrec \begin{array}[t]{l} 
  \id{Even} \uu \modulel \\
  \ \ \id{even} \ass \id{even} \uu  \fun x . (x \eql 0)~\OR~(\id{Odd}.\id{odd}~(x - 1)) \\
  \moduler, \\
  \id{Odd} \ex{n} \modulel \\
  \ \ \id{odd} \ass \id{odd} \uu  \fun x . (x > 0)~\AND~(\id{Even}.\id{even}~(x - 1)) \\
  \moduler
  \end{array} \\
\inlet~\ldots
\end{array}$$
(where $n$ is assumed to be the right size indication).  Notice
that the function definitions and the first module do not need to have
known sizes, since the only forward reference concerns the second
module $\id{Odd}$. 

Beyond recursive structures, it is desirable to encode recursive
functor applications, which appear in many practical uses of
recursive modules. For instance, consider the following example, taken
from the OCaml documentation~\cite[section 7.9]{OCaml-manual}.
\begin{verbatim}
        module A : sig
            type t = Leaf of string | Node of ASet.t
            val compare: t -> t -> int
          end = struct
            type t = Leaf of string | Node of ASet.t
            let compare t1 t2 = ... ASet.compare ... 
          end 
        and ASet : Set.S with type elt = A.t
                 = Set.Make(A)
\end{verbatim}
After erasing the type components of structures, we encode this
example in $\lambdaab$ by
$$\begin{array}[t]{l}
\letrec \begin{array}[t]{l}
  A \ass A \uu \modulel \\
  \ \ \id{compare} \ass \id{compare} \uu \ldots \id{ASet}.\id{compare} \ldots \\
  \moduler, \\
  \id{ASet} \ass \id{ASet} \ex{n} \id{Set}.\id{Make}~A 
\end{array} \\
\inlet~\ldots
\end{array}$$
(where $n$ is, again, assumed to be the right size indication).
This expression evaluates
correctly because $\id{Set}.\id{Make}$ only weakly depends on its argument.
The extension of this
encoding to a separate compilation setting does not raise the
problem of sizes we had for mixin modules: the sizes of ML
modules can be guessed from their types. However, the dependency
analysis remains difficult, and we are working on this issue. 

This section has demonstrated the expressive power of $\lambdaab$ by
showing encodings of mixin modules and recursive modules, which attests
its expressive power.  In order to show how to compile it to efficient
machine code, we now define a more elementary language 
called $\lambdaalloc$, into which we then translate $\lambdaab$. 

\section{The target language $\lambdaalloc$}
\label{section-lambdaalloc}

In this section, we define $\lambdaalloc$, a $\lambda$-calculus with
explicit heap. It was carefully engineered to map directly to an
abstract machine with a heap, and to enable efficient compilation to
machine code.  In particular, the heaps used in the semantics closely
correspond to machine-level heaps.  (This is apparent in the size
requirement for the update operation to work.)

\subsection{Syntax}

\begin{figure}
\begin{framed}
$$\begin{array}{llcll} 
\mbox{Variable:}
& x & \in & \Vars & \\
\mbox{Name:}
& \X & \in & \Namesfun & \vspace{8pt} \\
\mbox{Expression:}
& \E \in \Expr & \bnf& n & \textrm{Natural number} \\ 
& & \alt & x \alt \fun x .\E \alt \E~\E & \textrm{$\lambda$-calculus} \\ 
&& \alt & \letin{\B}{\E} & \textrm{Non-recursive definitions}\\
&& \alt & \record{\Ss} \alt \E \rsel \X & \textrm{Record operations}\\ 
&& \alt & \alloc \alt \update & \textrm{Heap operations} \vspace{8pt} \\ 
\mbox{Record row:}
&\Ss & \bnf & \emptysequence \alt (\X \eql \V, \Ss) & \\
\mbox{Binding:}
& \B & \bnf & \emptysequence \alt (\bd \eql \E, \B) &  \\
& \bd & \bnf & x \alt \wc & \textrm{Variable or wildcard} \vspace{8pt} \\
\mbox{Value:}
& \V \in \Values &\bnf& x \alt n  \\
\mbox{Stored value:}
& \Hv \in \VBlocks &\bnf& \fun x .\E \alt \alloc~n \alt 
        \record{\Sv} 
\vspace{8pt}\\
\mbox{Heap:}
& \Th \in \Heaps & \bnf & \emptysequence \alt x \eql \Hv, \Th \\
\mbox{Configuration:}
&\Conf & \bnf & \config{\Th}{\E} \\
\mbox{Evaluation answer:}
& \A \in \Answers &\bnf& \config{\Th}{\V} 
\end{array}$$

\end{framed}
\caption{Syntax of $\lambdaalloc$}
\label{figure-syntax-lambdaalloc}
\end{figure}%

The syntax of the target language $\lambdaalloc$ is presented in
Figure~\ref{figure-syntax-lambdaalloc}. It includes the
$\lambda$-calculus with natural numbers and non-recursive $\letseq$
binding. Note that a $\letseq$ definition $\bd \eql \E$ computes $\E$,
and then either binds the result (if $\bd$ is a variable) or ignores
it (if $\bd \meq \wc$).  The multiple value binding $\letin{\bd_1 \eql
  \E_1, \ldots, \bd_n \eql \E_n}{\E}$ should be understood as
$\letin{\bd_1 \eql \E_1}{\ldots \letin{\bd_n \eql \E_n}{\E}}.$ We
write $\emptysequence$ for the empty binding. Having a multiple
$\letseq$ binding contributes to make the equational theory of
$\lambdaalloc$ rich enough for the immediate in-place update scheme to
be correct.  Additionally, there are constructs for record operations
(creation and selection), and constructs for modeling the heap: an
allocation operator $\alloc$, and an update operator $\update$.

The semantics of $\lambdaalloc$ uses a notion of heap, which comes in
the form of a kind of global $\Letrec$.  A \intro{raw configuration}
$\Conf$ is a pair $\State{\Th}{\E}$ of a \intro{heap} $\Th$ and an
expression~$\E$. A heap is list of bindings $x \heql \Hv$, where the
\intro{stored value} $\Hv \in \VBlocks$ is either a function $\fun x .
\E$, or a record $\record{\Sv}$, or an application of the shape
$\alloc~ n$ for some natural number~$n$.  A \intro{value}~$\V$ is
either a natural number or a variable (but not a stored value).  An
evaluation \intro{answer} is a raw configuration of the shape
$\config{\Th}{\V}$.

Record rows $\Ss$, (resp.~bindings $\B$ and heaps $\Th$) are required
not to define the same name (resp.~variable) twice. We use for them
the same notations as for $\lambdaab$ record rows and bindings for
domain, codomain, concatenation, and so on. Observe that the wildcard
$\wc$ is not a variable, hence is not in the domain of bindings nor in
their free variables.

\begin{figure}
\begin{framed}
$$\begin{array}{l@{{} \meq {}}l@{\quad}l@{{} \meq {}}l}
\FV{n} & \emptyset &
\FV{\record{\Ss}} & \FV{\Ss} \\
\FV{x} & \ens{x} &
\FV{\E.\X} & \FV{\E} \\
\FV{\fun x . \E} & \FV{\E} \setminus \ens{x} &
\FV{\alloc} & \emptyset \\
\FV{\E_1\ \E_2} & \FV{\E_1} \cup \FV{\E_2} &
\FV{\update} & \emptyset \\
\FV{\letin{\B}{\E}} & \FV{\B, \wc \eql \E} \setminus \dom{\B} \\[2mm]
\FV{\emptysequence} & \emptyset &
\FV{\bd \eql \E,\B} & \FV{\E} \cup \FV{\B} \cup (\ens{\bd} \cap \Vars) \\
\FV{\Ss} & \Union{\X \in \dom{\Ss}}{\FV{\Ss(\X)}} &
\FV{\config{\Th}{\E}} & (\FV{\Th} \cup \FV{\E}) \setminus \dom{\Th} \\
\FV{\emptyheap} & \emptyset &
\FV{x \heql \Hv, \Th} & \ens{x} \cup \FV{\Hv} \cup \FV{\Th}
\end{array}$$
\end{framed}
\caption{Free variables in $\lambdaalloc$}
\label{figure-scope-alloc}
\end{figure}

\paragraph{Structural equivalence} Free variables are defined in
Figure~\ref{figure-scope-alloc}.  We call \intro{structural
  equivalence} the smallest equivalence relation including reordering
of heap bindings and renaming of bound variables.  We call
\intro{configurations} structural equivalence classes of raw
configurations. We write $\meq$ for equality of raw configurations and
$\aeq$ for equality of configurations.  We extend substitutions to
expressions and configurations in the standard way.  For defining
capture-avoiding substitution on expressions, the only non-trivial
case is $\letin{\B}{\E}$: the application of a substitution to an
expression of the shape $\letin{\bd_1 \eql \E_1, \ldots, \bd_n \eql
  \E_n}{\E}$ proceeds exactly as applying it to $\letin{\bd_1 \eql
  \E_1}{\ldots \letin{\bd_n \eql \E_n}{\E}}$.

Finally, the free variables of a substitution $\sub$ (any function
from variables to one of the syntactic classes) are defined by
$$\FV{\sub} \meq \Union{x \in \supp{\sub}}{\ens{x} \cup
                 \FV{\sub(x)}}.$$

\subsection{Dynamic semantics}

The semantics of $\lambdaalloc$ is defined by a \intro{reduction
  relation} $\reduct$, which, like that of $\lambdaab$, is first
defined as a relation over raw configurations, then
straightforwardly lifted to a relation over configurations.

\begin{figure}
\begin{framed}
$$\begin{array}[t]{l@{\quad}rrl}
\mbox{Lift context:} &
\Lcont &\bnf& \E~\trou \alt \trou~\V  \alt \trou \rsel \X \\[5pt]
\mbox{Nested lift context:} &
\Fcont &\bnf& \trou \alt \E~\Fcont \alt \Fcont~\V \alt \Fcont \rsel \X \\[5pt]
\mbox{Evaluation context:} &
\Cont &\bnf& \Fcont 
 \alt  \letin{\bd \eql \Fcont, \B}{\E}  \\[5pt]
\mbox{Allocation context:} &
\Acont &\bnf& \trou \alt \Acont~\E \alt \E~\Acont \alt \Acont \rsel \X 
              \alt \letin{\B_1, \bd \eql \Acont, \B_2}{\E} 
              \alt \letin{\B}{\Acont}
\end{array}$$
\end{framed}
\caption{Evaluation and allocation contexts of $\lambdaalloc$}
\label{figure-plain-evaluation-contexts-lambdaalloc}
\end{figure}

\begin{figure}
\begin{framed}
Alpha equivalence:
  \begin{mathpar}
%
\inferrule{
\Acont_2 \aeq \Acont'_2
}{
\appcontext{\Acont_1}{\Acont_2} \aeq \appcontext{\Acont_1}{\Acont'_2}
}
\and
%
\inferrule{
\E \aeq \E'
}{
\E\ \Acont \aeq \E'\ \Acont
}
\and
%
\inferrule{
\E \aeq \E'
}{
\Acont\ \E \aeq \Acont\ \E'
}
\and
%
\inferrule{
\B_1 \aeq \B_1' \\
(\letin{\B_2}{\E}) \aeq (\letin{\B_2'}{\E'})
}{
(\letin{\B_1, x \is \Acont, \B_2}{\E}) \aeq
(\letin{\B_1', x \is \Acont, \B_2'}{\E'})
}
\and
%
\inferrule{
\B \aeq \B'
}{
\letin{\B}{\Acont} \aeq
\letin{\B'}{\Acont}
}
\and
%
\inferrule{
\E \aeq \E'
}{
(\B_1, x \eql \E, \B_2) \aeq (\B_1, x \eql \E', \B_2)
}
\and
\end{mathpar}
Free variables:
$$\begin{array}{l@{{} \meq {}}l}
\FV{\trou} & \emptyset \\
\FV{\Acont\ \E} & \FV{\Acont} \cup \FV{\E} \\
\FV{\E\ \Acont} & \FV{\Acont} \cup \FV{\E}  \\
\FV{\Acont \rsel \X} & \FV{\Acont} \\
\FV{\letin{\B}{\Acont}} & \FV{\B} \cup \FV{\Acont} \\
\FV{\letin{\B_1, \bd \eql \Acont, \B_2}{\E}} & \!\!
\begin{array}[t]{l}
\FV{\B_1}  \cup \FV{\Acont} \cup \FV{\letin{\B_2}{\E}} \cup (\ens{\bd} \cap \Vars)
\end{array}
\end{array}$$
Captured variables:
$$\begin{array}{l@{{} \meq {}}l}
\Captcont{\trou} & \emptyset \\
\Captcont{\Acont\ \E} & \Captcont{\Acont} \\
\Captcont{\E\ \Acont} & \Captcont{\Acont} \\
\Captcont{\Acont \rsel \X} & \Captcont{\Acont} \\
\Captcont{\letin{\B}{\Acont}} & \dom{\B} \\
\Captcont{\letin{\B_1, \bd \eql \Acont, \B_2}{\E}} & \dom{\B_1} \cup \Captcont{\Acont}
\end{array}$$
\end{framed}
\caption{Structural equivalence of $\lambdaalloc$ allocation contexts}
\label{figure-scope-contexts-alloc}
\end{figure}

\subsubsection{The reduction relation} The reduction relation is
defined in Figures~\ref{figure-plain-evaluation-contexts-lambdaalloc},
\ref{figure-scope-contexts-alloc},
and~\ref{figure-computational-reduction-lambdaalloc}, using the
following hypothesis.

\begin{Hypothesis}[Size in $\lambdaalloc$]
  \label{hypothesis-size-alloc}
  We assume given a function $\Sizeallocfun$ from stored values to natural
  numbers such that
  \begin{itemize}
  \item for all $n$,
    $\Sizealloc{\alloc~n} \meq n$, and
  \item for all $\sub \in \Vars \to \Values$ and $\Hv$, $\Sizealloc{\sub(\Hv)} \meq
    \Sizealloc{\Hv}$.
  \end{itemize}
\end{Hypothesis}


The second condition follows the intuition that the size of a stored
value is determined by its top constructor, and is therefore invariant
under substitutions (which do not change the top constructor, only its
arguments). (It is also of technical use in the proof of correctness.)

The reduction rules are defined in
Figure~\ref{figure-computational-reduction-lambdaalloc}, using the
notions of contexts defined in
Figure~\ref{figure-plain-evaluation-contexts-lambdaalloc}, and the
scoping rules and functions of
Figure~\ref{figure-scope-contexts-alloc}.  

Rule \RuleBetaalloc is unusual in that it applies a heap allocated
function to an argument $\V$. The function must be a variable $x$
bound in the heap to a value $\fun y . \E$, and the result is $\sbst{y
  \repl \V}(\E)$. The reduction can take place in any
evaluation context $\Cont$.

Rule \RuleSelectalloc projects a name $\X$ out of a heap allocated
record $\record{\Sv}$ at variable $x$, returning $\Sv(\X)$.

Rule \RuleUpdatealloc copies the contents (the stored value) of a
variable to another variable. Both stored values must have exactly the
same size and the copied one must not have the shape $\alloc\ n$. This
condition may seem unnecessary, but it is used to prove that
faultiness is preserved by our translation.  Recall
that $\Th\where{x \heql \Hv}$ denotes $\Th$ where the binding for $x$
is replaced by $x \heql \Hv$.

As in $\lambdaab$, the evaluation of bindings is confined to the top
level of configurations.  This requires the \RuleLiftalloc rule, which
lifts a binding outside of a lift context $\Lcont$.

By rule \RuleIMalloc, if the first definition of the top-level binding
$\B$ is itself a binding $\letin{\B_1}{\E_1}$, then $\B_1$ is merged
with $\B$. 

Rule \RuleLetalloc describes the top-level evaluation of bindings.
Let $\sbst{\bd \repl \V}$ denote $\sbst{x \repl \V}$ if $\bd$ is a
variable $x$, and the identity substitution otherwise.  Once the first
definition is evaluated, if $\bd$ is a variable, then this
variable is replaced with the obtained value in the rest of the
expression; if $\bd \meq \wc$, evaluation proceeds directly.
When the binding becomes empty, it can be removed with rule
\RuleEmptyLetalloc.
\begin{figure}
\begin{framed}

\begin{mathpar}
\inferrule{
\Th(x) \meq \fun y . \E
}{
\State{\Th}{\Context{x~\V}} \reduct
\State{\Th}{\Context{\sbst{y \repl \V}(\E)}}
}
~~(\RuleBetaalloc)
\and
\inferrule{
\Th(x) \meq \record{\Sv}}{
\State{\Th}{\Context{x \rsel \X}}
\reduct
\State{\Th}{\Context{\Sv(\X)}}}
~~(\RuleSelectalloc)
\and
\inferrule{
\Th(y) \notin \ens{\alloc \ n \alt n \in \nat} \\
\Sizealloc{\Th(y)} \meq \Sizealloc{\Th(x)} \\
}{
\State{\Th}{\Context{\update~x~y}}
\reduct
\State{\Th \where{x \heql \Th(y)}}{\Context{\record{}}}
}
~~(\RuleUpdatealloc)
\and
\inferrule{
\dom{\B} \orth \FV{\Lcont}
}{
\State{\Th}{\Context{\Lcontext{\letin{\B}{\E}}}}
\reduct
\State{\Th}{\Context{\letin{\B}{\Lcontext{\E}}}}
}
~~(\RuleLiftalloc)
\and
\inferrule{
\dom{\B_1} \orth \ens{t} \cup \FV{\B_2} \cup \FV{\E_2}
}{
\config{\Th}{\letin{\bd \eql (\letin{\B_1}{\E_1}), \B_2}{\E_2}} \\\\
~~{} \reduct
\config{\Th}{\letin{\B_1, \bd \eql \E_1, \B_2}{\E_2}}}
~~(\RuleIMalloc)
\and
\inferrule{}{
\config{\Th}{\letin{\bd \eql \V, \B}{\E}} \\\\ {}
\reduct
\config{\Th}{\sbst{\bd \repl \V}(\letin{\B}{\E})}}
~~(\RuleLetalloc)
\and
\inferrule{}{
\config{\Th}{\letin{\emptysequence}{\E}} \reduct
\config{\Th}{\E}}
~~(\RuleEmptyLetalloc)
\and
\inferrule{
x \notin (\FV{\cosieve{\Th}{\ens{x}}} \cup \FV{\E})}{
\config{\Th}{\E} \reduct \config{\cosieve{\Th}{\ens{x}}}{\E}}
~~(\RuleGCalloc)
\and
\inferrule{
x \notin \FV{\Hv} \cup \FV{\Acont} \cup \FV{\Th} \\
\FV{\Hv} \orth \Captcont{\Acont}
}{
\config{\Th}{\Acontext{\Hv}}
\reduct
\config{\hfuone{x \heql \Hv}{\Th}}{\Acontext{x}}}
~~(\RuleAllocatealloc)
\and
\end{mathpar}
\end{framed}
\caption{Dynamic semantics of $\lambdaalloc$}
\label{figure-computational-reduction-lambdaalloc}
\end{figure}%

By rule \RuleGCalloc, when a heap binding is not used by any other
binding than itself, and not used by the expression either, it can be
removed.  This is formalized by requiring that the corresponding
variable $x$ be outside the set of free variables
$\FV{\cosieve{\Th}{\ens{x}}} \cup \FV{\E}$ of other heap bindings and
of the main expression.  This simple rule is here to model the garbage
collection step mentioned in the explanation of
Figure~\ref{figure-picture-inplace-updating-trick}: it allows
garbage-collecting the blocks obtained by evaluation of the
recursively-defined expressions once they have been copied to the
pre-allocated blocks.  A general garbage collection rule could detect
more kinds of dead data structures, in particular mutually dependent,
otherwise unused data structures.  This additional power is not needed
in this paper, so we do not have a general garbage collection rule.

Finally, rule \RuleAllocatealloc is one of the key points of
$\lambdaalloc$, by which a configuration of the shape
$\config{\Th}{\Acontext{\Hv}}$ evaluates to the configuration
$\config{\hfuone{x \heql \Hv}{\Th}}{\Acontext{x}}$, where~$x$ is a fresh
variable.  In particular, if $\Hv$ is $\alloc~ n$,
the evaluation allocates a dummy block of size $n$ on the heap.  This reduction
can happen in any \intro{allocation context}~$\Acont$.  Allocation
contexts cover all contexts of $\lambdaalloc$, except under
$\lambda$-abstractions. The idea is that a value can be allocated in
advance in the heap. For instance, given a configuration
$\config{\Th}{\letin{\B}{\Hv}}$, it is possible to allocate $\Hv$
before computing the binding, provided $\Hv$ does not use the
variables defined in $\B$.  The side condition $\FV{\Hv} \orth
\Captcont{\Acont}$ ensures this, where $\Captcont{\Acont}$ denotes the
set of binders located above the context hole in $\Acont$, here
$\dom{\B}$ (see Figure~\ref{figure-scope-contexts-alloc}).

\begin{Remark}[Non-determinism and evaluation order]
\label{remark-non-determinism-and-evaluation-order}
Unlike in $\lambdaab$, the reduction of $\lambdaalloc$ is not
deterministic because of rules \RuleGCalloc and \RuleAllocatealloc. 
Nevertheless, $\lambdaalloc$ remains close to an abstract machine,
which would simply implement a particular reduction strategy. 
Furthermore, this non-determinism makes the equational theory of
$\lambdaalloc$ rich enough for the correctness proof of
Section~\ref{section-correctness}. 

Although $\lambdaalloc$ is not deterministic, function applications
are evaluated from right-to-left, because of the lift contexts
$\trou~\V$ and $\E~\trou$.  This makes the presentation more concise,
since it avoids lift contexts of the shape $\alloc~\trou$,
$\update~\trou$, and $\update~x~\trou$, and explains why
$\lambdaab$ also evaluates its arguments from right to left. The
results of the paper can be adapted to a left-to-right evaluation
setting with some additional work. 
\end{Remark}

\subsubsection{Confluence and errors}

Since reduction in $\lambdaalloc$ is not deterministic, it is
important to make sure that it is confluent. In fact, we show that the
reduction relation is strongly commuting, which implies that it is
confluent by Hindley's lemma.

\begin{Lemma}[The reduction rules are strongly commuting]
  \label{lemma-strong-commutation}
  For all reduction rules $R_1, R_2$, and configurations $\Conf,
  \Conf_1, \Conf_2$, if $\Conf \extrarrow{R_1} \Conf_1$ and
  $\Conf \extrarrow{R_2} \Conf_2$, then there exists $\Conf'$ such that
  $\Conf_1 \extrarrow{R_2} \Conf'$ and $\Conf_2 \extrarrow{R_1}
  \Conf'$. 
\end{Lemma}

\begin{pf}
By case analysis on the possible pairs of reductions.  The reduction relation
without rules \RuleGCalloc and \RuleAllocatealloc is deterministic, so
we only have to examine the pairs involving at least one of these
rules. 
\end{pf}

A configuration is said to be \intro{faulty} if it reduces to a
configuration in normal form that is not in $\Answers$.  For a better
understanding of the semantics, we now characterize the set of faulty
configurations.



\begin{figure}
\begin{framed}
  \begin{center}
  \begin{tabular}{lp{0.4\textwidth}}
    \multicolumn{1}{c}{\textrm{Expression}} & Comments \\ \hline \\[-0.5ex]
    $\begin{array}[t]{l}
      \config{\emptyheap}{\\
        (\fun x . (x.\X.\Y)) \\
        ~~(\letin{y = \record{\Y \eql 0}}{\record{\X \eql y}})}
      \\[1em]
      \hspace{6em}\downarrow_+
    \end{array}$
    & Before applying rule \RuleBetaalloc, we must reduce the function
      and the argument to values. For this, we apply 
      (several possible orders) rules \RuleAllocatealloc (three times), 
      \RuleLetalloc and \RuleEmptyLetalloc.
    \\[1em]
    $\begin{array}[t]{l}
      \config{\left \{ \begin{array}{l}
          x_1 \heql \record{\Y \eql 0}, \\
          x_2 \heql \record{\X \eql x_1}, \\
          x_3 \heql \fun x . (x.\X.\Y)
          \end{array} \right \}}{\\
          x_3~x_2} \\
      \hspace{6em}\downarrow
    \end{array}$
    & We then apply rule \RuleBetaalloc (the heap $\Th$
      remains unchanged). 
    \\[3em]
    $\begin{array}[t]{l}
      \config{\Th}{x_2.\X.\Y} \\[.5em]
      \hspace{6em}\downarrow_+ 
    \end{array}$
    & We finally apply rule \RuleSelectalloc twice.
    \\
    $\begin{array}[t]{l}
      \config{\Th}{0}
    \end{array}$
    &
  \end{tabular}
  \end{center}
\end{framed}
\caption{An example of reduction in $\lambdaalloc$}
\label{figure-example-alloc-1}
\end{figure}


\begin{Proposition}[Faulty $\lambdaalloc$ configurations]
  A configuration is faulty iff it reduces to a configuration
  $\Conf$ in normal form such that:
\begin{itemize}
\item $\Conf \hts \config{\Th}{\Context{x~\V}}$, with either
  \begin{itemize}
  \item $x \notin \dom{\Th}$, or 
  \item $\Th(x)$ is not a function,
  \end{itemize}

  \item $\Conf \hts \config{\Th}{\Context{n~\V}}$, 

  \item or $\Conf \hts \config{\Th}{\Context{x.\X}}$, with either
    \begin{itemize}
    \item $x \notin \dom{\Th}$, or 
    \item $\Th(x)$ is not a record
      with field $\X$,
    \end{itemize}
  \item or $\Conf \hts \config{\Th}{\Context{n.\X}}$,

  \item or $\Conf \hts \config{\Th}{\appcontext{\Cont}{\alloc}}$ and
    $\Cont \neq \appcontext{\Acont'}{\trou~n}$, for all $\Acont',n$,

\item or $\Conf \hts \config{\Th}{\Context{\update~x~y}}$, with either
    \begin{itemize}
    \item $x$ or $y$ not in $\dom{\Th}$, or 
    \item $x$ and $y$ have different sizes, i.e.,
      $\Sizealloc{\Th(x)} \nmeq \Sizealloc{\Th(y)}$, or
    \item  $\Th(y)$ of the shape $\alloc\ n$,
    \end{itemize}

  \item or $\Conf \hts \config{\Th}{\Context{\update}}$ and $\Cont \neq
  \appcontext{\Cont'}{\trou~x~y}$, for all $\Cont',
  x, y$. 
\end{itemize}
\end{Proposition}

\begin{figure}
\begin{framed}
  \begin{center}
  \begin{tabular}{lc}
    \multicolumn{1}{c}{\textrm{Expression}} & Comments \\ \hline \\[-0.5ex]
    $\begin{array}[t]{l}
      \config{\emptyheap}{ \\
        \letseq~
        \begin{array}[t]{l}
          \id{odd} \eql \alloc~n, \\
          \id{even} \eql \fun x . 
          \begin{array}[t]{l}
            (x \eql 0)~\OR \\ 
            (\id{odd}~(x - 1)), 
          \end{array} \\
          \wc \eql 
          \begin{array}[t]{l}
            \update \id{odd} \\
            \fun x . 
            \begin{array}[t]{l}(x > 0)~\AND \\ 
              (\id{even}~(x - 1)), 
            \end{array} \\
          \end{array} \\
        \end{array} \\
        \inlet~\id{even}~56} \\[.5em]
      \hspace{6em}\downarrow_+ \\[.5em]
    \end{array}$
    & \begin{minipage}[t]{.4\textwidth}
      We pre-allocate a block for $\id{odd}$, evaluate $\id{even}$
      (which points to the dummy block), 
      then evaluate the definition of $\id{odd}$ and update the dummy
      block with it.
    \end{minipage} \\
    $\begin{array}[t]{l}
      \config{\left \{ 
          \begin{array}{l}
            x_1 \heql \alloc~n, \\
            x_2 \heql \fun x . 
            \begin{array}[t]{l}
              (x \eql 0)~\OR \\ 
              (x_1~(x - 1)), 
            \end{array} \\
            x_3 \heql \fun x . 
            \begin{array}[t]{l}(x > 0)~\AND \\ 
              (x_2~(x - 1))
            \end{array}
          \end{array} \right \}}{\\
        \letseq~
        \begin{array}[t]{l}
          \wc \eql \update~x_1~x_3
        \end{array} \\
        \inlet~x_2~56} \\[.5em]
      \hspace{6em}\downarrow_+ \\[.5em]
    \end{array}$
    & \begin{minipage}{.4\textwidth}
      First, the two evaluated heap blocks defining $\id{odd}$ and
      $\id{even}$ are allocated, yielding $x_1$ and $x_2$, 
      respectively. Then, the second argument of $\update$ is allocated, 
      yielding $x_3$.
    \end{minipage} \\
    $\begin{array}[t]{l}
      \config{\left \{ 
          \begin{array}{l}
            x_1 \heql \fun x . 
            \begin{array}[t]{l}(x > 0)~\AND \\ 
              (x_2~(x - 1)),
            \end{array} \\
            x_2 \heql \fun x . 
            \begin{array}[t]{l}
              (x \eql 0)~\OR \\ 
              (x_1~(x - 1)), 
            \end{array}
          \end{array} \right \}}{\\
        x_2~56}
    \end{array}$
    & \begin{minipage}{.4\textwidth}
      Now $x_1$ is updated with $x_3$, which can then be 
      garbage-collected, and the evaluation can proceed
      with the two expected mutually recursive functions.
    \end{minipage} \\
  \end{tabular}
  \end{center}
\end{framed}
\caption{Mutually recursive functions in $\lambdaalloc$ (compare with Figure~\ref{figure-example-3})}
\label{figure-example-alloc-2}
\end{figure}


\subsection{Examples}

Figure~\ref{figure-example-alloc-1} exemplifies the evaluation of a
function application in $\lambdaalloc$.  The function selects the $\Y$
field of the $\X$ field of its argument. However, in $\lambdaalloc$,
neither the function nor the argument are considered values. The
evaluation of the argument $(\letin{y \eql \record{\Y \eql 0}}{\record{\X
    \eql y}})$ involves two heap allocations: first, $x_1 \heql
\record{\Y \eql 0}$ is allocated; then, we apply rules \RuleLetalloc
and \RuleEmptyLetalloc; finally, we allocate $x_2 \heql \record{\X
  \eql x_1}$.  The evaluation of the function $\fun x . (x.\X.\Y)$
involves one heap allocation $x_3 \heql \fun x . (x.\X.\Y)$.  The
executed expression is then $x_3~x_2$, which reduces in one step to
$x_2.\X.\Y$, and then in two steps to $0$.

Figure~\ref{figure-example-alloc-2} shows the evaluation of a mutually
recursive function definition.  It is the $\lambdaalloc$ analogue of
the example shown earlier in Figure~\ref{figure-example-3}.

\section{Compilation}
\label{section-compilation}

\subsection{The standard translation}
\label{subsection-std-translation}

\begin{figure}
\begin{framed}

$\begin{array}{l@{\quad}lcl}
\mbox{Translation of expressions:} &
\transl{x} &\teq& x \\
&\transl{\fun x . \e} &\teq& 
        \fun x . \transl{\e} \\
&\transl{\e_1~\e_2} &\teq& \transl{\e_1}~\transl{\e_2} \\
&\transl{\record{\s}} &\teq&\record{\s} \\
&\transl{\e \rsel \X} &\teq& \transl{\e} \rsel \X \\
&\transl{\letrecin{\bb}{\e}} &\teq& 
\letin{\Dum{\bb}, \Up{\bb}}{\transl{\e}} \\[1em]
\mbox{Pre-allocation of bindings:} &
\Dum{\emptysequence} &\teq& \emptysequence \\
&\Dum{x \ex{n} \e, \bb} &\teq& (x \eql \alloc~ n, \Dum{\bb}) \\
&\Dum{x \uu \e, \bb} &\teq& \Dum{\bb} \\[1em]
\mbox{Computation of bindings:} &
\Up{\emptysequence} &\teq& \emptysequence \\
&\Up{x \ex{n} \e, \bb} &\teq& (\wc \eql (\update~x~\transl{\e}), \Up{\bb}) \\
&\Up{x \uu \e, \bb} &\teq& (x \eql \transl{\e}, \Up{\bb})
\end{array}$

\end{framed}
\caption{Standard translation from $\lambdaab$ to $\lambdaalloc$}
\label{figure-translation}
\end{figure}

We now define a translation from $\lambdaab$ to $\lambdaalloc$ that
straightforwardly implements the in-place update trick. This
translation, called the \intro{\STD} translation, is defined in
Figure~\ref{figure-translation}.  

The translation is straightforward for variables, functions,
applications, and record operations.  The translation of a binding
$\bb$ is the concatenation of two $\lambdaalloc$ bindings.  The
first binding $\Dum{\bb}$ is called the \intro{pre-allocation}
binding, and gives instructions to allocate dummy blocks on the heap
for definitions of known sizes. The second binding $\Update{\bb}$ is
called the \intro{update} binding.  It evaluates the definitions and
either updates the previously pre-allocated dummy blocks for
definitions of known sizes, or simply binds the result for definitions
of unknown sizes. 

\begin{Example}
  The \STD translation of the first expression of
  Figure~\ref{figure-example-3} is (part of) the first configuration
  of Figure~\ref{figure-example-alloc-2}:
      $$\begin{array}[t]{l}
        \letrec \left(
        \begin{array}{l} 
          \id{even} \uu \fun x . \begin{array}[t]{l}(x \eql 0)~\OR \\ (\id{odd}~(x - 1)), \end{array} \\
          \id{odd} \ex{n} \fun x . \begin{array}[t]{l}(x > 0)~\AND \\ (\id{even}~(x - 1)) \end{array}
        \end{array} \right)
        \inlet~\id{even}~56
      \end{array}$$
is translated to
  $$\begin{array}[t]{l}
        \letseq~ \left(
        \begin{array}{l}
          \id{odd} \eql \alloc~n, \\
          \id{even} \eql \fun x . 
          \begin{array}[t]{l}
            (x \eql 0)~\OR \\ 
            (\id{odd}~(x - 1)), 
          \end{array} \\
          \wc \eql 
          \begin{array}[t]{l}
            \update \id{odd} \\
            (\fun x . 
            \begin{array}[t]{l}(x > 0)~\AND \\ 
              (\id{even}~(x - 1))), 
            \end{array} \\
          \end{array} \\
        \end{array} \right)
        \inlet~\id{even}~56
    \end{array}$$
\end{Example}

\begin{Remark}[Restriction on forward references in $\lambdaab$]
\label{remark-explaining-the-restriction-on-forward-references-in-lambdaab}
The standard translation crucially relies on the fact that $\lambdaab$
forbids forward references to definitions of unknown sizes: such
forward references, after translation, would produce references to
unbound variables.  For example, consider the
illegal binding $x \uu y, y \uu \e$. Its pre-allocation pass is empty,
and it is translated as $x \eql y, y \eql \transl{e}$, where $y$ is unbound.
(Recall that $\lambdaalloc$ bindings do not have a recursive scope.) 
\end{Remark}

For any reduction rule $R$, write $\reductby{R}$ for the set
of pairs of expressions or configurations that are instances of $R$.

\begin{Proposition}
\label{proposition-values-std-top}
For all $\val \in \values \setminus \Vars$, there exist $\Th, x$ such that
$$\config{\emptyheap}{\transl{\val}} \xrightarrow{\textsc{\scriptsize Alloc${}_{\kwd{a}}$}}
\config{\Th}{x}$$
\end{Proposition}

\begin{pf}
  By case analysis on $\val$. 
\end{pf}

From now on, we assume that the notions of size in $\lambdaab$ and
$\lambdaalloc$ are coherent, in the following sense.

\begin{Hypothesis}[Size]
\label{hypothesis-size}
For all $\Th, x$, and $\val \in \values \setminus \Vars$, if
$\config{\emptyheap}{\transl{\val}} \reduct^* \config{\Th}{x}$, then
$\Sizeab{\val} \meq \Sizealloc{\Th(x)}$.
\end{Hypothesis}

Our main result is:
\begin{Theorem}[Correctness]\label{theorem-correctness}
  For all $\e$, if $\e$ reduces to an answer, loops, or is faulty
  in $\lambdaab$ then so does $\transl{\e}$ in $\lambdaalloc$.
\end{Theorem}

The rest of the paper is devoted to proving this theorem.  This raises
several difficulties, which we explain before actually delving into
the proof.

\subsection{Overview of difficulties}
\label{subsection-overview-of-difficulties}

A natural approach to proving the correctness of our translation is to use a
simulation argument: if $e \reduct e'$ in $\lambdaab$, then
$\transl{e} \reduct^+ \transl{e'}$; moreover, if $e$ is an answer,
$\transl{e}$ should be an answer as well.  However, both properties
fail, for reasons illustrated in the 
following examples. 

\begin{Example}[Administrative reductions]
\label{example-administrative-reductions}
Consider $\e \aeq \fun x . x$. Its translation is $\E \aeq \fun x . x$,
which is not an answer. An allocation has to be performed in
order to reduce it to the answer $\config{\heap{y \heql \fun x . x}}{y}$. 
In general, the translation of a $\lambdaab$ value reduces in a
finite number of \RuleAllocatealloc steps to a $\lambdaalloc$
answer. 
\end{Example}

\begin{Example}[More administrative reductions]
\label{example-more-administrative-reductions}
Consider $\e_1 \aeq \letrecin{y \ex{n} \fun x . x}{\e_2},$
where $n \meq \Sizeab{\fun x . x}$.  If $\e_2 \ccontraction \e'_2$, then
$\e_1$ reduces to $\e'_1 \aeq \letrecin{y \ex{n} \fun x . x}{\e'_2}$ in
$\lambdaab$.  However, the translations of $\e_1$ and $\e'_1$ are
\begin{eqnarray*}
\transl{\e_1} & \teq & \letin{y \eql \alloc~ n, \wc \eql \update~ y~(\fun x.x)}{\transl{\e_2}}
\\
\transl{\e'_1} & \teq & \letin{y \eql \alloc ~n, \wc \eql \update~ y~(\fun x.x)}{\transl{\e'_2}}
\end{eqnarray*}
and $\transl{\e_1}$ does not reduce to $\transl{\e'_1}$ in $\lambdaalloc$:
it is generally not possible to reduce $\transl{\e_2}$ until the
enclosing $\letseq$ has been fully evaluated. So, if evaluation in
$\lambdaab$ occurs under a size-respecting binding, then in the
compiled code the evaluation of this binding requires a finite number
of \RuleAllocatealloc, \RuleUpdatealloc, \RuleLetalloc, \RuleEmptyLetalloc, and 
\RuleGCalloc steps, which are exactly
the same in $\transl{\e_1}$ and $\transl{\e'_1}$. 
\end{Example}

In order to deal with these administrative reductions, we will
introduce another translation function, called the \intro{\TOP
  translation}, which performs them on the fly. This is directly
inspired by Plotkin's \emph{colon translation}~\cite{Plotkin75}. However,
there are other complications that we now illustrate, writing $\TT{\e}$
for the \TOP translation.

\begin{Example}[Granularity]
\label{example-granularity}
Consider $\e \aeq (\letrecin{x \uu \fun y .y}{x~z})$.  It reduces by
rule \RuleSubstab to $\e' \aeq (\letrecin{x \uu \fun y .y}{(\fun y .
  y)~z})$, and then by rule \RuleBetaab to $\e'' \aeq (\letrecin{x \uu
  \fun y .y}{\letrecin{y \uu z}{y}})$.  Remark that rule
\RuleSubstab duplicates $\fun y. y$, which is not innocent w.r.t. the
translation: $\TT{\e}$ does not reduce to $\TT{\e'}$. Thus, rule
\RuleSubstab alone is not simulated. In the compiled code, abstracting
over the administrative reductions, there is no substitution: rule
\RuleBetaalloc is applied directly, fetching the value of $x$ from the
heap. Initially, we have something like $\config{\Th}{x'~z}$, where
$\Th(x') \aeq \fun y . y$, which reduces in one step to
$\config{\Th}{z}$.
\end{Example}

\begin{Example}[Beta and the top-level binding]
\label{example-beta-and-the-top-level-binding}
From Example~\ref{example-granularity}, one could expect that although
rule \RuleSubstab is not exactly simulated, the combination of rules
\RuleSubstab and \RuleBetaab is. This is not the case, because rule
\RuleBetaab leaves a fully evaluated binding right where the
subreduction happened, which is not necessarily at top-level.
Consider again Example~\ref{example-granularity}: we have seen that
$\TT{\e}$ is a configuration of the shape $\config{\Th}{x'~z}$, where
$\Th(x') \aeq \fun y . y$, which reduces in one step to $\Conf \aeq
\config{\Th}{z}$.  However, after applying \RuleSubstab and
\RuleBetaab to~$\e$, we obtain $\e'' \aeq (\letrecin{x \uu \fun y
  .y}{\letrecin{y \uu z}{y}})$, where the inner $\letrec$ is not at
top level. Hence, $\TT{\e''}$ is $\config{\Th}{\letin{y \eql z}{y}}$,
which is different from $\Conf$.  Nevertheless, applying \RuleEMab to
$\e''$, we obtain $\e''' \aeq \letrecin{x \uu \fun y .y, y \uu z}{y}$,
whose \TOP translation is exactly $\Conf$. More generally, it turns
out that enough reduction sequences consisting of applications of
\RuleSubstab, \RuleBetaab, and a combination of \RuleLiftab,
\RuleIMab, and \RuleEMab are simulated by $\TTfun$.
\end{Example}

\begin{Example}[Stuttering reductions]
\label{example-dummy-reductions}
In some cases, we have $\e \reduct \e'$, but $\TT{\e} \aeq \TT{\e'}$. 
For instance, consider $\e$ of the shape $\e \aeq
\letrecin{\bv}{\letrecin{x \uu (\letrecin{\bb}{\e_1})}{\e_2}}$. By
rule \RuleEMab, $\e$ reduces to $\e' \aeq \letrecin{\bv, x \uu
  (\letrecin{\bb}{\e_1})}{\e_2}$.  In fact, in both cases, $\TTfun$
translates $\bv$ on the fly, so that $\TT{\e} \aeq \TT{\e'}$. Thus, the
preservation of non-termination is not trivial. 
\end{Example}

\begin{Example}[Lifting and allocation]
\label{example-lifting}
Let $\bb \aeq (y \uu (\fun x_2 . x_2)\ z)$ and consider $\e \aeq (\fun x_1
. x_1)~(\letrecin{\bb}{y})$, which reduces by rule \RuleLiftab to $\e'
\aeq \letrecin{\bb}{(\fun x_1 . x_1)~y}$.  Anticipating again the
definition of $\TTfun$ below, in $\e$, $\fun x_1 . x_1$ appears at
top-level, and is therefore allocated on the fly, but not $\fun x_2
. x_2$, so we obtain
$$\Conf \aeq \TTranslid{\e} \aeq
          \config{\heap{x \heql \fun x_1 .  x_1}}{
                  x~(\letin{y \eql (\fun x_2 . x_2)~z}{y})}.$$
On the other hand, in $\e'$, $\fun x_2.x_2$
appears at top-level, but not $\fun x_1 . x_1$, which lies below
a not fully evaluated binding, so we have 
$$\Conf' \aeq \TTranslid{\e'} \aeq 
\config{\heap{x' \heql \fun x_2 . x_2}}{
        \letin{y \eql x'~z}{(\fun x_1 . x_1)~y}}.$$
      Thus, some \RuleAllocatealloc reductions performed in $\TT{\e}$
      are not performed in $\TT{\e'}$.  Here, $\Conf$ reduces by
      \RuleLiftalloc and \RuleAllocatealloc to $\config{\heap{x \heql \fun x_1 . x_1,
          x' \heql \fun x_2 .  x_2}}{\letin{y \eql x'~z}{x~y}}$,
      which can be reached from $\Conf'$ by 
      \RuleAllocatealloc. 
\end{Example}

\subsection{Overview of the correctness proof}
\label{subsection-overview-of-the-correctness-proof}

Here is how we deal with these difficulties. First,
Example~\ref{example-lifting} shows that no small-step simulation
holds, so we adopt a less accurate notion of observation, namely
evaluation answers and non-termination:
\begin{itemize}
\item if $\e$ reduces to an answer $\answer$, then its translation
  reduces to some $\lambdaalloc$ answer related to
  $\answer$;
\item if $\e$ reduces infinitely, then so does its translation. 
\end{itemize}

In order to prove this result, we consider some of the reduction rules of
$\lambdaab$ and $\lambdaalloc$ as structural, i.e., not counting as
proper reduction steps. This eliminates almost all the difficulties
and preserves our notion of observation. The only remaining difficulty is
that of Example~\ref{example-granularity}, which we cannot solve in
the same way.  Indeed, we neither want \RuleSubstab nor \RuleBetaab
and \RuleSelectab to be considered structural, as we now explain.
First, deeming \RuleBetaab structural would prevent us from proving
that non termination is preserved (and doing so for \RuleSelectab is
thus only a partial, unsatisfactory solution).  Furthermore, the
equational theory of $\lambdaalloc$ is not rich enough to equate
$\TT{\e}$ and $\TT{\e'}$ when $\e \reductby{\RuleSubstab} \e'$.
Indeed, this would involve a currently forbidden duplication
(``unsharing'') of a stored value.  It seems possible to extend
$\lambdaalloc$ in a meaningful way, so as to support unsharing of
stored values in some cases. It also seems possible to modify the
semantics of $\lambdaab$ to avoid duplication before rules \RuleBetaab
and \RuleSelectab. However, the spirit of this article is to keep the
source and target languages as standard as possible, which rules out
these solutions.  Our solution is to consider bigger steps as atomic
in $\lambdaab$: we consider atomic a sequence of applications of
\RuleSubstab, followed by an application of \RuleBetaab or
\RuleSelectab, followed by possible applications of \RuleLiftab, and
terminated by a possible application of \RuleIMab or \RuleEMab (to
lift a possible binding created by application of \RuleBetaab and
merge it with the top-level binding).

We now outline the main steps of the proof, detailed in
Section~\ref{section-correctness}.

\newcommand{\TopLevelTranslation}{The \TOP translation}
\newcommand{\QuotientingLambdaalloc}{Quotient of $\lambdaalloc$}
\newcommand{\QuotientingLambdaab}{Quotient of $\lambdaab$}
\newcommand{\Correctness}{Correctness}

\paragraph{\TopLevelTranslation}
We start by defining the \TOP translation $\TTfun$, based on an
enriched notion of context in $\lambdaalloc$, which lends itself
better than the \STD translation to a simulation argument. 

\paragraph{\QuotientingLambdaalloc}
Then, we consider $\lambdaallocs$, defined as $\lambdaalloc$ modulo
rules \RuleUpdatealloc, \RuleLetalloc, \RuleEmptyLetalloc,
\RuleGCalloc, and \RuleAllocatealloc. These rules are strongly
normalizing, and we define a faithful translation from $\lambdaallocs$
to $\lambdaalloc$, by taking normal forms as representatives of
equivalence classes.  Furthermore, the translations $\translfun$ and
$\TTfun$ are well-defined from $\lambdaab$ to $\lambdaallocs$, by
composition with the canonical injection from $\lambdaalloc$ to
$\lambdaallocs$.  Define $\eqalloc$ as the equality in
$\lambdaallocs$, i.e., the equality of equivalence classes.  We then
show two crucial properties gained by taking the quotient. First, we
abstract over the administrative reductions: for any $\e$,
$\transl{\e} \eqalloc \TT{\e}$. Second, we make the translation
compositional: for all $\e$, $\cont$, $\TT{\context{\e}} \eqalloc
\Appcontext{\TT{\cont}}{\transl{\e}}$.  This addresses the problems
illustrated by
Examples~\ref{example-administrative-reductions},
\ref{example-more-administrative-reductions},
and~\ref{example-lifting}.

\paragraph{\QuotientingLambdaab}
Then, we modify the notion of evaluation of $\lambdaab$ by merging
rule \RuleSubstab with the immediately following rules.  We obtain a
language where, instead of first copying the value of a variable and
then reducing, we perform the reduction exactly as in $\lambdaalloc$
by fetching the value from the heap, applying the appropriate rule,
and, in the case of beta reduction, merging the obtained binding with
the top-level one (all this in one step).  This language correctly
simulates $\lambdaab$, since it reaches the same values, diverges on the
same expressions, and goes wrong on the same expressions. This addresses
the problems described in Examples~\ref{example-granularity}
and~\ref{example-beta-and-the-top-level-binding}.  Then, we quotient
the obtained language by rule \RuleEMab. This gives a
language called $\lambdaabsem$ which also simulates $\lambdaab$,
eliminating the issue raised by
Example~\ref{example-dummy-reductions}. 

\paragraph{\Correctness}
Finally, $\TTfun$, as a function from $\lambdaabsem$ to
$\lambdaallocs$, yields a simulation.  Writing $\reductsem$ for
reduction in $\lambdaabsem$, $\reducta$ for reduction in
$\lambdaallocs$, $\injectionfun$ for the injection from $\lambdaab$
into $\lambdaabsem$, and $\repr{\Conf}$ for the normal form of $\Conf$
modulo rules \RuleUpdatealloc, \RuleLetalloc, \RuleEmptyLetalloc,
\RuleGCalloc, and \RuleAllocatealloc, the proof may be summarized as
in Figure~\ref{figure-summary} (for observation of evaluation
answers), where the dotted arrows and equal signs corresponds to
intermediate results.

\begin{figure}
\begin{framed}
\xymatrix{
&\e \ar[r]^>{*} \ar[d]_{\injectionfun} 
          & \answer  \ar[d]^{\injectionfun} \\
& \e \ar@{.>}[r]^>{*}_>{\scriptstyle{\overline{\circ}}} \ar@/_1pc/[dl]_{\translfun} \ar[d]_{\TTfun}
          & \answer \ar@/^1pc/[dr]^{\translfun} \ar[d]^{\TTfun} \\
\config{\emptyheap}{\transl{\e}} \ar@{:}[r]_>{\overline{\kwd{a}}} \ar[dr]_{\reprfun}
    & \TT{\e} \ar[d]_{\reprfun} \ar@{.>}[r]^>{*}_>{\overline{\kwd{a}}} 
          & \TT{\answer} \ar[d]^{\reprfun} 
          & {\config{\emptyheap}{\transl{\answer}}} \ar@{:}[l]^<{\overline{\kwd{a}}} \ar[dl]^{\reprfun} \\
& \Conf \ar@{.>}[r]^>{*} & \A
}
\end{framed}
\caption{Summary of the proof (for observation of evaluation answers)}
\label{figure-summary}
\end{figure}
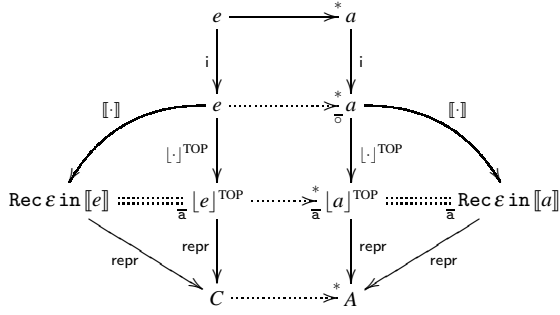

\section{Correctness}
\label{section-correctness}
\subsection{\TopLevelTranslation}
\subsubsection{Overview}

We first define the \TOP translation from $\lambdaab$ to
$\lambdaalloc$.  We start with a simple example. 

\begin{Example}
The \TOP translation of the first expression of Figure~\ref{figure-example-3}
is the last configuration of Figure~\ref{figure-example-alloc-2}:
      $$\begin{array}[t]{l}
        \letrec \left(
        \begin{array}{l} 
          \id{even} \uu \fun x . \begin{array}[t]{l}(x \eql 0)~\OR \\ (\id{odd}~(x - 1)), \end{array} \\
          \id{odd} \ex{n} \fun x . \begin{array}[t]{l}(x > 0)~\AND \\ (\id{even}~(x - 1)) \end{array}
        \end{array} \right)
        \inlet~\id{even}~56
      \end{array}$$
  gives 
  $$\begin{array}[t]{l}
      \config{\left(
          \begin{array}{l}
            x_1 \heql \fun x . 
            \begin{array}[t]{l}(x > 0)~\AND \\ 
              (x_2~(x - 1)),
            \end{array} \\
            x_2 \heql \fun x . 
            \begin{array}[t]{l}
              (x \eql 0)~\OR \\ 
              (x_1~(x - 1)), 
            \end{array}
          \end{array} \right)}{
        x_2~56}
    \end{array}$$
\end{Example}

Roughly, we consider three levels in the translated expression~$\e$.  
\begin{description}
\item[\TTOP] The first level consists of the (possibly empty) fully
  evaluated part $\bv$ of the top-level binding of $\e$, if any.  At
  this level, $\TTfun$ performs all administrative reductions, as
  previewed in Examples~\ref{example-administrative-reductions}
  and~\ref{example-more-administrative-reductions}.  Hence, the \TOP
  translation maps $\bv$ to a pair of a heap and a substitution,
  representing the heap after evaluation of the \STD translation of
  $\bv$, plus the successive substitutions produced by this
  evaluation. For instance, if $\bv \aeq (y \ex{n} \fun x .  \ldots y
  \ldots)$, then its \STD translation is $y \eql \alloc~n, \wc \eql
  \update~y~(\fun x . \ldots y \ldots)$.  Its \TOP translation gives
  directly what we would obtain after performing the admninistrative
  reductions, i.e., the heap $y' \heql \fun x .  \ldots y' \ldots$ and
  the substitution $\sbst{y \repl y'}$. 
\item[\AALLOC] After $\bv$, we expect $\TTfun$ to map answers to
  answers. Thus, we also want administrative reductions to be
  performed on the fly. The difference with the previous level is that
  (although we could do it) we do not perform administrative
  reductions on $\letrec$'s. Indeed, it is not necessary, and it would
  lead to considering more rules as administrative in $\lambdaalloc$.
  For example, consider an expression of the shape $\e \aeq
  (\letrecin{\bv}{(\letrecin{\bb}{\e_1})~\e_2})$, with $\dom{\bb}
  \orth \FV{\bv} \cup \FV{\e_2}$. This expression reduces by rules
  \RuleLiftab and \RuleEMab to $\e' \aeq (\letrecin{\bv,
    \bb}{\e_1~\e_2})$.  The purpose of performing the administrative
  reductions on the fly is to abstract over some reduction rules that
  are considered administrative in $\lambdaalloc$, because they have
  no equivalent in $\lambdaab$. Here, rule \RuleLiftab does have an
  equivalent in $\lambdaab$, so we may avoid the administrative
  reductions for $\bb$ in $\TT{\e}$ (and perform them for $\TT{\e'}$).
  
  Thus, for translating after $\bv$, we must define a translation
  function different from $\TTfun$, but nevertheless performing some
  administrative reductions. This is the purpose of the \intro{\ALLOC}
  translation $\Translfun$. In fact, except for the $\letrec$ case,
  $\TTfun$ and $\Translfun$ perform exactly the same administrative
  reductions, and we define $\TTfun$ in terms of $\Translfun$.
\item[\SSTD] In other the parts of $\e$, where no administrative
  reductions are to be performed, we apply $\translfun$.
\end{description}

\subsubsection{The \ALLOC translation}
\label{subsubsection-alloc-translation}

Let us now formally define $\Translfun$.  The idea is to translate the
evaluated part of the input expression into a proper $\lambdaalloc$
evaluation context, performing the administrative reductions on the
fly.  When the not-yet-evaluated parts of the expression are reached,
the \STD translation is used. For instance, given a function
application $\e_1~\e_2$, where $\e_2$ is not a value, one can consider
that the current evaluation point is inside $\e_2$, and therefore that
$\e_1$ has remained untouched. So, we will use $\transl{\e_1}$ and
$\Transl{\e_2}$.  The function $\Transl{\cdot}$ is defined in
Figure~\ref{figure-alloc-translation}.

\begin{Definition}[Locations and substitutions]
\label{definition-locations}
We choose a set $\Locs \subset \Vars$ of \intro{locations}, ranged
over by $\loc$, such that $\Locs$ and $\Vars \setminus \Locs$ are both
infinite. We consider only $\lambdaab$ raw expressions whose
free and bound variables are in $\Vars \setminus \Locs,$ which is from
now on ranged over by $x$.  We consider only $\lambdaalloc$ raw
configurations $\config{\Th}{\E}$ such that $\dom{\Th} \subseteq
\Locs$ and locations are never bound in $\E$ or the right-hand sides
of $\Th$. From now on, we also call \intro{substitutions}, ranged over
by $\sub$, functions from $\Vars$ to $\Vars$ whose support is disjoint
from $\Locs$.  Composition of these substitutions is well defined as
mere function composition.  We call \intro{variable allocations} such
substitutions that are furthermore injective on their support and
whose cosupport only contains locations.  We denote them by $\va$
(final sigma).
\end{Definition}

We stress in passing that the cosupport, and free variables of
substitutions stay the same, e.g., cosupport may contain locations.

We then define $\Transl{\cdot}$ as a function from $\lambdaab$
equivalence classes to $\lambdaalloc$ configurations (it is obviously
well defined).

As for the \STD translation, variables are translated into themselves.
A function $\fun x .  \e$ is translated to $\fun x .  \transl{\e}$,
but the result is allocated on the heap, at a fresh location $\loc$:
$\config{\heap{\loc \heql \fun x .  \transl{\e}}}{\loc}$. The
translation of records is similar.  For translating function
application, we use a new notation: given two heaps $\Th_1$ and
$\Th_2$ such that $\dom{\Th_1} \orth \dom{\Th_2}$, we write
$\hfu{\Th_1}{\Th_2}$ for their concatenation, which is a heap again.
If the argument part is not a value, then it is translated with
$\Translfun$, while the function is translated with $\translfun$.  If
the argument is a value, then both parts are translated with
$\Translfun$.  The translation of a record selection $\e \rsel \X$
consists of translating $\e$ with $\Translfun$ and then selecting the
field $\X$.  Finally, a binding $\letrecin{\bb}{\e}$ is translated to
$\config{\emptyheap}{\transl{\letrecin{\bb}{\e}}}$.

\begin{figure}
\begin{framed}

$\begin{array}{lcll}
\Transl{x} &\teq& \config{\emptyheap}{x} \\
\Transl{\fun x . \e} &\teq& 
  \config{\heap{\loc \heql \fun x . \transl{\e}}}{\loc} \\
\Transl{\record{\s}} &\teq& 
  \config{\heap{\loc \heql \record{\s}}}{\loc} \\
\Transl{\e~\val} &\teq& \config{\hfu{\Th_1}{\Th_2}}{\E~\V} &
\mbox{if {}} \left \{
\begin{array}{l}
\Transl{\e} \teq \config{\Th_1}{\E} \\
\Transl{\val} \teq \config{\Th_2}{\V}
\end{array} \right .  \vspace{8pt} \\
\Transl{\e_1~\e_2} &\teq&
\config{\Th}{\transl{\e_1}~\E} &
\mbox{if {}} \left \{
\begin{array}{l}
\e_2 \notin \values  \\
\Transl{\e_2} \teq \config{\Th}{\E}
\end{array} \right .  \vspace{8pt} \\
\Transl{\e \rsel \X} &\teq& \config{\Th}{\E \rsel \X} & \mbox{if {}} \Transl{\e} \teq \config{\Th}{\E} \\
\Transl{\letrecin{\bb}{\e}} &\teq& \config{\emptyheap}{\transl{\letrecin{\bb}{\e}}} 
\end{array}$

\end{framed}
\caption{The \ALLOC translation from $\lambdaab$ to $\lambdaalloc$}
\label{figure-alloc-translation}
\end{figure}

\subsubsection{Generalized contexts} 
\label{subsubsection-generalized-contexts}
Given an expression $\e$, in order to calculate $\TT{\e}$, we will
decompose $\e$ into its top-level binding $\bb_1$ and the rest of the
expression $\e_1$, and the result will be the translation of $\e_1$,
put in some context representing $\bb_1$, written $\TT{\bb_1}$, which
is defined in Figure~\ref{figure-top-translation}
(Section~\ref{subsubsection-top-translation}), using notions defined
in Sections~\ref{subsubsection-generalized-contexts}
to~\ref{subsubsection-properties-genconts}.  The binding is divided
into its evaluated part $\bv$ and the rest $\bb$, which can be empty,
but does not begin with a size-respecting definition.  We start by
giving an informal account of the handling of $\bv$ and $\bb$, which
leads us to the definition of a generalized notion of context in
$\lambdaalloc$.

Let us first explain the translation of the unevaluated part $\bb$.
In $\translfun$, the $\Dummyfun$ function produces instructions for
allocating dummy blocks.  In the \TOP translation, these blocks are
directly allocated by the function $\TDumfun$ (see
Section~\ref{subsubsection-top-translation}), which returns the heap
of dummy blocks and the substitution replacing variables with the
corresponding locations. As a first example, given a binding $\bb \aeq
(x_1 \uu \e_1, x_2 \ex{n} \e_2)$, $\TDum{\bb}$ essentially returns a heap
$\heap{\loc_2 \heql \alloc~n}$ and the substitution $\sbst{x_2 \repl
  \loc_2}$. This corresponds to the fact that after the pre-allocation
pass (as generated by the \STD translation), the update pass takes
place under this heap and substitution.

In $\translfun$, the $\Updatefun$ function produces instructions to
either update a dummy block with the translation of the definition, or
to perform the binding implied by the definition.  In $\TTfun$, the
only difference is that the first definition in $\bb$ is translated
with $\Translfun$, while the remaining ones -- still considered to lie
past the current evaluation point -- are translated with $\translfun$.
This is done by function $\TUpfun$ (see
Section~\ref{subsubsection-top-translation}). On the previous example,
if $\Transl{\e_1} \aeq \config{\Th_1}{\E_1}$, then $\TUp{\bb}$
essentially returns the heap $\Th_1$ and the binding $x_1 \eql \E_1,
\wc \eql \update~x_2~\transl{\e_2}$.  Under the substitution returned
by $\TDumfun$, the second definition becomes
$\update~\loc_2~\transl{\e_2}$, as expected.


Now, what should be the \TOP translation, written $\Transltopid{\bv}$,
of the evaluated part $\bv$?  As mentioned above, this translation yields
a heap and a substitution. The translation of definitions is
relatively natural, but it is difficult to assemble the results in a
coherent manner. First, consider a single definition $x \is \val$. The
\ALLOC translation of $\val$ is an answer, of the shape
$\config{\Th}{\V}$.  It is thus clear that the generated heap and
substitution should be $\Th$ and $\sbst{x \repl \V}$, respectively.

The next question is how to assemble the results obtained for each
definition.  First, we remark that in the absence of forward
references, substitutions should be composed from right to left. For
instance, on a binding like $\bv \aeq (x_1 \uu x_0, x_2 \uu x_1)$, the
generated substitution must be $\sbst{x_1 \repl x_0} \rond \sbst{x_2
\repl x_1}$, and not the converse. Thus, definitions can be altered by
previous definitions, which may have replaced some variables with
other values. 

However, because of forward references in $\lambdaab$ bindings, the
translated definitions may also have to be altered by subsequent
definitions. For instance, consider the binding $\bv \aeq (x_1 \uu
x_2, x_2 \ex{n} \fun x . x),$ where $n \meq \Sizeab{\fun x .  x}$.
The \TOP translation turns $\bv$ into a heap and a substitution.  The
translation of the first definition consists of the heap $\Th_1 \aeq
\emptyheap$ and the substitution $\sub \meq \sbst{x_1 \repl x_2}$, so
that subsequent occurrences of $x_1$ are replaced with $x_2$.  Then,
we translate the second definition.  This gives $\Th_2 \aeq
(\heap{\loc_2 \heql \fun x . x})$ and $\va \meq \sbst{x_2 \repl
  \loc_2}$, for some fresh location $\loc_2$.  Naively, one could
think that the substitution corresponding to the whole binding should
be the right-to-left composition of the obtained substitutions. But
this is wrong, since the obtained substitution would be $\sub \rond
\va.$ Under this substitution, a call to $x_1$ becomes $\sbst{x_1
  \repl x_2}(\sbst{x_2 \repl \loc_2}(x_1)) \meq x_2,$ while it should
rather be directed to $\loc_2$.  This example illustrates that
variable allocations performed by the translation are expected to
alter previous forward references to them, which possibly appear as
substitutions.

This leads us to define a new notion of context in $\lambdaalloc$,
called \intro{generalized context}, in terms of which we define the
translation of bindings. The functions $\TDumfun, \TUpfun$, and
$\Transltopfun$ will be defined as returning generalized contexts,
which makes their uniform treatment easier. Basically, the idea of
generalized contexts is that they contain a heap, an allocation
context, and two substitutions, rather than one. This allows
distinguishing variable allocations $x \repl \loc$, which might alter
previous translations, from normal substitutions $x \repl y$, which
may not.  Basically, only definitions of known size can alter previous
translations, because they are the only ones that can be forward
referenced.  Furthermore, crucially, we will require that the normal
substitution of a generalized context be \intro{one-way}, in the
following sense.

\begin{Definition}[One-way substitution]
  A substitution $\sub$ (with implicitly $\supp{\sub} \orth \Locs$, by
  Definition~\ref{definition-locations}) is \intro{one-way} iff
  $\supp{\sub} \orth \cosupp{\sub}$.
\end{Definition}

From the informal explanations above, it should sound natural that the
unknown size definitions of the shape $x \uu y$ generate one-way
substitutions. Indeed, they only ``go left'' in the binding, and no
binding may define, say $x \uu y, y \uu x$, because of the syntactic
restriction on forward references.

Let us first prove the following easy lemmas on substitutions.

\begin{Lemma}
\label{lemma-idempotent-one-way}
  For all one-way substitutions $\sub$, $\sub \rond \sub \meq \sub$.
\end{Lemma}

\begin{pf}
  For all variable $x$, either $x \notin \supp{\sub}$, and then both
  sides are equal to $x$, or $x \in \supp{\sub}$, but then $\sub(x)
  \in \cosupp{\sub}$, which by hypothesis implies $\sub(x) \notin
  \supp{\sub}$, hence $\sub(\sub(x)) \meq \sub(x)$, as expected.
\end{pf}

\begin{Lemma}
\label{lemma-easy-substitution-composition}
For all substitutions $\sub_1$ and $\sub_2$, 
\begin{itemize}
\item $\supp{\sub_1 \rond \sub_2} \subseteq \supp{\sub_1} \cup
\supp{\sub_2}$,  and
\item $\cosupp{\sub_1 \rond \sub_2} \subseteq \cosupp{\sub_1} \cup
\cosupp{\sub_2}$.
\end{itemize}
\end{Lemma}

\begin{pf}
  The first point is point is easy by contradiction.

  For the second point, assume $x \in \cosupp{\sub_1 \rond
    \sub_2}$. There is some $y \neq x$ such that $(\sub_1 \rond
  \sub_2)(y) \meq x$. 
  
  Let $z \meq \sub_2(y)$. If $x \notin \cosupp{\sub_1}$, then $z \meq
  x$, so $\sub_2(y) \meq x$, and $x \in \cosupp{\sub_2}$.

\end{pf}

\begin{Lemma}
\label{lemma-commute-subst}
  For all substitutions $\sub_1$ and $\sub_2$, if $\FV{\sub_1} \orth
  \supp{\sub_2}$, then $\sub_1 \rond \sub_2 \meq \sub_1 (\sub_2) \rond
  \sub_1$.
\end{Lemma}

\begin{pf}
  Let $x \in \Vars$.
  \begin{itemize}
  \item If $x \in \supp{\sub_2}$, then $x \notin \FV{\sub_1}$, so
    $(\sub_1(\sub_2) \rond \sub_1)(x) \meq (\sub_1(\sub_2))(x)$, which is 
    the expected result.
  \item Otherwise, if $x \in \supp{\sub_1}$, then $\sub_1(x) \in
    \cosupp{\sub_1}$, so $\sub_1(x) \notin \supp{\sub_2}$, hence
    $(\sub_1(\sub_2) \rond \sub_1)(x) \meq \sub_1(x) \meq (\sub_1
    \rond \sub_2)(x).$
  \end{itemize}
\end{pf}

\begin{Lemma}
\label{lemma-commute-dupl}
  For all $\sub, \va$, if $\supp{\sub} \orth \supp{\va}$, then
  $\va(\sub) \rond \va \meq \va \rond \sub \rond \va$.
\end{Lemma}

\begin{pf}
  Let $x \in \Vars$. 
  \begin{itemize}
  \item If $\va(x) \in \supp{\sub}$, then both sides are equal to
    $(\va \rond \sub \rond \va)(x)$.
  \item Otherwise, since $\va$ is one-way, $\va \rond \va \meq \va$,
    so $(\va(\sub) \rond \va)(x) \meq \va(x) \meq (\va \rond \va)(x)
    \meq (\va \rond \sub \rond \va)(x).$
  \end{itemize}
\end{pf}

\begin{Corollary}
  \label{corollary-va-sub-commute}
  For all $\va, \sub$, if $\supp{\va} \orth \supp{\sub}$, then
  $\va \rond \sub \meq \va \rond \sub \rond \va$.
\end{Corollary}

\begin{pf}
  By Lemmas~\ref{lemma-commute-subst} and~\ref{lemma-commute-dupl},
  since obviously $\supp{\va} \orth \supp{\sub}$ implies
  $\FV{\va} \orth \supp{\sub}$.
\end{pf}

Then, we have the following obvious, but useful result.

\begin{Proposition}
\label{proposition-sub-acont}
  For all $\Acont, \sub, \E$, if $\FV{\sub} \orth \Captcont{\Acont}$,
  then $\sub(\appcontext{\Acont}{\E}) \aeq
  \appcontext{(\sub(\Acont))}{\sub(\E)}$.
\end{Proposition}

\begin{Corollary}
  \label{corollary-va-acont-commute}
  For all $\Acont, \sub, \E$, if $\sub \rond \sub \meq \sub$ and
  $\FV{\sub} \orth \Captcont{\Acont}$, then $\sub(\Acontext{\E}) \aeq
  \sub(\Acontext{\sub(\E)})$.
\end{Corollary}

\begin{pf}
By Proposition~\ref{proposition-sub-acont},
$\sub(\appcontext{\Acont}{\E}) \aeq
  \appcontext{(\sub(\Acont))}{\sub(\E)} \aeq
  \appcontext{(\sub(\Acont))}{(\sub \rond \sub)(\E)}
\aeq
  \sub(\appcontext{\Acont}{\sub(\E)})$.
\end{pf}

Then, we give the following sufficient condition for the composition
of two one-way substitutions to be one-way. We recall that
$\FV{\sub_1} \orth \supp{\sub_2}$ means
  \begin{itemize}
  \item $\supp{\sub_1} \orth \supp{\sub_2}$ and 
  \item $\cosupp{\sub_1} \orth \supp{\sub_2}$.
  \end{itemize}

\begin{Lemma}
\label{lemma-one-way-substitution-composition}
For all one-way substitutions $\sub_1$ and $\sub_2$, if $\FV{\sub_1}
\orth \supp{\sub_2}$, then 
\begin{itemize}
\item $\sub_1 \rond \sub_2$ is one-way,
\item $\supp{\sub_1 \rond \sub_2} \meq \supp{\sub_1} \cup
\supp{\sub_2}$, 
\item $\cosupp{\sub_1} \subseteq \cosupp{\sub_1 \rond \sub_2}$.
\end{itemize}
\end{Lemma}

\begin{pf}	
  We prove that $\sub_1 \rond \sub_2$ is one-way by contradition. Let
  $\sub \meq \sub_1 \rond \sub_2$ and assume the existence of $x \in
  \cosupp{\sub} \cap \supp{\sub}$, i.e., the existence of $x, y$, and
  $z$, such that
  \begin{itemize}
  \item $\sub(x) \meq y$ with $x \neq y$, and
  \item $\sub(z) \meq x$ with $x \neq z$.
  \end{itemize}
  Let then $x' \meq \sub_2(x)$ and $z' \meq \sub_2(z)$, so that we
  informally have:
  
  \expandafter\ifx\csname xgraph\endcsname\relax
   \csname newbox\expandafter\endcsname\csname xgraph\endcsname
\fi
\ifx\graphtemp\undefined
  \csname newdimen\endcsname\graphtemp
\fi
\expandafter\setbox\csname xgraph\endcsname
 =\vtop{\vskip 0pt\hbox{%
    \graphtemp=.5ex
    \advance\graphtemp by 0.067in
    \rlap{\kern 0.100in\lower\graphtemp\hbox to 0pt{\hss $x$\hss}}%
    \graphtemp=.5ex
    \advance\graphtemp by 0.067in
    \rlap{\kern 1.533in\lower\graphtemp\hbox to 0pt{\hss $x'$\hss}}%
    \graphtemp=.5ex
    \advance\graphtemp by 0.067in
    \rlap{\kern 2.967in\lower\graphtemp\hbox to 0pt{\hss $y$\hss}}%
    \special{pn 8}%
    \special{pa 1331 42}%
    \special{pa 1431 67}%
    \special{fp}%
    \special{pa 1331 92}%
    \special{pa 1431 67}%
    \special{fp}%
    \special{pa 200 67}%
    \special{pa 1431 67}%
    \special{fp}%
    \graphtemp=\baselineskip
    \multiply\graphtemp by -1
    \divide\graphtemp by 2
    \advance\graphtemp by .5ex
    \advance\graphtemp by 0.067in
    \rlap{\kern 0.817in\lower\graphtemp\hbox to 0pt{\hss $\sub_2$\hss}}%
    \special{pa 2764 42}%
    \special{pa 2864 67}%
    \special{fp}%
    \special{pa 2764 92}%
    \special{pa 2864 67}%
    \special{fp}%
    \special{pa 1633 67}%
    \special{pa 2864 67}%
    \special{fp}%
    \graphtemp=\baselineskip
    \multiply\graphtemp by -1
    \divide\graphtemp by 2
    \advance\graphtemp by .5ex
    \advance\graphtemp by 0.067in
    \rlap{\kern 2.250in\lower\graphtemp\hbox to 0pt{\hss $\sub_1$\hss}}%
    \graphtemp=.5ex
    \advance\graphtemp by 0.467in
    \rlap{\kern 0.100in\lower\graphtemp\hbox to 0pt{\hss $z$\hss}}%
    \graphtemp=.5ex
    \advance\graphtemp by 0.467in
    \rlap{\kern 1.533in\lower\graphtemp\hbox to 0pt{\hss $z'$\hss}}%
    \graphtemp=.5ex
    \advance\graphtemp by 0.467in
    \rlap{\kern 2.967in\lower\graphtemp\hbox to 0pt{\hss $x$\hss}}%
    \special{pa 1331 442}%
    \special{pa 1431 467}%
    \special{fp}%
    \special{pa 1331 492}%
    \special{pa 1431 467}%
    \special{fp}%
    \special{pa 200 467}%
    \special{pa 1431 467}%
    \special{fp}%
    \graphtemp=\baselineskip
    \multiply\graphtemp by -1
    \divide\graphtemp by 2
    \advance\graphtemp by .5ex
    \advance\graphtemp by 0.467in
    \rlap{\kern 0.817in\lower\graphtemp\hbox to 0pt{\hss $\sub_2$\hss}}%
    \special{pa 2764 442}%
    \special{pa 2864 467}%
    \special{fp}%
    \special{pa 2764 492}%
    \special{pa 2864 467}%
    \special{fp}%
    \special{pa 1633 467}%
    \special{pa 2864 467}%
    \special{fp}%
    \graphtemp=\baselineskip
    \multiply\graphtemp by -1
    \divide\graphtemp by 2
    \advance\graphtemp by .5ex
    \advance\graphtemp by 0.467in
    \rlap{\kern 2.250in\lower\graphtemp\hbox to 0pt{\hss $\sub_1$\hss}}%
    \hbox{\vrule depth0.533in width0pt height 0pt}%
    \kern 3.067in
  }%
}%
\showgraph

  \begin{itemize}
  \item If $x \neq x'$ and $x \neq z'$, then $x \in \supp{\sub_2} \cap
    \cosupp{\sub_1}$ which is impossible by hypothesis.
  \item If $x \meq x'$ and $x \neq z'$, then $x \in \supp{\sub_1} \cap
    \cosupp{\sub_1}$, which is impossible because $\sub_1$ is one-way.
  \item If $x \meq z'$ and $x \neq x'$, then $x \in \supp{\sub_2} \cap
    \cosupp{\sub_2}$, which is impossible because $\sub_2$ is one-way.
  \item If $x \meq x'$ and $x \meq z'$, then $\sub(z) \meq y$ which is
    impossible since $\sub(z) \meq x$ and $x \neq y$.
  \end{itemize}

The second point is proved as follows.
\begin{itemize}
\item By Lemma~\ref{lemma-easy-substitution-composition},
  $\supp{\sub_1 \rond \sub_2} \subseteq \supp{\sub_1} \cup
  \supp{\sub_2}$;
\item If $x \in \supp{\sub_1} \cup \supp{\sub_2}$ but $x \notin
  \supp{\sub_1 \rond \sub_2}$, i.e., $\sub_1(\sub_2(x)) \meq x$, then
  let $y \meq \sub_2(x)$. If $x \meq y$, then $\sub_1(x) \meq
  \sub_1(y) \meq x$, so $x$ is neither in $\supp{\sub_1}$ nor in
  $\supp{\sub_2}$, which contradicts $x \in \supp{\sub_1} \cup
  \supp{\sub_2}$. Otherwise, we have $y \neq x$, which implies $x \in
  \supp{\sub_2}$, and furthermore and $(\sub_1 \rond \sub_2)(x) \meq
  \sub_1(y) \meq x$, so $x \in \cosupp{\sub_1} \cap \supp{\sub_2}$, a
  contradiction.  
\end{itemize}

Let us now prove the third point: let $x \in \cosupp{\sub_1}$. There
exists $y \neq x$ such that $\sub_1(y) \meq x$.  By hypothesis, this
$y$ is not in $\supp{\sub_2}$, so $(\sub_1 \rond \sub_2)(y) \meq x$,
and $x \in \cosupp{\sub_1 \rond \sub_2}$.
\end{pf}

\begin{Definition}[Generalized contexts]
  A \intro{generalized context} is a 4-tuple of a heap $\Th$, an
  allocation context $\Acont$, a substitution $\sub$, and a variable
  allocation $\va$, written $\Gcont \bnf
  \GCC{\Th}{\Acont}{\sub}{\va}$, such that
  \begin{itemize}
  \item $\sub$ is one-way,
  \item $\supp{\sub} \orth \supp{\va}$, 
    \item $\Acont$ and the range of $\Th$ do not have any free location, 
    \item $\cosupp{\sub} \cap \Locs \subseteq \dom{\Th},$
    \item and $\cosupp{\va} \subseteq \dom{\Th}$.
    \end{itemize}

    A \intro{generalized evaluation context} is a generalized context
    whose allocation context is an evaluation context, i.e., a
    generalized context of the shape $\GCC{\Th}{\Cont}{\sub}{\va}$.

    The generalized contexts generated by the translation of
    size-respecting bindings will have $\trou$ as their allocation
    context.  We call such generalized contexts \intro{\extheaps}, and
    write them $\BB$.

    The generalized contexts generated by dummy allocation of bindings
    will have the shape $\GCC{\Th}{\trou}{\identity}{\va}$. We call
    such generalized contexts \intro{\allocations}, and write them
    $\BBd$.

    Also, we define the syntactic sugar $\GCC{\Th}{\B}{\sub}{\va}$, which, if
    $\B$ is not empty, denotes
    $\GCC{\Th}{\letin{\B}{\trou}}{\sub}{\va}$, and
    otherwise denotes $\GCC{\Th}{\trou}{\sub}{\va}$.
    Further, bindings $\B$ are implicitly coerced to generalized
    contexts
    $\GCC{\emptyheap}{\B}{\identity}{\identity}$.
    Finally, we simply write $\sub$ for
    $\GCC{\emptyheap}{\trou}{\sub}{\identity}$, and define
    $\Substof{\GCC{\Th}{\Acont}{\sub}{\va}} \meq \va \rond \sub$ and
    $\Contof{\GCC{\Th}{\Acont}{\sub}{\va}} \aeq \Acont$.

\end{Definition}

Notes: $\dom{\Th}$ contains only locations, and is thus inherently
disjoint from $\supp{\sub}$ and $\supp{\va}$. Also, recall that every
evaluation context $\Cont$ is also an allocation context $\Acont$.

\begin{figure}
  \begin{framed}
$$\begin{array}{l@{{} \meq {}}l}
\FV{\GCC{\Th}{\Acont}{\sub}{\va}} & (\FV{\Th} \cup \FV{\Acont} \cup \FV{\sub} \cup \FV{\va}) \setminus \dom{\Th} \\
\Captcont{\GCC{\Th}{\Acont}{\sub}{\va}} & \Captcont{\Acont} \cup \supp{\sub} \cup \supp{\va}
\end{array}$$
\end{framed}
  \caption{Free variables and captured variables for generalized contexts}
  \label{figure-free-vars-gen-contexts}
\end{figure}

Next, we define structural equivalence on generalized contexts, using
the definition of free variables in
Figure~\ref{figure-free-vars-gen-contexts}. For helping the intuition
we also define the captured variables for generalized contexts. The
intuition behind structural equivalence of generalized contexts is
that the locations bound in $\dom{\Th}$ may be renamed freely, since
they may only be mentioned in the cosupport of $\sub$ and $\va$.
Formally, structural equivalence is defined in
Figure~\ref{figure-scope-gen-contexts-alloc}, as the least equivalence
relation respecting the rules. The first rule says that
$\alpha$-equivalence on expressions, heaps, and stored values is
included; the second rule says that a location in the heap may be
renamed, provided it does not clash with another one.

\begin{figure}
\begin{framed}
  \begin{mathpar}
%
\inferrule{
\Th \aeq \Th' \\
\Acont \aeq \Acont' \\
\sub \aeq \sub' \\
\va \aeq \va'
}{
\GCC{\Th}{\Acont}{\sub}{\va} \aeq
\GCC{\Th'}{\Acont'}{\sub'}{\va'} 
}
\and
%
\inferrule{
\loc' \notin \dom{\Th} \\
\sub' \meq  \sbst{\loc \repl \loc'} (\sub)  \\
\va' \meq    \sbst{\loc \repl \loc'}  (\va)
}{
\GCC{(\Th, \loc \eql
      \Hv)}{\Acont}{\sub}{\va}
\aeq 
\GCC{(\Th, \loc' \eql
      \Hv)}{
      \Acont}{
      \sub'}{
      \va'}
}
\end{mathpar}
\end{framed}
\caption{Structural equivalence of generalized contexts}
\label{figure-scope-gen-contexts-alloc}
\end{figure}

\begin{Example}
\label{example-translation-definitions}
Consider the bindings $\bv \aeq (x_1 \uu x_0, x_2 \uu x_4,
x_3 \uu x_1, x_4 \ex{n} \fun x . x)$, which is an interleaving of
previous examples, and $\bb \aeq (x_5 \ex{n} x_2)$. 

Via $\Ttopfun$, each definition in $\bv$ yields a generalized
context:
\begin{itemize}
\item
$x_1$ yields $\Gcont_{11} \aeq \GCC{\emptyheap}{\trou}{\sbst{x_1 \repl x_0}}{\identity}$, 
\item 
$x_2$ yields $\Gcont_{12} \aeq \GCC{\emptyheap}{\trou}{\sbst{x_2 \repl x_4}}{\identity}$, 
\item
$x_3$ yields $\Gcont_{13} \aeq \GCC{\emptyheap}{\trou}{\sbst{x_3 \repl x_1}}{\identity}$, and
\item
$x_4$ yields $\Gcont_{14} \aeq \GCC{\heap{\loc \heql \fun x . x}}{\trou}{\identity}{\sbst{x_4 \repl \loc}}$. 
\end{itemize}

The not yet evaluated binding $\bb$ yields the heap $\Th \aeq
\heap{\loc' \heql \alloc~n}$ and the variable allocation $\va \meq
\sbst{x_5 \repl \loc'}$ by function $\TDumfun$, which we write
$\Gcont_2 \aeq \GCC{\Th}{\trou}{\identity}{\va}$. 

Via $\TUpfun$, $\bb$ yields the heap $\Th' \aeq \emptyheap$ and
the binding $\B \aeq (\wc \eql \update~x_5~x_2)$, which we write $\Gcont_3
\aeq \GCC{\Th'}{\letin{\B}{\trou}}{\identity}{\identity}$. Note that this
is the only use of generalized contexts using allocation contexts (here
$\letin{\B}{\trou}$) which are not evaluation contexts. 
\end{Example}

\subsubsection{Composition of generalized contexts}

We then need a notion of composition of generalized contexts, in order
to assemble the pieces of our translation. The guiding intuition here
is that when composing two generalized contexts $\Gcont_1 \aeq
\GCC{\Th_1}{\Acont_1}{\sub_1}{\va_1}$ and $\Gcont_2 \aeq
\GCC{\Th_2}{\Acont_2}{\sub_2}{\va_2}$, we want the result to be
well-defined and equal to $$\GCC{\hfu{\Th_1}{\Th_2}}{
  \appcontext{\Acont_1}{\Acont_2} }{ \sub_1 \rond \sub_2 }{
  \funion{\va_1}{\va_2}},$$ but we also want the following two
equations to hold for any composable $\Gcont_1$ and $\Gcont_2$, and 
for any expression $\E$:
  $$((\funion{\va_1}{\va_2}) \rond \sub_1 \rond \sub_2)(\hfu{\Th_1}{\Th_2})
    \aeq \begin{array}[t]{l}
        \hfu{((\funion{\va_1}{\va_2}) \rond \sub_1)(\Th_1)}{\\
      (\va_1 \rond \va_2(\sub_1) \rond \va_2 \rond \sub_2)(\Th_2)
      })
      \end{array} 
      $$
      and 
      $$
      ((\funion{\va_1}{\va_2}) \rond \sub_1 \rond
      \sub_2)(\appcontext{\Acont_1}{\appcontext{\Acont_2}{\E}}) \aeq
      ((\funion{\va_1}{\va_2}) \rond
      \sub_1)(\appcontext{\Acont_1}{(\va_2 \rond
        \sub_2)(\appcontext{\Acont_2}{\E})}).$$

    In these equations, the left member is (part of) what we get by
    applying the result of the composition to $\E$, as defined below
    (Definition~\ref{definition-generalized-context-application}). The
    right member describes how we would like the four substitutions to
    interact.  For instance, $\sub_2$ is a standard substitution, which
    does not affect upper levels of context: in both left members, it
    does not act on the components of $\Gcont_1$. On the other hand
    $\va_2$ has to affect them, but should not be shortcut by
    $\Acont_1$ and $\sub_1$, which explains why we require that it
    still affects the variables in $\appcontext{\Acont_2}{\E}$ and
    $\Th_2$ before, respectively, $\Acont_1$ and $\sub_1$, which come
    first in the left members. 

    We use the following definition, which natural except for the
    domain of definition: two generalized contexts $\Gcont_1 \aeq
    \GCC{\Th_1}{\Acont_1}{\sub_1}{\va_1}$ and $\Gcont_2 \aeq
    \GCC{\Th_2}{\Acont_2}{\sub_2}{\va_2}$ are \intro{composable},
    written $\composable{\Gcont_1}{\Gcont_2}$, iff
    \begin{itemize}
    \item the four substitutions $\sub_1, \va_1, \sub_2$, and $\va_2$
      have pairwise disjoint supports,
    \item $\supp{\sub_2} \orth \FV{\Gcont_1}$,
    \item $\Captcont{\Acont_1} \orth \FV{\sub_2} \cup \FV{\va_2}$.
    \end{itemize}
    These conditions are oviously preserved by structural equivalence,
    which justifies the definition of composability on equivalence
    classes of generalized contexts.
    
    We then state:
\begin{Definition}[Composition of generalized contexts]
\label{definition-composition-of-generalized-contexts}
For all such composable generalized contexts $\Gcont_1$ and $\Gcont_2$
define their \intro{composition} $\Gcont_1 \rondtopbv \Gcont_2$
by $\Gcont_1 \rondtopbv \Gcont_2 \aeq \GCC{\hfu{\Th_1}{\Th_2} }{
  \appcontext{\Acont_1}{\Acont_2} }{ \sub_1 \rond \sub_2 }{
  \funion{\va_1}{\va_2}},$ provided $\dom{\Th_1} \orth \dom{\Th_2}$
(which can always be reached by structural equivalence).
\end{Definition}

\begin{Proposition}
  The conditions for being composable are equivalent to
  \begin{itemize}
  \item $\FV{\sub_1} \orth \supp{\sub_2}$,
  \item $\supp{\va_1} \orth \supp{\va_2}$,
  \item $\supp{\sub_1} \cup \supp{\sub_2} \orth \supp{\va_1} \cup
    \supp{\va_2}$,
  \item $\supp{\sub_2} \orth \FV{\Th_1} \cup \FV{\Acont_1}$,
  \item $\cosupp{\sub_2} \orth \Captcont{\Acont_1}$,
  \item $\supp{\va_2} \orth \supp{\sub_1} \cup \Captcont{\Acont_1}$.
  \end{itemize}
\end{Proposition}
\begin{pf}
  Easy check.
\end{pf}

In the light of this, the conditions for composability may be
understood as follows:
\begin{itemize}
\item The first three items ensure that the result is a well-formed
  generalized context. Well, actually they do a bit more: they use the
  sufficient condition of
  Lemma~\ref{lemma-one-way-substitution-composition}, requiring
  $\FV{\sub_1} \orth \supp{\sub_2}$ and $\supp{\sub_1} \cup
  \supp{\sub_2} \orth \supp{\va_1} \cup \supp{\va_2}$ instead of
  requiring $\sub_1 \rond \sub_2$ to be one-way, and $\supp{\sub_1
    \rond \sub_2} \orth \supp{\va_1} \cup \supp{\va_2}$ to hold.  But
  this more restrictive requirement allows an easy proof of weak
  associativity for composition of generalized contexts, and is
  general enough for our purposes.
\item The fourth item ensures that $\sub_2$ does not affect $\Th_1$
  and $\Acont_1$.
\item The fifth item ensures that $\sub_2$ is not shortcut by
  $\Acont_1$ (i.e., $\sub_2(\appcontext{\Acont_1}{\ldots} ) \aeq
  \appcontext{\sub_2(\Acont_1)}{\sub_2(\ldots)}$, which by the
  previous point is in fact $\appcontext{\Acont_1}{\sub_2(\ldots)}$).
\item The sixth item ensures that $\va_2$ is not shortcut by $\sub_1$
  and $\Acont_1$.
\end{itemize}

\begin{Example}
  Consider again $\bv$ from
  Example~\ref{example-translation-definitions}.  Its \TOP translation
  is $\Gcont_{11} \rondtopbv \Gcont_{12} \rondtopbv \Gcont_{13}
  \rondtopbv \Gcont_{14}$, which is exactly $\Gcont_1 \aeq
  \GCC{\Thbv}{\trou}{\subbv}{\vabv}$, with the heap $\Thbv \aeq \heap{\loc
    \heql \fun x . x}$, the variable allocation $\vabv \meq \sbst{x_4
    \repl \loc}$, and the substitution $\subbv \meq \sbst{x_1,x_3 \repl
    x_0, x_2 \repl x_4}$. Note that the rest of the translation
  ensures that variable allocations are always applied after
  substitutions, so that $x_2$ will eventually be redirected to
  $\loc$. 
\end{Example}

We noww prove useful sufficient conditions for composability and
associativity.  They use the following notation for, respectively, the
\intro{unknown size} and \intro{known size} variables of a binding
$\bb$:
\begin{itemize}
\item $\UV{\bb} \meq \ens{x \alt \exists \e, (x \uu \e) \in \bb}$,
\item $\KV{\bb} \meq \ens{x \alt \exists n, \e, (x \ex{n} \e) \in \bb}$.
\end{itemize}

\begin{Proposition}
  \label{proposition-forward-refs-macro}
  If $(\bb_1, \bb_2)$ is syntactically correct, then $\FV{\bb_1} \orth
  \UV{\bb_2}$.
\end{Proposition}

\begin{Definition}
  For all generalized contexts $\Gcont \aeq
  \GCC{\Th}{\Acont}{\sub}{\va}$, and correct bindings $\bb$, 
  we say that $\bb$ \intro{justifies} $\Gcont$, and write 
  $\justifies{\bb}{\Gcont}$, iff:
  \begin{itemize}
  \item $\FV{\Gcont} \subseteq \FV{\bb}$, and more specifically,
    \item $\supp{\sub} \subseteq \UV{\bb}$, 
    \item $\supp{\va} \subseteq \KV{\bb}$.
    \end{itemize}
\end{Definition}

\begin{Lemma}
\label{lemma-justify-compose}
  Assume a correct binding of the shape $(\bb_1, \bb_2)$ and two
  generalized contexts $\Gcont_i \aeq
  \GCC{\Th_i}{\Acont_i}{\sub_i}{\va_i}$, such that
  $\justifies{\bb_i}{\Gcont_i}$, for $i \meq 1, 2$.  If moreover
  $\Captcont{\Acont_1} \meq \emptyset$, then
  $\composable{\Gcont_1}{\Gcont_2}$ and 
  $\justifies{\bb_1, \bb_2}{\Gcont_1 \rondtopbv \Gcont_2}$.
\end{Lemma}

\begin{pf}
  First, the four involved substitutions have as supports the domains
  of pairwise disjoint parts of $\bb_1, \bb_2$, hence have pairwise
  disjoint supports. Furthermore, since $\bb_1, \bb_2$ is correct,
  $\bb_1$ makes no (forward) reference to $\UV{\bb_2}$, hence
  $\supp{\sub_2} \orth \FV{\Gcont_1}$. Thus, $\Captcont{\Acont_1}$
  being empty, we have $\composable{\Gcont_1}{\Gcont_2}$.
  
  Furthermore, we have 
  $$\FV{\Gcont_1 \rondtopbv \Gcont_2} \subseteq
  \FV{\Gcont_1} \cup \FV{\Gcont_2} \subseteq
  \FV{\bb_1} \cup \FV{\bb_2} \meq \FV{\bb_1, \bb_2}.$$
  
  By a similar reasoning on substitutions, we obtain
  $\justifies{\bb_1, \bb_2}{\Gcont_1 \rondtopbv \Gcont_2}$ as desired.
\end{pf}

\begin{Lemma}
\label{lemma-justify-assoc}
  Assume a correct binding of the shape $(\bb_1, \bb_2, \bb_3)$ and
  three generalized contexts $\Gcont_i \aeq
  \GCC{\Th_i}{\Acont_i}{\sub_i}{\va_i}$, such that
  $\justifies{\bb_i}{\Gcont_i}$, for $i \meq 1, 2, 3$.
  If moreover $\Captcont{\Acont_i} \meq \emptyset$, for $i \meq 1, 2$,
    then $(\Gcont_1 \rondtopbv \Gcont_2) \rondtopbv \Gcont_3$ and
    $\Gcont_1 \rondtopbv (\Gcont_2 \rondtopbv \Gcont_3)$ are defined
    and equal.
\end{Lemma}

\begin{pf}
  By the previous Lemma, we obtain $\composable{\Gcont_1}{\Gcont_2}$
  and $\justifies{\bb_1, \bb_2}{\Gcont_1, \Gcont_2}$, hence by the
  same Lemma, $\composable{(\Gcont_1 \rondtopbv \Gcont_2)}{\Gcont_3}.$
  Symmetrically, $\composable{\Gcont_1}{(\Gcont_2 \rondtopbv
    \Gcont_3)}.$ Thus, both sides are defined at the same
  time. Equality is then a simple check.
\end{pf}

\subsubsection{Generalized context application}
We have seen that the top-level binding will be translated as a
generalized context.  We will then fill the context hole with the
translation of the rest of the expression, using generalized context
application, which we now define.

\begin{Definition}[Generalized context application]
\label{definition-generalized-context-application}
For every generalized context $\Gcont \aeq
\GCC{\Th}{\Acont}{\sub}{\va}$ and configuration $\Conf \aeq
\config{\Th'}{\E}$, let $\Appcontext{\Gcont}{\Conf} \aeq (\config{
  (\va \rond \sub)(\hfu{\Th}{\Th'})}{(\va \rond \sub)(\Acontext{\E})
})$ be the \emph{application} of $\Gcont$ to $\Conf$, provided
$\dom{\Th'} \orth \dom{\Th} \cup \cosupp{\sub} \cup \cosupp{\va}$
(which may always be reached by structural equivalence).
\end{Definition}

\begin{Example}
  Consider again the binding $(\bv, \bb)$ from
  Example~\ref{example-translation-definitions}.  Its translation is
  $\Gcont_2 \rondtopbv \Gcont_1 \rondtopbv \Gcont_3$, which is exactly
  $\Gcont \aeq \GCC{\Th_0}{\Acont_0}{\subbv}{\va \rond \vabv}$, with
  \begin{itemize}
  \item the heap
    $\Th_0 \aeq \left (
      \begin{array}{l}
        \loc \heql \fun x . x, \\
        \loc' \heql \alloc~n
      \end{array}
    \right )$
  \item and the context $\Acont_0 \aeq \letin{\B}{\trou}$. 
  \end{itemize}

  If, for instance, $\Gcont$ is filled with a configuration
  $\config{\Th}{\E}$, if the conditions for the generalized context
  application are met, we get $\Appcontext{\Gcont}{\config{\Th}{\E}}
  \aeq
  \config{\sub(\hfu{\Th_0}{\Th})}{\sub(\Appcontext{\Acont_0}{\E})}$, where
  $\sub \meq (\va \rond \vabv \rond \subbv) \meq
  \left (
      \begin{array}{l}
        x_5 \repl \loc', \\
        x_1,x_3 \repl x_0, \\
        x_2 \repl \loc, \\
        x_4 \repl \loc
      \end{array}
    \right ).$
\end{Example}

We now prove the equations that had motivated the definition of
generalized context composition.

\begin{Lemma}
  For all composable generalized contexts $\Gcont_i \aeq
  \GCC{\Th_i}{\Acont_i}{\sub_i}{\va_i}$ and configuration
  $\config{\Th}{\E}$, if $\dom{\Th} \orth \dom{\hfu{\Th_1}{\Th_2}}$,
  then $$\Appcontext{(\Gcont_1 \rondtopbv \Gcont_2)}{\config{\Th}{\E}} \aeq
  \begin{array}[t]{l}
    \config{
      \begin{array}[t]{l}
        \hfu{((\va_1 \fu \va_2) \rond \sub_1)(\Th_1) }{\\
          (\va_1 \rond \va_2(\sub_1) \rond \va_2 \rond \sub_2)(\hfu{\Th_2}{\Th})}
      \end{array} \\
    }{
      ((\va_1 \fu \va_2) \rond \sub_1)(\appcontext{\Acont_1}{(\va_2 \rond \sub_2)(
        \appcontext{\Acont_2}{\E})})}.
  \end{array}
$$
\end{Lemma}

\begin{pf}
  First, by $\composable{\Gcont_1}{\Gcont_2}$, we have 
  $\FV{\Th_1} \orth \supp{\sub_2}$, so 
  $((\va_1 \fu \va_2) \rond \sub_1 \rond \sub_2)(\Th_1) \aeq
  ((\va_1 \fu \va_2) \rond \sub_1)(\Th_1)$.

  Furthermore, $\va_1 \fu \va_2 \meq \va_1 \rond \va_2$. But by
  $\composable{\Gcont_1}{\Gcont_2}$ again, we have $\supp{\va_2} \orth
  \supp{\sub_1}$. So by Lemma~\ref{lemma-commute-subst}, $\va_2 \rond
  \sub_1 \meq \va_2(\sub_1) \rond \va_2$. This gives $((\va_1 \fu
  \va_2) \rond \sub_1 \rond \sub_2)(\Th_2, \Th) \aeq (\va_1 \rond
  \va_2(\sub_1) \rond \va_2 \rond \sub_2)(\hfu{\Th_2}{\Th})$.

  Now, by composability again, $\Captcont{\Acont_1} \orth \FV{\sub_2}
  \cup \FV{\va_2}$. But $\FV{\va_2 \rond \sub_2} \subseteq \FV{\sub_2}
  \cup \FV{\va_2}$, so $\Captcont{\Acont_1} \orth \FV{\va_2 \rond
    \sub_2}$. By Proposition~\ref{proposition-sub-acont} and the
  above, this yields $(\va_2 \rond \sub_2)
  (\appcontext{\Acont_1}{\appcontext{\Acont_2}{\E}}) \aeq
  \appcontext{((\va_2 \rond \sub_2)\Acont_1)}{(\va_2 \rond
    \sub_2)(\appcontext{\Acont_2}{\E})}$.

  But $\supp{\sub_2} \orth \FV{\Acont_1}$ and $\va_2 \rond \va_2 \meq
  \va_2$, so $\appcontext{((\va_2 \rond \sub_2)\Acont_1)}{(\va_2 \rond
    \sub_2)(\appcontext{\Acont_2}{\E})} =
  \appcontext{(\va_2(\Acont_1))}{(\va_2 \rond \va_2 \rond
    \sub_2)(\appcontext{\Acont_2}{\E})}$, hence by
  Proposition~\ref{proposition-sub-acont} again, this is equal to
  $\va_2(\appcontext{\Acont_1}{(\va_2 \rond
    \sub_2)(\appcontext{\Acont_2}{\E})})$.  Thus, $((\va_1 \fu \va_2)
  \rond \sub_1 \rond \sub_2)
  (\appcontext{\Acont_1}{\appcontext{\Acont_2}{\E}})$ is indeed
  equal to 
  $(\va_1 \rond \va_2 (\sub_1) \rond \va_2) (\appcontext{\Acont_1}{(\va_2 \rond
    \sub_2)(\appcontext{\Acont_2}{\E})})$, which
  is in turn equal to 
  $((\va_1 \fu \va_2) \rond \sub_1) (\appcontext{\Acont_1}{(\va_2 \rond
    \sub_2)(\appcontext{\Acont_2}{\E})})$, as desired.
\end{pf}

\subsubsection{Weak composition of generalized contexts}

Although the notion of composition of generalized contexts is needed
to properly translate size-respecting bindings, it is somewhat
inconvenient to reason with.  For instance, the usual equation
$\Appcontext{(\Gcont_1 \rondtopbv \Gcont_2)}{\Conf} \aeq
\Appcontext{\Gcont_1}{\Appcontext{\Gcont_2}{\Conf}}$ obviously does
not hold in general: the variable allocation of $\Gcont_2$ may affect
$\Gcont_1$ in $\Appcontext{(\Gcont_1 \rondtopbv \Gcont_2)}{\Conf}$,
but not in $\Appcontext{\Gcont_1}{\Appcontext{\Gcont_2}{\Conf}}$.
Nevertheless, when $\Gcont_1$ and $\Gcont_2$ stem from distinct
bindings with no defined variable in common, we recover more standard
properties. More generally, we define the following notions of
\intro{context interference} and \intro{weak composition}, which take
advantage of such cases in the following sense: weak composition has
more standard properties than $\rondtopbv$, and coincides with it when
the considered contexts do not interfere. We first define weak
composition and show that it satisfies the equation above.

\begin{Definition}[Weak composition of generalized contexts]
  Given $\Gcont_i \aeq \GCC{\Th_i}{\Acont_i}{\sub_i}{\va_i}$, for $i
  \meq 1,2$, if $\composable{\Gcont_1}{\Gcont_2}$ is defined, we
  define $\Gcont_1 \rond \Gcont_2 \aeq
  \GCC{(\hfu{\Th_1}{\Th_2})}{\appcontext{\Acont_1}{\Acont_2}}{(\va_1
    \rond \sub_1 \rond \va_2 \rond \sub_2)}{\identity}$.
\end{Definition}

\begin{Proposition}
\label{proposition-lift-context-application}
  For all $\Gcont_1, \Gcont_2$, and $\Conf$, $\Appcontext{(\Gcont_1
    \rond \Gcont_2)}{\Conf} \aeq
  \Appcontext{\Gcont_1}{\Appcontext{\Gcont_2}{\Conf}}$, when the
  former is defined. 
\end{Proposition}

\begin{pf}
  By definition of weak composition and generalized context application. 
\end{pf}

Now, we define context interference, and state the expected result. 

\begin{Definition}[Context interference]
  Given two generalized contexts $\Gcont_i \aeq
  \GCC{\Th_i}{\Acont_i}{\sub_i}{\va_i}$, for $i \meq 1,2$, let us say
  that the pair (the order matters) $(\Gcont_1, \Gcont_2)$
  \intro{interferes} iff $\supp{\va_2}$ intersects $\FV{\sub_1}$, so
  that $\va_2 \rond \sub_1$ and $\sub_1 \rond \va_2$ are not
  necessarily equal. 
\end{Definition}

\begin{Proposition}
  If $\supp{\va} \orth \FV{\sub}$, then $(\va \rond \sub) \meq (\sub \rond \va)$.
\end{Proposition}

\begin{pf}
  Since $\va$ is a variable allocation, $\cosupp{\va} \orth
  \supp{\sub}$, by which the result follows.
\end{pf}

\begin{Proposition}
\label{proposition-interfere}
For all $\Gcont_1$ and $\Gcont_2$, $\Gcont_1 \rondtopbv \Gcont_2$ and
$\Gcont_1 \rond \Gcont_2$ are defined at the same time, and if
$(\Gcont_1, \Gcont_2)$ does not interfere, then $(\Gcont_1 \rondtopbv
\Gcont_2) \aeq (\Gcont_1 \rond \Gcont_2).$
\end{Proposition}

\begin{pf}
  By definition of composition and interference. 
\end{pf}

\subsubsection{Preservation of some reductions inside generalized
  contexts}
\label{subsubsection-properties-genconts}
In this section, before presenting the top-level translation, we
collect two small results about preservation of certain reductions
inside certain generalized contexts.

First, we remark that not every reduction is preserved inside
generalized contexts, since for instance, rules \RuleLetalloc and
\RuleEmptyLetalloc are only valid at top-level.  However, inside
\extheaps, reduction is preserved. Note that every \allocation is
\anextheap, so the following result also applies to \allocations.

\begin{Proposition}
\label{proposition-weak-context-reduction}
For all $\Conf_1,\Conf_2$, and $\Pcont$, if $\Conf_1 \reduct \Conf_2$, then
$\Pcontext{\Conf_1} \reduct \Pcontext{\Conf_2}$. 
\end{Proposition}

\begin{pf}
  By case analysis on the applied rule. 
\end{pf}

Moreover, we note that rule \RuleAllocatealloc is preserved by
generalized context application.

\begin{Proposition}
\label{proposition-allocate-in-context}
For all generalized contexts $\Gcont$, and configurations
$\Conf$ and $\Conf'$, if $\Conf \xrightarrow{\textsc{\scriptsize
    Alloc${}_{\kwd{a}}$}} \Conf'$, then $\Gcontext{\Conf}
\xrightarrow{\textsc{\scriptsize Alloc${}_{\kwd{a}}$}} \Gcontext{\Conf'}$.
\end{Proposition}

\begin{pf}
  Follows from composability of allocation contexts: for all
  $\Acont_1$ and $\Acont_2$, $\appcontext{\Acont_1}{\Acont_2}$ is an
  allocation context.
\end{pf}

\subsubsection{The \TOP translation} 
\label{subsubsection-top-translation}
We now present the \TOP translation $\TTfun$, defined in
Figure~\ref{figure-top-translation}, as a function from $\lambdaab$
($\alpha$-equivalence classes of) expressions to $\lambdaalloc$
configurations (it is well defined).  As explained above, \extheaps
are used to record the already translated definitions along the
translation of top-level bindings, preserving the distinction between
variable allocations $\va$ and ordinary substitutions $\sub$.
Variable allocations that must alter previous translations are those
generated by the translation of a ${}\ex{n}{}$ definition, since only
those can be forward referenced.

We first define the \TOP translation without checking the validity of
the involved generalized context compositions. They are checked
shortly afterwards.

The \TOP translation handles the size-respecting part of top-level
bindings with the function $\Transltopfun$.  This function expects a
size-respecting binding. When its argument is the empty binding, it
returns the empty \extheap. For non-empty bindings, the definitions
are translated as sketched above.  For a definition of unknown size $x
\uu \val$, $\val$ is translated by $\Transl{\cdot}$ to
$\config{\Th}{\V}$, and is included in the translation as the \extheap
$\Topbv{\Th}{\sbst{x \repl \V}}{\identity}$.  A definition of known
size $x \ex{n} \val$ is translated into a heap and a variable
allocation: $\val$ has a translation of the shape
$\config{\Th}{\loc}$, and it is included in the translation of $\bv$
as $\GCC{\Th}{\trou}{\identity}{\sbst{x \repl \loc}}$.  The \TOP
translation of an evaluated binding is the composition of the
translations of its definitions. If the result is some
$\Topbv{\Th}{\sub}{\va}$, then the variable allocation is applied
after the ordinary substitution, which allows the correct treatment of
forward references, as sketched in
Section~\ref{subsubsection-generalized-contexts}.

\begin{figure}
\begin{framed}
$\begin{array}{lcll}
\TT{\letrecin{\bv}{\e}} &\teq& \appcontext{\TT{\bv}}{\Transl{\e}} \\
\TT{\letrecin{\bb}{\e}} &\teq& 
\multicolumn{2}{l}{
\appcontext{\TT{\bb}}{\config{\emptyheap}{\transl{\e}}} 
\hfill \mbox{if $\bb$ is not size-respecting}} \\
\TTransl{\Pcont}{\e} &\teq& 
\multicolumn{2}{l}{
\Transl{\e} \hfill
 \mbox{if $\e$ is not of the form $\letrecin{\bb}{\e'}$}} \vspace{8pt}\\
\TTransl{\trou}{\bv,\bb} &\teq& \TDum{\bb} \rondtopbv \Transltop{\trou}{{\bv}} \rondtopbv \TUp{\bb}
& \\[.3em]
\multicolumn{4}{l}{\hspace{14.5em}\mbox{\begin{tabular}[t]{l} 
      where $\bb$ does not begin with a \\[.3em]
      size-respecting definition.
    \end{tabular} }}
\end{array}$\ligne
$\begin{array}{lcll}
\Transltop{\BB}{x \uu \val} &\teq& \Topbv{\Th}{\sbst{x \repl \V}}{\identity} &
  \mbox{\hbox{} if $\Transl{\val} \teq \config{\Th}{\V}$} \\
\Transltop{\BB}{x \ex{n} \val} &\teq& \Topbv{\Th}{\identity}{\sbst{x \repl \loc}} &
  \mbox{\begin{tabular}[t]{l}
        if $\Transl{\val} \teq \config{\Th}{\loc}$ \\
        and $\Sizeab{\val} \meq n$
        \end{tabular}}
\end{array}$ \vspace{8pt} \\
$\begin{array}{lcll}
\Transltop{\BB}{\emptysequence} &\teq& \Topbv{\emptyheap}{\identity}{\identity} \\
\Transltop{\BB}{x \is \val, \bw} &\teq& 
{\Transltop{\BB}{x \is \val}} \rondtopbv {\Transltop{\BB \rondtopbv \Transltop{\BB}{x \is \val}}{\bw}}\\
\end{array}$ \ligne
$\begin{array}{lcll}
\TDum{\emptysequence} &\teq& \TDumbb{\emptyheap}{\identity} \\
\TDum{x \uu \e, \bb} &\teq& \TDum{\bb}
& \\
\TDum{x \ex{n} \e, \bb} &\teq& (\TDumbb{\heap{\loc \eql \alloc n}}{\sbst{x \repl \loc}}) \rond \TDum{\bb}
\end{array}$  \ligne
$\begin{array}{lcll}
\TUp{\emptysequence} &\teq& \GCC{\emptyheap}{\trou}{\identity}{\identity}
& \hspace{.6cm}{}\\
\TUp{x \uu \e, \bb} &\teq& 
\multicolumn{2}{l}{\TUpbb{\Th}{(x \eql \E, \Up{\bb})}}\\
&&
\multicolumn{2}{r}{\mbox{if 
  $\Transl{\e} \teq \config{\Th}{\E}$}} \\
\TUp{x \ex{n} \e, \bb} &\teq& 
\TUpbb{\Th}{(\wc \eql (\update \, x~\E), \Up{\bb})}
& \\
&&
\multicolumn{2}{r}{\mbox{if 
$\Transl{\e} \teq \config{\Th}{\E}$}}
\end{array}$

\end{framed}
\caption{The \TOP translation from $\lambdaab$ to $\lambdaalloc$}
\label{figure-top-translation}
\end{figure}

The two other functions, $\TDumfun$ and $\TUpfun$, are defined as
announced in the beginning of
Section~\ref{subsubsection-generalized-contexts}. The three functions
return generalized contexts: $\TDumfun$ returns \anallocation
$\GCC{\Th}{\trou}{\identity}{\va}$, $\TUpfun$ returns
$\TUpbb{\Th}{\B}$, which (by the notation of
Section~\ref{subsubsection-generalized-contexts}) is
$\GCC{\Th}{\letin{\B}{\trou}}{\identity}{\identity}$ if $\B \neq
\emptysequence$, and $\GCC{\Th}{\trou}{\identity}{\identity}$
otherwise.

In case the whole binding $(\bv, \bb)$ is evaluated (i.e., $\bb$ is
empty), the contexts for pre-allocation and update, $\TDum{\bb}$ and
$\TUp{\bb}$ are empty, and $\TT{\letrecin{\bv, \bb}{\e}}$ is
$\Transl{\e}$, put in the context $\Transltopid{\bv}$.  Otherwise,
$\TT{\letrecin{\bv, \bb}{\e}}$ is $\config{\emptyheap}{\transl{\e}}$,
put in the context $\TDum{\bb} \rondtopbv \Transltopid{\bv} \rondtopbv
\TUp{\bb}$.  Notice that there is no context interference, since the
innermost one, $\TUp{\bb}$, does not have any variable allocation, and
the outermost one, $\TDum{\bb}$, has no substitution (but only a
variable allocation). So, we could equivalently use $\TDum{\bb} \rond
\Transltopid{\bv} \rond \TUp{\bb}$.  We have two easy results on
answers and faulty terms:

\begin{Proposition}\label{prop-answers}
  The function $\TTfun$ maps answers to answers.
\end{Proposition}
\begin{pf}
  A simple case inspection.
\end{pf}

\begin{Proposition}\label{prop-faulty}
  For $\e_0$ of one of the shapes in $(3)$ of
  Proposition~\ref{proposition-faulty-ab}, $\TT{\e_0}$ is faulty.
\end{Proposition}
\begin{pf}
  By case inspection.
\end{pf}

Finally, we prove that all the generalized context compositions we use
are well-defined and associative.

\begin{Proposition}
\label{proposition-all-is-well}
  For all $\bb$, $x$, $\val$, and $\bv$, the following hold
\begin{center}
\begin{tabular}{lcr} $\justifies{\bb}{\TDum{\bb}}$, & \hfill &
 $\justifies{\bb}{\TUp{\bb}}$, \\
 $\justifies{x \is \val}{\Ttop{x \is \val}}$, & \hfill and \hfill &
 $\justifies{\bv}{\Ttop{\bv}}$.
\end{tabular}
\end{center}
\end{Proposition}

As a corollary, all the possible generalized contexts resulting from
the \TOP translation may be composed (by Lemma~\ref
{lemma-justify-compose}) in an associative fashion (by
Lemma~\ref{lemma-justify-assoc}).  This justifies the absence of
parentheses in the definition.

\subsection{\QuotientingLambdaalloc}

In this section, we relate the three translation functions
$\translfun$, $\Translfun$, and $\TTfun$: we show that their results
are equivalent modulo the rules \RuleUpdatealloc, \RuleLetalloc,
\RuleEmptyLetalloc, \RuleGCalloc, and \RuleAllocatealloc. So, letting
$\lambdaallocs$ be the quotient of $\lambdaalloc$ modulo these rules,
we obtain that they are equal as functions from $\lambdaab$ to
$\lambdaallocs$.  Then, we study the compositionality of this
function.

\begin{Definition}[$\lambdaallocs$]
  Define $\eqalloc$ as the smallest equivalence relation over
  $\lambdaalloc$ containing the rules \RuleUpdatealloc, \RuleLetalloc,
  \RuleEmptyLetalloc, \RuleGCalloc, and \RuleAllocatealloc.  Let
  $\lambdaallocs$ be the set of $\eqalloc$-equivalence classes.  Let
  reduction in $\lambdaallocs$, written $\reducta$, be defined by the
  rules:
\begin{mathpar}
\inferrule{
\Conf_1 \eqalloc \Conf'_1 \\
\Conf'_1 \reductby{R} \Conf'_2 \\
\Conf'_2 \eqalloc \Conf_2
}{
\Conf_1 \reducta \Conf_2
}
\end{mathpar}
where $R$ ranges over the other rules (\RuleBetaalloc,
\RuleSelectalloc, \RuleLiftalloc, and \RuleIMalloc).

Define ${\eqallocr} \subseteq {\eqalloc}$ to be $\lambdaalloc$
convertibility by rule \RuleAllocatealloc.

\end{Definition}

We obtain that $\lambdaallocs$ and $\lambdaalloc$ behave identically:

\begin{Lemma}\label{lemma-alloc-allocs}
  For all $\Conf$, $\Conf$ reduces to an answer, loops, or is faulty
  in $\lambdaalloc$ iff it does in $\lambdaallocs$.
\end{Lemma}
\begin{pf}
  We show the following:
  \begin{enumerate}
\item If $\Conf \reduct^* \A$, then $\Conf \reducta^* \repr{\A}$.
   The reduction sequence in $\lambdaalloc$ is one in $\lambdaallocs$
   where some steps become equalities.

 \item Conversely, a reduction sequence to an answer in
   $\lambdaallocs$ corresponds to a sequence of reductions and
   anti-reductions in $\lambdaalloc$, which by strong commutation
   (Lemma~\ref{lemma-strong-commutation}) lead to a sequence of
   reductions.

\item If $\Conf$ loops in $\lambdaalloc$, then $\Conf$ also loops in
  $\lambdaallocs$, because any infinite reduction sequence
  involves an infinite number of rules \RuleBetaalloc and 
  \RuleSelectalloc.

\item Conversely, we obtain from any infinite reduction sequence in
  $\lambdaallocs$ an infinite sequence of reductions and
  anti-reductions in $\lambdaalloc$, with an infinite number of
  \RuleBetaalloc and \RuleSelectalloc reductions and no such
  anti-reduction.  By strong commutation, this yields an infinite
  reduction sequence in $\lambdaalloc$.

\item Finally, faulty configurations are the same in both calculi.
\end{enumerate}

\end{pf}

\subsubsection{Equating the three translations}

We first show that the three translations coincide as functions to
$\lambdaallocs$. First, we have the following for the \ALLOC
translation. 

\begin{Proposition}
\label{proposition-top-translation-of-values}
For all $\val$, $\Transl{\val} \aeq
\TTransl{\Pcont}{\val}$. 
\end{Proposition}

\begin{pf}
  By definition of $\TTfun$. 
\end{pf}

\begin{Proposition}
\label{proposition-translation-of-values-reduces-to-TOP}
For all $\val$, $(\config{\emptyheap}{\transl{\val}})
\eqallocr \Transl{\val}$. 
\end{Proposition}

\begin{pf}
  Trivial for variables. For other values, apply
  Proposition~\ref{proposition-values-std-top}. 
\end{pf}

\begin{Proposition}
\label{proposition-standard-translation-reduces-to-semi-top-translation}
For all $\e$, $(\config{\emptyheap}{\transl{\e}}) \eqallocr
\Transl{\e}$. 
\end{Proposition}

\begin{pf}
  By induction on $\e$, using
  Propositions~\ref{proposition-translation-of-values-reduces-to-TOP}
  and~\ref{proposition-allocate-in-context}. 
\end{pf}

Consider now the \TOP translation of bindings.  It splits the bindings
in two, cutting at the first non-size-respecting~or non-evaluated
definition. But of course, one could split at another point, provided
the first part is size-respecting. Indeed, the first part is given as
an argument to the $\Transltopfun$ function, which is defined only on
size-respecting, evaluated bindings, whereas the second part is given
as an argument to the $\TDumfun$ and $\TUpfun$ functions, which work
as well on value and non-value definitions.

\begin{Definition}[Partial translation]
  For all $\bb \aeq (\bv,\bb')$, let the \intro{partial translation of
    $\bb$ up to $\bv$} be $\TDum{\bb'} \rond \Transltopid{\bv} \rond
  \TUp{\bb'}$. 
\end{Definition}

The partial translation of $\bb$ up to $\bv$ is its \TOP
translation, computed as if $\bb'$ did not begin with a
size-respecting definition. In fact, any partial translation is
$\eqalloc$-equivalent to the \TOP translation, as we now show, using
the following properties of the functions $\TDumfun$ and $\TUpfun$,
and of substitution. 

\begin{Proposition}
\label{proposition-update-tupdate}
For all $\Conf \aeq (\config{\emptyheap}{\E})$, and $\bb$,
$\Appcontext{\Update{\bb}}{\Conf} \eqallocr
\Appcontext{\TUp{\bb}}{\Conf}$, using the notation of
Section~\ref{subsubsection-generalized-contexts} for coercing bindings
to generalized contexts.
\end{Proposition}

\begin{pf}
  By
  Propositions~\ref{proposition-standard-translation-reduces-to-semi-top-translation}
  and~\ref{proposition-allocate-in-context}.
\end{pf}

\begin{Proposition}
\label{proposition-dummy}
For all $\bb, \B$, and $\E$, 
$$\config{\emptyheap}{\letin{\Dummy{\bb}, \B}{\E}} \eqalloc
\Appcontext{\TDum{\bb}}{\config{\emptyheap}{\letin{\B}{\E}}}.$$
\end{Proposition}

\begin{pf}
  By induction on $\bb$ and rules \RuleAllocatealloc and \RuleLetalloc. 
\end{pf}

\begin{Proposition}
\label{proposition-semi-commuting-atomic-substitution}
For all $\V, \sub$, and $x \notin \FV{\sub}$, 
$\sbst{x \repl \sub(\V)} \rond \sub \meq \sub \rond \sbst{x \repl \V}$. 
\end{Proposition}

The key lemma (\ref{lemma-one-step}) then states that the in-place
update machinery indeed computes the expected recursive definition.
Hypothesis~\ref{hypothesis-size} is crucial here, ensuring that the
update is valid.

\begin{Lemma}
\label{lemma-one-step}
For all $\Conf \aeq \config{\emptyheap}{\E}$ and size-respecting $\bvo
\aeq (\bv, x \is \val)$, it holds that
$$
\begin{array}[c]{l}
\Appcontext{(\TDum{x \is \val} \rond \Transltopid{\bv} \rond \TUp{x \is \val, \bb})}{\Conf} \\
{} \eqalloc \Appcontext{(\Transltopid{\bvo} \rond \Update{\bb})}{\Conf}, 
\end{array}
$$
using the notation of
Section~\ref{subsubsection-generalized-contexts} for coercing bindings
to generalized contexts.
\end{Lemma}

\begin{pf}

Let $\BBdx \qet \TDum{x \is \val} \teq \Topbv{\Thdx}{\identity}{\vadx}$
and $\BBbv \qet \Transltopid{\bv} \teq \Topbv{\Thbv}{\subbv}{\vabv}$.
Let also 
$\Conf_1 \aeq \Appcontext{\BBdx \rond \BBbv \rond \TUp{x \is \val, \bb}}{\Conf}$
and $\Conf_2 \aeq \Appcontext{(\Transltopid{\bvo} \rond \Update{\bb})}{\Conf}$. 

First, we have 
$\Transltopid{\bvo} \aeq \BBbv \rondtopbv \Transltop{\BBbv}{x \is \val}$. 

Then, we proceed by case analysis on $x \is \val$. 
\begin{itemize}

\item $(x \is \val) \aeq (x \ex{n} \val)$ with $\Sizeab{\val} \meq n$.
  Then, $\val$ is not a variable. Thus, by definition of $\Translfun$,
  $\Transl{\val}$ has the shape $\config{\loc \eql \Hv}{\loc}$, for
  some $\loc \notin \FV{\Hv}$.  By $\alpha$-equivalence, we may choose
  another fresh location $\loc'$ such that $\Transl{\val} \aeq
  (\config{\loc' \eql \Hv}{\loc'})$. It then holds that $\Thdx \aeq
  (\heap{\loc \heql \alloc~ n})$ and $\vadx \meq \sbst{x \repl \loc}$,
  for some $\loc$.

  Let $\sub_1 \meq (\vadx \fu \vabv) \rond \subbv$. We have:
  $$\begin{array}{rl}
  \Conf_1 \aeq & 
(\config{{\loc \heql \alloc n}, {\loc' \eql \sub_1(\Hv)}, \sub_1(\Thbv) \\ & }{
  \letin{\wc \eql \update~ \loc~\loc', \sub_1(\Update{\bb}) }{\sub_1(\E)}}) \\
  {} \eqalloc &
(\config{{\loc \heql \sub_1(\Hv)}, {\loc' \eql \sub_1(\Hv)}, \sub_1(\Thbv) \\ & }{
  \letin{\sub_1(\Update{\bb}) }{\sub_1(\E)}}) \\
\multicolumn{2}{r}{\mbox{(by rules \RuleUpdatealloc and \RuleLetalloc),}}
\end{array}$$
because $\Sizealloc{\sub_1(\Hv)} \meq \Sizealloc{\Hv} \meq \Sizeab{\val} \meq n \meq \Sizealloc{\alloc n}$, 
by Hypotheses~\ref{hypothesis-size-alloc} and~\ref{hypothesis-size}. 
But then, $\loc'$ is unused, so the obtained configuration reduces 
by rule \RuleGCalloc to
  $$\begin{array}{l}
    \Conf'_1 \aeq  
    \config{{\loc \heql \sub_1(\Hv)}, \sub_1(\Thbv)}{
      \sub_1(\letin{\Update{\bb} }{\E})} \\
    {} \aeq \Appcontext{(\Topbv{\heap{\loc \eql \Hv, \Thbv}}{\subbv}{(\vadx \fu \vabv)})}
    {\config{\emptyheap}{\letin{\Update{\bb}}{\E}}} \\
    {} \aeq \Appcontext{(\Transltopid{\bvo} \rond \Update{\bb})}{\Conf} \aeq \Conf_2.
  \end{array}$$

\item $(x \is \val) \aeq (x \uu \val)$. Then, $\BBdx \qet \Topbv{\emptyheap}{\identity}{\identity}$. 
  Let $\Transl{\val} \teq \config{\Thv}{\V}$. 
  We have $\Transltop{\BBbv}{x \is \val} \aeq \Topbv{\Thv}{\sbst{x \repl \V}}{\identity}$. 

  Now, let $\Th_1 \aeq \hfu{\Thv}{\Thbv}$ and $\sub_1 \meq \vabv \rond \subbv$.  
  We have $\Conf_1 \aeq \config{\sub_1(\Th_1)}{\sub_1(\letin{x \eql
  \V, \Update{\bb}}{\E})}$.  By rule \RuleLetalloc, we
  have 
  $$\Conf_1 \eqalloc \config{\sub_1(\Th_1)
  }{ \sbst{x \repl \sub_1(\V)}(\sub_1(\letin{\Update{\bb}}{\E}))}.$$ 
  
  But $\bb_1$ may not contain forward references to definitions of
  unknown size, so the definitions of $\bvo$ can not depend on $x$.
  So, $\sub_1(\Th_1) \aeq \sbst{x \repl
    \sub_1(\V)}(\sub_1(\Th_1))$, and moreover, by
  Proposition~\ref{proposition-semi-commuting-atomic-substitution}, we
  have $\sbst{x \repl \sub_1(\V)} \rond \sub_1 \meq \sub_1 \rond
  \sbst{x \repl \V}$.  So, the obtained configuration is equal to
  $\config{(\sub_1 \rond \sbst{x \repl
        \V})(\Th_1)}{(\sub_1 \rond \sbst{x
        \repl \V})(\letin{\Update{\bb}}{\E})},$ which is $\Conf_2$, since $\sub_1
  \rond \sbst{x \repl \V} \meq \vabv \rond (\subbv \rond \sbst{x \repl
    \V})$.

\end{itemize}

\end{pf}

We then obtain the following. 

\begin{Lemma}
\label{lemma-finish-translation-of-bindings-bis}
For all $\bv,\bvo, \bb$, and $\Conf \aeq \config{\emptyheap}{\E}$, if
$(\bv, \bvo)$ is size-respecting, then $\Appcontext{(\TDum{\bw} \rond
  \Transltopid{\bv} \rond \TUp{\bw, \bb})}{\Conf} \eqalloc
\Appcontext{(\Transltopid{\bv, \bw} \rond \TUp{\bb})}{\Conf}.  $
\end{Lemma}

\begin{pf}
  By induction on $\bw$. The base case is obvious. For the induction
  step, assume that $\bw \aeq (x \is \val, \bvi)$. We have $\TDum{\bw}
  \aeq \hfu{\TDum{x \is \val}}{\TDum{\bvi}}$. By
  Lemma~\ref{lemma-one-step}, 
  $$\begin{array}{l}
    \Appcontext{(\TDum{x \is \val} \rond \Transltopid{\bv} \rond \TUp{x \is \val, \bvi, \bb})}{\Conf} \\
    {}  \eqalloc \Appcontext{(\Transltopid{\bv, x \is \val} \rond \Update{\bvi, \bb})}{\Conf} \\
    {}  \eqalloc \Appcontext{(\Transltopid{\bv, x \is \val} \rond \TUp{\bvi, \bb})}{\Conf} 
    \ \ \ \ \  \textrm{(by Proposition~\ref{proposition-update-tupdate})}.    
  \end{array}$$
  This obviously gives (using Proposition~\ref{proposition-lift-context-application})
$$\begin{array}{l}
\Appcontext{(\TDum{x \is \val, \bvi} \rond \Transltopid{\bv} \rond \TUp{x \is \val, \bvi, \bb})}{\Conf}  \\
{} \aeq \Appcontext{\TDum{\bvi}}{\Appcontext{(\TDum{x \is \val} \rond \Transltopid{\bv} \rond \TUp{x \is \val, \bvi, \bb})}{\Conf}}  \\
{} \eqalloc \Appcontext{(\TDum{\bvi} \rond \Transltopid{\bv, x \is \val} \rond \TUp{\bvi, \bb})}{\Conf}.
\end{array}$$
By induction hypothesis, we obtain 
$$\begin{array}{l}
  \Appcontext{(\TDum{\bvi} \rond \Transltopid{\bv, x \is \val} \rond \TUp{\bvi, \bb})}{\Conf} \\
{} \eqalloc \Appcontext{(\Transltopid{\bv, \bw} \rond
  \TUp{\bb})}{\Conf},
\end{array}$$ which gives the expected result.
\end{pf}

\begin{Lemma}
For all $\bb$ and $\E$, 
$$\config{\emptyheap}{\letin{\Dummy{\bb}, \Update{\bb}}{\E}} \eqalloc
  \Appcontext{\TT{\bb}}{\config{\emptyheap}{\E}}.$$
\end{Lemma}

\begin{pf}
  First, if $\bb$ is empty, then the results holds by application of
  rule \RuleEmptyLetalloc, which is included in $\eqalloc$.

  Otherwise, we have
  $$\begin{array}{l}
    \config{\emptyheap}{\letin{\Dummy{\bb}, \Update{\bb}}{\E}} \\
    {} \eqalloc 
    \Appcontext{\TDum{\bb}}{\config{\emptyheap}{\letin{\Update{\bb}}{\E}}} \hfill \mbox{  (By Proposition~\ref{proposition-dummy})}\\
    {} \aeq
    \Appcontext{\TDum{\bb}}{\Appcontext{\Update{\bb}}{\Conf}} \hfill \mbox{ (For $\Conf \aeq \config{\emptyheap}{\E}$)} \\
    {} \eqalloc 
    \Appcontext{\TDum{\bb}}{\Appcontext{\TUp{\bb}}{\Conf}} \hfill \mbox{ (By Proposition~\ref{proposition-update-tupdate})} \\
    {} \eqalloc 
    \Appcontext{(\TDum{\bb} \rond \TUp{\bb})}{\Conf} \hfill \mbox{ (By Proposition~\ref{proposition-lift-context-application}).}
  \end{array}$$

  Now, $\bb$ may be decomposed as $\bb \aeq (\bvo, \bbo)$, where $\bbo$ does not begin with
  a size-respecting definition.
  By Lemma~\ref{lemma-finish-translation-of-bindings-bis} with $\bv \meq \emptysequence$, 
  we have 
  $$\Appcontext{(\TDum{\bvo} \rond \TUp{\bvo, \bbo})}{\Conf} \eqalloc \Appcontext{(\Transltopid{\bvo} \rond \TUp{\bbo})}{\Conf},$$
  which gives
  $$\begin{array}{l}
    \Appcontext{\TDum{\bb} \rond \TUp{\bb})}{\Conf} \\
    {} \aeq \Appcontext{\TDum{\bvo, \bbo} \rond \TUp{\bvo, \bbo})}{\Conf} \\
    {} \aeq \Appcontext{\TDum{\bbo}}{\Appcontext{(\TDum{\bvo} \rond \TUp{\bvo, \bbo})}{\Conf}} 
    \hfill \mbox{(By Proposition~\ref{proposition-lift-context-application})}\\
    {} \eqalloc \Appcontext{\TDum{\bbo}}{\Appcontext{(\Transltopid{\bvo} \rond \TUp{\bbo})}{\Conf}} \hfill 
      \mbox{(By Proposition~\ref{proposition-weak-context-reduction})} \\
    {} \aeq \Appcontext{(\TDum{\bbo} \rond \Transltopid{\bvo} \rond \TUp{\bbo})}{\Conf} 
    \hfill \mbox{(By Proposition~\ref{proposition-lift-context-application})}\\
    {} \aeq \Appcontext{\TT{\bb}}{\Conf} \hfill \mbox{(By definition of $\TTfun$)}.
  \end{array}$$

\end{pf}

\begin{Corollary}
\label{corollary-relate}
For all $\bb$ and $\e$, 
  $\config{\emptyheap}{\transl{\letrecin{\bb}{\e}}} \eqalloc
  \Appcontext{\TTranslid{\bb}}{\config{\emptyheap}{\transl{\e}}}.$
\end{Corollary}

Finally, the following lemma states that the three translations
$\translfun$, $\Translfun$, and $\TTfun$ are equal as functions from
$\lambdaab$ to $\lambdaallocs$.

\begin{Lemma}
  For all $\e$, it holds that $(\config{\emptyheap}{\transl{\e}}) \eqalloc \Transl{\e}
  \eqalloc \TT{\e}.$
\end{Lemma}

\begin{pf}
  Proposition~\ref{proposition-standard-translation-reduces-to-semi-top-translation}
  directly implies $(\config{\emptyheap}{\transl{\e}}) \eqalloc \Transl{\e}$.
  
  To prove $\Transl{\e} \eqalloc \TT{\e}$, we proceed by case analysis
  on $\e$.  If $\e$ is not of the shape $\letrecin{\bb}{\e'}$, then
  the result follows by definition of $\TTfun$. Otherwise, by
  Corollary~\ref{corollary-relate}, we have $\Transl{\e} \aeq
  \config{\emptyheap}{\transl{\letrecin{\bb}{\e'}}} \eqalloc
  \Appcontext{\TT{\bb}}{\config{\emptyheap}{\transl{\e'}}}$, so we
  just have to prove
  $\Appcontext{\TT{\bb}}{\config{\emptyheap}{\transl{\e'}}} \eqalloc
  \TT{\letrecin{\bb}{\e'}}$.  If $\bb$ is not size-respecting, then
  the result holds by definition of $\TTfun$.  Otherwise, we have
  $\TT{\letrecin{\bb}{\e'}} \aeq \Appcontext{\TT{\bb}}{\Transl{\e'}}$.
  But by
  Proposition~\ref{proposition-standard-translation-reduces-to-semi-top-translation},
  $(\config{\emptyheap}{\transl{\e'}}) \eqallocr \Transl{\e'}$, so by
  Proposition~\ref{proposition-allocate-in-context} we obtain
  $\Appcontext{\TT{\bb}}{\Transl{\e'}} \eqallocr
  \Appcontext{\TT{\bb}}{\config{\emptyheap}{\transl{\e'}}}$, which
  gives the expected result.
\end{pf}

\subsubsection{Compositionality}

For proving that the evaluation of an expression in $\lambdaab$
corresponds to the evaluation of its translation in $\lambdaalloc$, we
seek compositionality properties of our translations.  The \STD
translation is obviously compositional, in the following sense. 
\begin{Definition}[Standard translation of contexts]
\label{definition-std-translation-of-contexts}
Define $\transl{\cont}$ by extension of $\translfun$ on expressions,
with $\transl{\trou} \teq \trou$. 
\end{Definition}

\begin{Proposition}
For all $\cont$ and $\e$, 
$\transl{\context{\e}} \aeq \Appcontext{\transl{\cont}}{\transl{\e}}.$
\end{Proposition}

\begin{pf}
  By trivial induction on $\cont$. 
\end{pf}

\begin{figure}
\begin{framed}
$\begin{array}{lclp{0.1\textwidth}}
\MkGen{\config{\Th}{\Fcont}} & \aeq &
                    \GCC{\Th}{\Fcont}{\identity}{\identity} \\ 
\TTranslid{\fcont} & \teq & \MkGen{\Transl{\fcont}} & {} \\
\TTranslid{\letrecin{\bv}{\fcont}} & \teq & \TT{\bv} \rond \MkGen{\Transl{\fcont}} \\
\TTranslid{\letrecin{\bv, x \is \fcont, \bb}{\e}} & \teq & \\
\multicolumn{4}{r}{
\TDum{x \is \fcont, \bb} \rond \Transltopid{\bv} \rond \TUpe{x}{\is}{\fcont}{\bb}{\transl{\e}}} \\
\TUpe{x}{\is}{\fcont}{\bb}{\E} & \teq & 
\GCC{\Th}{\letin{\bd \eql \Fcont, \B}{\E}}{\identity}{\identity} \\
\multicolumn{4}{r}{\mbox{if $\TUp{x \is \fcont, \bb} \aeq \GCC{\Th}{(\bd \eql \Fcont, \B)}{\identity}{\identity}$
}}
\end{array}$
\end{framed}
\caption{Top-level translation of contexts from $\lambdaab$ to $\lambdaalloc$}
\label{figure-top-translation-of-contexts}
\end{figure}

However, we have seen that $\translfun$ does not lend itself to a
simulation argument, so we consider the compositionality of $\TTfun$.
We obtain below in Corollary~\ref{corollary-weak-compositionality}
that for all expressions $\e$ and evaluation contexts $\cont$,
$\TT{\context{\e}} \eqalloc \Appcontext{\TT{\cont}}{\Transl{\e}}$.
This is not exactly what one could have hoped for ($\TT{\context{\e}}
\eqalloc \Appcontext{\TT{\cont}}{\TT{\e}}$) but it will be enough to
prove correctness of our translation.

Figure~\ref{figure-top-translation-of-contexts} defines
$\MkGen{\config{\Th}{\Fcont}}$ as the obvious generalized context made
of $\Th$ and $\Fcont$, using the fact that nested lift contexts are
allocation contexts.  Then, define $\Translfun$ on nested lift
contexts by extension of $\Translfun$ on expressions: we consider
$\trou$ not to be a value, and put $\Transl{\trou} \teq
\config{\emptyheap}{\trou}$. 
For any $\fcont$, the translation $\TTranslid{\fcont}$ has 
the shape $\config{\Th}{\Fcont}$ for some $\Th$ and $\Fcont$.
This gives

\begin{Proposition}
\label{proposition-is-a-generalized-context}
  For all $\fcont$, $\MkGen{\Transl{\fcont}}$ is a generalized evaluation
  context. 
\end{Proposition}
\begin{pf}
  By induction on $\fcont$ and case analysis on lift contexts. 
\end{pf}

Figure~\ref{figure-top-translation-of-contexts} then defines the
translation of evaluation contexts.  The translation $\TDum{x \is
  \fcont, \bb}$ has no hole; $\TUp{x \is \fcont, \bb}$ has two: one
from $\fcont$, plus one for the body of the returned $\letseq$-binding
(present for any $\TUp{\bb}$). The special notation
$\TUpe{x}{\is}{\fcont}{\bb}{\E}$ fills the latter with $\E$.  We
obtain the following immediately.

\begin{Proposition}
  For all $\cont$, $\TT{\cont}$ is a generalized evaluation
  context whose variable allocation is $\identity$. 
\end{Proposition}
\begin{pf}
  By case analysis on $\cont$ and
  Proposition~\ref{proposition-is-a-generalized-context}, $\TT{\cont}$
  is a generalized evaluation context.  By case analysis, we then
  prove that its variable allocation is $\identity$. If $\cont$ is a
  nested lift context, then by definition of $\MkGenfun$, it is the
  case.  In both other cases, $\TT{\cont}$ is defined as the weak
  composition of more than one generalized evaluation context, which
  by definition has $\identity$ as its variable allocation.
\end{pf}

This allows stating the expected compositionality result. 
\begin{Lemma}
\label{lemma-compositionality-for-evaluation-contexts}
For $\cont \aeq \trou$ and all $\e$,
$\TTranslid{\context{\e}} \aeq \Appcontext{\TT{\cont}}{\TT{\e}}$. 

\noindent
For $\cont \naeq \trou$ and all $\e$, if $\e \notin \values$,
then
$\TTranslid{\context{\e}} \aeq \Appcontext{\TTranslid {\cont}}{\Transl{\e}}.$

\noindent
For all $\cont$ and $\val$, ${\TTranslid{\context{\val}}} \eqalloc
{\Appcontext{\TTranslid{\cont}}{\Transl{\val}}}$. 
\end{Lemma}

\begin{pf}
  The first point is trivial. The second is obtained for lift contexts
  first (by a simple case analysis), then for nested lift contexts by
  straightforward induction, and finally by case analysis on $\cont$. 
  As for the last point, if $\trou$ is replaced with a value, it may
  permit the \ALLOC and \TOP translations to perform more
  administrative reductions, as for instance in contexts of the shape
  $\e~\trou$. The proof uses
  Lemma~\ref{lemma-finish-translation-of-bindings-bis}. 
\end{pf}

\begin{Corollary}[Weak compositionality]
\label{corollary-weak-compositionality}
For all $\cont$ and $\e$, $\TT{\context{\e}} \eqalloc
\Appcontext{\TT{\cont}}{\Transl{\e}}$. 
\end{Corollary}

Full compositionality does not hold: $\TT{\context{\e}}$ is not always
$\eqalloc$-equivalent to $\Appcontext{\TT{\cont}}{\TT{\e}}$. The
reason is because $\eqalloc$ does not include rule \RuleIMalloc, as
shown by taking $\cont \aeq (\letrecin{x \uu \trou}{x})$ and $\e \aeq
(\letrecin{y \uu \record{\X \eql z}, z' \uu y.\X}{y})$.  In this case,
$\TT{\context{\e}} \aeq (\config{\emptyheap}{ \letin{x \eql (\letin{y
      \eql \record{\X \eql z}, z' \eql y.\X}{y})}{x}})$, and
$\Appcontext{\TT{\cont}}{\TT{\e}} \aeq (\config{\loc \eql \record{\X
    \eql z}}{\letin{x \eql (\letin{z' \eql \loc.\X}{\loc})}{x}})$.  An
application of \RuleIMalloc is needed in order to relate them.

Further quotienting $\lambdaallocs$ by this rule might lead to full
compositionality.  Moreover, we think that it would preserve the good
properties of the translation. In particular, $\lambdaab$ reductions
by rule \RuleIMab cannot be infinite, so non-termination would remain
correctly simulated by the translation. However, this is not needed
to complete our correctness proof, so we did not investigate it.

\subsection{\QuotientingLambdaab}

We now define $\lambdaabs$, based on the following notions of binding
scraping $\scrapev{\bb}{\vs}{x}$ and context scraping
$\scrapecont{\cont}{\val}$.  The intuition is that
$\scrapecont{\cont}{x}$ does much the same work as iterating the rule
\RuleSubstab until a non-variable value is found. Below, we use it to
replace rule \RuleSubstab, in the case where $\cont$ is dereferencing.
In such cases, if there is no non-variable value for $x$, then
$\context{x}$ is faulty, so we do not have to consider it.
Technically, $\scrapecont{\cont}{x}$ is then undefined.

\begin{Example}
  Let $\bv \aeq (x \uu \fun x'.x', y \uu x)$ and consider $\e \aeq
  (\letrecin{\bv}{(y~\record{})}).$ In order to reduce to $\e' \aeq
  (\letrecin{\bv}{((\fun x'.x')~\record{})})$, $\e$ takes two
  \RuleSubstab steps. In $\lambdaabs$, we will directly replace $y$
  with $\scrapecont{(\letrecin{\bv}{\trou})}{y}$, which is $\fun
  x'.x'$, and perform the \RuleBetaab step on-the-fly.
\end{Example}

\begin{Definition}[Binding scraping]
\label{definition-scrape-bindings}
For all sets $\vs$ of variables, bindings $\bb$ (not necessarily
size-respecting), and variables $x \in \dom{\bb}$, define binding
scraping recursively by: $$\begin{array}[t]{lcll}
  \scrapev{\bb}{\vs}{x} &\aeq& \bb(x) & \mbox{if $\bb(x) \notin \Vars$ or $\bb(x) \in \Vars \setminus \dom{\bb}$} \\
  \scrapev{\bb}{\vs}{x} &\aeq& \vcycle & \mbox{if $\bb(x) \in \dom{\bb} \cap \vs$} \\
  \scrapev{\bb}{\vs}{x} &\aeq& \scrapev{\bb}{(\ens{x} \cup \vs)}{\bb(x)}
  & \mbox{if $\bb(x) \in \dom{\bb} \setminus \vs$.}
\end{array}$$

For all such $\bb$ and $x$, if $\scrapev{\bb}{\emptyset}{x} \naeq \vcycle$, define $\scrape{\bb}{x} \aeq
\scrapev{\bb}{\emptyset}{x}$. 
\end{Definition}

\begin{Definition}[Context scraping]
\label{definition-accessible-binding}

Define
$\begin{array}[t]{lcl}
\scrapecont{\cont}{x} &\aeq& \scrape{(\Binding{\cont})}{x} \mbox{ if $x \in \dom{\Binding{\cont}}$} \\
\scrapecont{\cont}{\val} &\aeq& \val \mbox{ if $\val \notin \Vars$ or $\val \in \Vars \setminus \dom{\Binding{\cont}}$.} 
\end{array}$ 

\end{Definition}

Let us now prove elementary properties of binding scraping. 

\begin{Lemma}
  Binding scraping is well-defined, i.e., for all $\bb$ and $\vs$,
  $\scrapevfun{\bb}{\vs}$ is a total function.
\end{Lemma}
\begin{pf}
Let the measure $\measurefunction$ be defined from
pairs of a  binding and a set of variables to
natural numbers by
$\measure{\bb, \vs} \meq \cardinal{\dom{\bb} \setminus \vs}$, the
cardinality of $\dom{\bb} \setminus \vs$. 

First, we notice that if $\measure{\bb, \vs} \meq 0$, then binding
scraping immediately returns, on any variable $x \in \dom{\bb}$. 
Indeed, if $\bb(x) \notin \Vars$, it returns $\bb(x)$.  Otherwise, if
the variable $\bb(x)$ is in $\dom{\bb}$, then it is also in $\vs$, so
$\scrapev{\bb}{\vs}{x} \meq \vcycle$, and if the variable $\bb(x)$ is not in
$\dom{\bb}$, then $\scrapev{\bb}{\vs}{x} \meq \bb(x)$. 

Then, as the measure decreases by $1$ at each recursive call, we
conclude that $\scrapev{\bb}{\vs}{x}$ is well-defined for any $x \in
\dom{\bb}$. 
\end{pf}

\begin{Proposition}
  \label{proposition-scraped-var-notin-dombv}
  For all $\bb, \vs, y,$ and $x \in \dom{\bb}$, if $\scrapev{\bb}{\vs}{x} \aeq y$, then
  $y \notin \dom{\bb}$.
\end{Proposition}

\begin{pf}
  By induction on the proof of $\scrapev{\bb}{\vs}{x} \aeq y$.  The
  base case is when $\bb(x) \aeq y$ and $y \notin \dom{\bb}$, which
  gives immediately the expected result.  The induction step is when
  there exists $z \in \dom{\bb} \setminus \vs$ such that
  $\scrapev{\bb}{\ens{x} \cup \vs}{z} \aeq y$.  By induction
  hypothesis, this gives $y \notin \dom{\bb}$ as expected.
\end{pf}

We now define $\lambdaabs$ as having the same expressions as
$\lambdaab$, but a different reduction relation, written $\reducts$.

\begin{Definition}
\label{definition-merging}
Let the \intro{merging} $\mergecont{\cont}{\bb}{\e}$ of $\letrecin{\bb}{\e}$ into the context $\cont$
be defined as

\begin{itemize}
\item $\letrecin{\bb}{\e}$ if $\cont \aeq \trou$,
\item and otherwise the result of normalizing
  $\context{\letrecin{\bb}{\e}}$ w.r.t.~rule
  \RuleContextab/\RuleLiftab, plus, if $\cont$ had a top-level
  binding, applying \RuleIMab or \RuleEMab once.
\end{itemize}

\end{Definition}

  Note that the capture-avoiding side conditions of
  Definition~\ref{definition-merging} are always satisfiable by bound
  variable renaming.  


Then, we define $\reducts$ relatively to $\reduct$ by removing rules
\RuleBetaab, \RuleSelectab, and \RuleSubstab, and adding the following
three rules:
  \begin{mathpar}
    \inferrule[\RuleBetaabs]{
      \scrapeval{\cont}{\val_0} \meq \fun y . \e
    }{
      \context{\val_0~\val} \reducts \mergecont{\cont}{y \uu \val}{\e}
    }
    \and
    \inferrule[\RuleSelectabs]{
      \scrapeval{\cont}{\val_0} \meq \record{\sv}
    }{
      \context{\val_0.\X} \reducts \context{\sv(\X)}
    }
    \and
    \inferrule[\RuleUpdateabs]{
      \scrape{\bv}{y} \meq \val \\ \Sizeab{\val} \meq n
    }{
      \letrecin{\bv,x \ex{n} y,\bb}{\e} \reducts \letrecin{\bv, x \ex{n} \val,\bb}{\e} .
    } 
  \end{mathpar}

  Observe that all $\lambdaab$ rules are simulated in $\lambdaabs$,
  except rule \RuleSubstab. Indeed, rules \RuleBetaab and
  \RuleSelectab are special cases of rules \RuleBetaabs and
  \RuleSelectabs. Rule \RuleSubstab, albeit not directly simulated,
  yields a simulation w.r.t. our obserables: evaluation answers,
  non-termination, and faultiness, as we now show.

  \begin{Lemma}
\label{lemma-pre-scrape}
    For all $\dcont$ and $\bv \aeq \Binding{\dcont}$, 
    for all $x \in \Vars$, $\val \notin \Vars$, and
    finite sets of variables $\vs$, if $x \notin \vs$ and
    $\scrapev{\bv}{\vs}{x} \aeq \val$, then there exists a
    value $\val'$ such that $\dcont(x) \aeq \val'$ and
    $\dcontext{\val'} \reduct^* \dcontext{\val}$.
  \end{Lemma}

  \begin{pf}
    We proceed by induction on the proof of
  $\scrapev{\bv}{\vs}{x} \aeq \val$.

\begin{itemize} \item
  If $\bv(x) \in \Vars \setminus \dom{\bb}$, then
  $\scrapev{\bv}{\vs}{x} \naeq \val$, contradiction.

\item  
  If $\bv(x) \aeq \val$, then, taking
  $\val' \aeq \val$, we have $\dcont(x) \aeq \bv(x) \aeq \val$ and
  $\dcontext{\val} \reduct^* \dcontext{\val}$ by reflexivity, as
  expected.

\item
  If $\bv(x) \in \dom{\bb} \cap \vs$, then $\scrapev{\bv}{\vs}{x} \aeq \vcycle$, contradiction.

\item If $\bv(x) \in \dom{\bb} \setminus \vs$, let $y \meq \bv(x)$.  We
  know $\scrapev{\bv}{\ens{x} \cup \vs}{y} \aeq \val$, so $y \notin
  \ens{x} \cup \vs$.  Thus, by induction hypothesis, there exists a
  $\val''$ such that $\dcont(y) \aeq \val''$ and $\dcontext{\val''}
  \reduct^* \dcontext{\val}$.  But then, $\dcontext{y} \reduct
  \dcontext{\val''} \reduct^* \dcontext{\val}$. So, taking $\val' \aeq
  y$, we obtain $\dcont(x) \aeq \val'$ and $\dcontext{\val'} \reduct^*
  \dcontext{\val}$.

\end{itemize}  

  \end{pf}

\begin{Lemma}
  \label{lemma-scrape}
 For all $\dcont,x,$ and $\val \notin \Vars$, if $\dcontext{x}
  \reduct^+ \dcontext{\val}$ then $\scrapecont{\dcont}{x} \aeq \val$. 
\end{Lemma}

\begin{pf}
  By Lemma~\ref{lemma-pre-scrape}, there exists $\val'$ such that
  $\dcont(x) \aeq \val'$ and $\dcontext{\val'} \reduct^* \dcontext{\val}$, 
  which immediately gives the expected result.
\end{pf}

Finally:

\begin{Lemma}
\label{lemma-encoding-lambdaab-lambdaabs}
For all $\e$, if $\e$ reduces to an answer, loops, or is faulty
in $\lambdaab$, then so it does in $\lambdaabs$.
\end{Lemma}
\begin{pf}
  The lemma says:
  \begin{enumerate}
  \item if $\e \reduct^* \answer$, then $\e \reducts^* \answer$;
  \item if $\e$ loops in $\lambdaab$, i.e., there exists an infinite
    reduction sequence starting from $\e$, then $\e$ also loops in
    $\lambdaabs$;
  \item if $\e$ is faulty in $\lambdaab$, then it is also faulty in
    $\lambdaabs$. 
  \end{enumerate}

  First, consider a reduction sequence from $\e$ to a normal form
  $\e_1$ in $\lambdaab$. We prove by induction on its length that
  \begin{itemize}\item 
    if $\e_1$ is an answer, then $\e$ reduces to an answer in
    $\lambdaabs$, and
  \item if $\e_1$ has the shape $\dcontext{\val}$, i.e., $\e$ is
    faulty in $\lambdaab$, then $\e$ is faulty in $\lambdaabs$ too.
  \end{itemize}
  The base case is trivial. For the induction step, if the first
  reduction step in the given sequence is not \RuleSubstab, then it is
  simulated in $\lambdaabs$, so we get the expected result by
  induction hypothesis. Otherwise, $\e \aeq \dcontext{x}$ and the
  first step has the shape $\dcontext{x} \reduct \dcontext{\val}$,
  with $\val \hts \dcont(x)$. Consider the maximal subsequence of the 
  given reduction sequence having the shape
  $$\dcontext{x} \aeq \dcontext{\val_0} \xrightarrow{\RuleSubstab}
  \dcontext{\val_1} \xrightarrow{\RuleSubstab} \ldots
  \xrightarrow{\RuleSubstab} \dcontext{\val_n}$$ with each $\val_i \naeq
  \val_{i+1}$, i.e., rule \RuleSubstab applies each time at $\dcont$.
  Thus, $n > 0$, and for $i < n$, $\val_i$ is a variable.
  
  Now, if $\dcontext{\val_n}$ is an answer, then $\dcont$ has the
  shape $\letrecin{\bcontof{\ex{m}}}{\val'}$ with $\Sizeab{\val_n} \meq
  m$, hence $\e \reducts \e_1$ by rule \RuleUpdateabs.
  
  Otherwise, if $\dcontext{\val_n}$ is not an answer, but is actually
  $\e_1$, i.e., is in normal form, then $\dcontext{x}$ is in normal
  form in $\lambdaabs$ and not an answer, hence faulty in both
  calculi.

  Otherwise, if $\dcont$ has the shape
  $\letrecin{\bcontof{\ex{n'}}}{\e'}$ for some $n'$,
  $\bcontof{\ex{n'}}$, and $\e'$, then $\e \reducts
  \dcontext{\val_n}$ by rule \RuleUpdateabs in $\lambdaabs$, and we
  conclude by induction hypothesis.

  Otherwise, $\dcontext{\val_n}$ further reduces by one of \RuleBetaab
  and \RuleSelectab, and then the corresponding rule (\RuleBetaabs or
  \RuleSelectabs) applies in $\lambdaabs$, and we again conclude by
  induction hypothesis.

  Finally, to show that non-termination is preserved, given an
  infinite reduction sequence in $\lambdaab$, build one in
  $\lambdaabs$ by the same algorithm: if the first step is not
  \RuleSubstab, then it is simulated directly, otherwise, consider the
  maximal subsequence of \RuleSubstab steps.
\end{pf}

We now quotient $\lambdaabs$ by \RuleEMab and \RuleUpdateabs to obtain
$\lambdaabsem$.

\begin{Definition}[$\lambdaabsem$]
  Define $\eqabsem$ as the smallest equivalence relation over
  $\lambdaabs$ containing the rules \RuleEMab and \RuleUpdateabs.  Let
  the terms of $\lambdaabsem$ be the set of $\eqabsem$-equivalence
  classes.  Let reduction in $\lambdaabsem$, written $\reductsem$, be
  defined by the rules:
\begin{mathpar}
\inferrule{
\e_1 \eqabsem \e'_1 \\
\e'_1 \reductbyin{R}{\overline{\circ}} \e'_2 \\
\e'_2 \eqabsem \e_2
}{
\e_1 \reductsem \e_2
}
\end{mathpar}
where $R$ ranges over the other rules (\RuleLiftab, \RuleContextab,
\RuleIMab, \RuleBetaabs, and \RuleSelectabs).
\end{Definition}

We now show that $\lambdaabsem$ simulates $\lambdaabs$, and that
$\TTfun$ remains well-defined as a function from $\lambdaabsem$ to
$\lambdaallocs$.  For this, we prove that rules \RuleEMab and
\RuleUpdateabs preserve $\TTfun$ (modulo $\eqalloc$, which
$\lambdaallocs$ is quotiented by) and that infinite $\lambdaabs$
reductions can not exclusively contain \RuleEMab or \RuleUpdateabs
reductions.  For each of these two rules, we start by defining a
measure, and show that each rule makes its measure strictly decrease,
and preserves the \TOP translation.  We start with \RuleEMab.

\begin{Definition}[Number of $\letrec$ nodes]
$\Nbletrec{\e}$ is the number of $\letrec$ nodes in $\e$. 
\end{Definition}

\begin{Lemma}[External merging]
\label{lemma-external-merging-letrec}
For all $\e$ and $\e'$, if $\e \xrightarrow{\textsc{\scriptsize EM${}_{\circ}$}} \e'$, then
$\Nbletrec{\e} > \Nbletrec{\e'}$ and $\TTranslid{\e} \eqalloc
\TTranslid{\e'}$. 
\end{Lemma}

\begin{pf} 
$\begin{array}[t]{ll}
\mbox{Let} & \e \aeq \letrecin{\bv}{\letrecin{\bb}{\e_0}} \\
\mbox{and} & \e' \aeq \letrec\ \bv, \bb\ \inlet\ \e_0. 
\end{array}$

Obviously, $\Nbletrec{\e} > \Nbletrec{\e'}$. Furthermore, we have
$$\TTranslid{\e} \teq 
\Appcontext{\Transltopid{\bv}}{\config{\emptyheap}{\letin{\Dummy{\bb},\Update{\bb}}{\transl{\e_0}}}}.$$ 
If $\bb$ is empty, after applying rule \RuleEmptyLetalloc (which is in
$\eqalloc$), the configuration becomes
$\Appcontext{\Transltopid{\bv}}{\config{\emptyheap}{\transl{\e_0}}}$, which is
$\eqalloc$-equivalent to
$\Appcontext{\Transltopid{\bv}}{\Transl{\e_0}} \qet \TT{\e'}$.

Otherwise, using Proposition~\ref{proposition-weak-context-reduction},
and after checking that $\Transltopid{\bv}$ is \anextheap, 
we have 
$$\begin{array}{l}
  \TT{\e} \aeq \Appcontext{\Transltopid{\bv}}{\config{\emptyheap}{\letin{\Dum{\bb}, \Update{\bb}}{\transl{\e_0}}}} 
  \\[.5em]
  {} \eqalloc
  \Appcontext{(\Transltopid{\bv} \rond
    \TDum{\bb})}{\config{\emptyheap}{\letin{\Update{\bb}}{\transl{\e_0}}}} \\
  \multicolumn{1}{r}{\textrm{(by Propositions~\ref{proposition-weak-context-reduction},~\ref{proposition-lift-context-application}, and~\ref{proposition-dummy})}} \\[.5em]
  {} \eqalloc \Appcontext{(\TDum{\bb} \rond
    \Transltopid{\bv})}{\config{\emptyheap}{\letin{\Update{\bb}}{\transl{\e_0}}}} \\
  \multicolumn{1}{r}{\textrm{(by Proposition~\ref{proposition-interfere}, since $\dom{\bb} \orth \dom{\bv} \cup
      \FV{\bv}$)}} \\[.5em]
  {} \eqalloc  \Appcontext{(\TDum{\bb} \rond \Transltopid{\bv} \rond
    \TUp{\bb})}{\config{\emptyheap}{\transl{\e_0}}} \\
  \multicolumn{1}{r}{\textrm{(by
      Propositions~\ref{proposition-weak-context-reduction},~\ref{proposition-lift-context-application}, and~\ref{proposition-update-tupdate})}}. 
\end{array}$$

But then, let $\bb \aeq (\bvo, \bb')$ with $\bb'$ not beginning with a size-respecting definition.
We have
$$\begin{array}{l}
  \Appcontext{(\TDum{\bb} \rond \Transltopid{\bv} \rond
    \TUp{\bb})}{\config{\emptyheap}{\transl{\e_0}}} \\
  {} \aeq
  \Appcontext{(\TDum{\bvo, \bb'} \rond \Transltopid{\bv} \rond
      \TUp{\bvo, \bb'})}{\config{\emptyheap}{\transl{\e_0}}} \\
  {} \aeq
  \Appcontext{(\TDum{\bvo} \rond \TDum{\bb'} \rond \Transltopid{\bv} \rond
      \TUp{\bvo, \bb'})}{\config{\emptyheap}{\transl{\e_0}}} \\
  {} \aeq
  \Appcontext{(\TDum{\bb'} \rond \TDum{\bvo} \rond \Transltopid{\bv} \rond
      \TUp{\bvo, \bb'})}{\config{\emptyheap}{\transl{\e_0}}} \\
  \multicolumn{1}{r}{\textrm{(by Proposition~\ref{proposition-interfere})}} \\[.5em]
  {} \aeq
  \Appcontext{\TDum{\bb'}}{\Appcontext{(\TDum{\bvo} \rond \Transltopid{\bv} \rond
      \TUp{\bvo, \bb'})}{\config{\emptyheap}{\transl{\e_0}}}} \\
  \multicolumn{1}{r}{\textrm{(by Proposition~\ref{proposition-lift-context-application})}} \\[.5em]
  {} \eqalloc
  \Appcontext{\TDum{\bb'}}{\Appcontext{(\Transltopid{\bv, \bvo} \rond
      \TUp{\bb'})}{\config{\emptyheap}{\transl{\e_0}}}} \\
  \multicolumn{1}{r}{\textrm{(by Lemma~\ref{lemma-finish-translation-of-bindings-bis}
      and Proposition~\ref{proposition-weak-context-reduction})}} \\[.5em]
  {} \aeq
  \Appcontext{(\TDum{\bb'} \rond \Transltopid{\bv, \bvo} \rond
      \TUp{\bb'})}{\config{\emptyheap}{\transl{\e_0}}} \\
  \multicolumn{1}{r}{\textrm{(by Proposition~\ref{proposition-lift-context-application}).}}
\end{array}$$

But if $\bb'$ is not empty, then this last configuration is exactly
$\TT{\e'}$.  Otherwise, if $\bb'$ is empty, then $(\TDum{\bb'} \rond
\Transltopid{\bv, \bvo} \rond \TUp{\bb'}) \aeq \Transltopid{\bv, \bvo}$
is \anextheap. But by
Propositions~\ref{proposition-standard-translation-reduces-to-semi-top-translation}
and~\ref{proposition-weak-context-reduction}, 
$$\begin{array}{l}
\Appcontext{\Transltopid{\bv, \bvo}}{\config{\emptyheap}{\transl{\e_0}}}
{} \eqalloc
\Appcontext{\Transltopid{\bv, \bvo}}{\Transl{\e_0}} 
\aeq \TT{\e'},
\end{array}$$
which gives the expected result.
\end{pf}

Next, we define a measure that strictly decreases by application of
rule \RuleUpdateabs. 

\begin{Definition}
  We define $\CoUpdatefun$ as follows:
$$\begin{array}[t]{rcll}
  \CoUpdate{\letrecin{\bv,\bb}{\e}} &\meq& \cardinal{\dom{\bb}} &
  \mbox{where $\bb$ does not begin with a} \\
 & & & \mbox{size-respecting definition.} \\[.5em]
  \CoUpdate{\e} &\meq& 0 & \mbox{if $\e$ does not begin with $\letrec$.}
\end{array}$$
\end{Definition}

The lemma for rule \RuleUpdateabs requires the following properties of
the \TOP translation, about how variables can be accessed in the
translation of a binding. 

\begin{Lemma}
\label{lemma-translation-of-access}
For all $x,\val,\Th,\sub,\va,$ and $\bv$, if $\Transltopid{\bv} \teq
\Topbv{\Th}{\sub}{\va}$ and $\scrape{\bv}{x} \aeq \val$, then there exist
$\Thv$ and $\V$ such that $\Transl{\val} \teq \config{\Thv}{\V}$, $(\va
\rond \sub)(x) \meq \V$ and $\Thv \subset \Th$. 
\end{Lemma}

\begin{pf}
  We prove more generally that for all $x,\val,\Th,\sub,\va,\vs$ and
  $\bv$, if $\Transltopid{\bv} \teq \Topbv{\Th}{\sub}{\va}$ and
  $\scrapev{\bv}{\vs}{x} \aeq \val$, then there exist $\Thv$ and $\V$
  such that $\Transl{\val} \teq \config{\Thv}{\V}$, $(\va \rond
  \sub)(x) \meq \V$ and $\Thv \subset \Th$.

  We proceed by induction on the proof of $\scrapev{\bv}{\vs}{x} \aeq
  \val$.  

  The base case amounts to proving the result with the
  additional hypothesis that $\bv(x) \aeq \val$. For this, we decompose
  $\bv$ into $\bvo,x \is \val, \bvi$.  
  By definition of
  $\Transltopfun$, 
  we have $\Transltopid{\bv} \aeq (\Transltopid{\bvo} \rondtopbv \Transltopid{x \is \val}
  \rondtopbv \Transltopid{\bvi})$.
  Let $\Transltopid{\bvo} \aeq \Topbv{\Tho}{\subo}{\vao}$,
  and $\Transltopid{\bvi} \aeq \Topbv{\Thi}{\subi}{\vai}$.

  \begin{itemize}
  \item If $\val$ is a variable $y$, then $\Transltopid{x \is \val}
    \aeq \Topbv{\emptyheap}{\sbst{x \repl y}}{\identity}$.  Let
    $\Thv \aeq \emptyheap$ and $\V \aeq y$.  We have $\sub \meq (\subo
    \rond \sbst{x \repl y} \rond \subi)$ and $\va \meq (\vao \rond
    \vai)$.  Furthermore, we know $x \notin \dom{\subi}$, and by
    Proposition~\ref{proposition-scraped-var-notin-dombv}, $y \notin
    \dom{\bv}$. This also gives $y \notin \dom{\va} \cup \dom{\sub}$,
    because $(\dom{\va} \cup \dom{\sub}) \subseteq \dom{\bv}$.  Thus
    $(\va \rond \sub)(x) \meq (\va \rond \subo)(y) \meq y$, as
    expected.

  \item If $\val$ is not a variable, then $\Transltopid{x \is \val}
    \aeq \Topbv{\Thv}{\identity}{\sbst{x \repl \loc}}$, with
    $\Transl{\val} \aeq \config{\Thv}{\loc}$. Take $\V \meq \loc$.  We
    have $\va \meq (\vao \rond \sbst{x \repl \loc} \rond \vai)$ and
    $\sub \meq (\subo \rond \subi)$.  But we know $x \notin \dom{\sub}
    \cup \dom{\vai} \cup \dom{\vao}$, so $(\va \rond \sub)(x) \meq
    (\vao \rond \sbst{x \repl \loc})(x) \meq \loc \meq \V$ as
    expected.
     
   \end{itemize}

   For the induction step, assume $\bv(x) \aeq y$ and
   $\scrapev{\bv}{\ens{x} \cup \vs}{y} \aeq \val$.  Then, $\bv$ has
   the shape $(\bvo, x \is y, \bvi)$ for some $\bvo, \bvi$.  By
   definition of $\Transltopfun$, we have $\Transltopid{\bv} \aeq
   (\Transltopid{\bvo} \rondtopbv \Transltopid{x \is y} \rondtopbv
   \Transltopid{\bvi})$.  Let $\Transltopid{\bvo} \aeq
   \Topbv{\Tho}{\subo}{\vao}$, and $\Transltopid{\bvi} \aeq
   \Topbv{\Thi}{\subi}{\vai}$.  We know that $\Th \aeq
   (\hfu{\Tho}{\Thi})$, $\sub \meq (\subo \rond \sbst{x \repl y} \rond
   \subi)$ and $\va \meq (\vao \rond \vai)$.  By induction hypothesis,
   there exist $\Thv$ and $\V$ such that $\Transl{\val} \aeq
   \config{\Thv}{\V}$, $\Thv \subseteq \Th$, and $(\va \rond \sub)(y)
   \meq \V$. Thus, there only remains to prove that $(\va \rond
   \sub)(y) \meq (\va \rond \sub)(x)$.
   \begin{itemize}
   \item If $y \in \dom{\bvi}$, then, since $\bv$ contains a forward
     reference from $x$ to $y$, $y$ has a known size indication in
     $\bv$.  So, $y \in \dom{\vai}$, hence $y \notin \dom{\sub}$.
     Thus, $(\va \rond \sub)(y) \meq \va(y) \meq (\va \rond \subo
     \rond \sbst{x \repl y})(x) \meq (\va \rond \sub)(x)$.
   \item If $y \notin \dom{\bvi}$, then $(\va \rond \sub)(y) \meq (\va
     \rond \subo)(y) \meq (\va \rond \subo \rond \sbst{x \repl y})(x)
     \meq (\va \rond \sub)(x).$
   \end{itemize}
\end{pf}

\begin{Lemma}
\label{lemma-translation-of-scraping}
For all $\cont,x,\Th,\Cont, \sub$, and $\val \notin \Vars
$, if
$\TTranslid{\cont} \teq \CC{\Th}{\Cont}{\sub}$, and
$\scrapecont{\cont}{x} \aeq \val$, then there exist $\Thv$ and $\loc$
such that $\Transl{\val} \teq \config{\Thv}{\loc}$, $\sub(x) \meq \loc$
and $\Thv \subset \Th$.
\end{Lemma}

\begin{pf}
  By case analysis on $\cont$.  First, if $\cont$ is a nested lift
  context, then $\val \aeq x$ and $x \notin \dom{\sub}$, which gives
  the expected result.

  If $\cont \aeq (\letrecin{\bv}{\fcont})$, then $\TT{\cont} \aeq
  \Transltopid{\bv} \rond \MkGen{\Transl{\fcont}}$.  But by
  definition, $\MkGen{\Transl{\fcont}} \aeq
  \GCC{\Th'}{\Fcont}{\identity}{\identity}$ for some $\Th'$ and
  $\Fcont$. So, $\sub \meq \Substof{\Transltopid{\bv}}$, and we
  conclude by Lemma~\ref{lemma-translation-of-access}.

  If $\cont \aeq (\letrecin{\bv, y \is \fcont, \bb}{\e})$, then
  $\TT{\cont} \aeq (\TDum{y \is \fcont, \bb} \rond \Transltopid{\bv}
  \rond \TUpe{y}{\is}{\fcont}{\bb}{\transl{\e}})$.  Let $\TDum{y \is
    \fcont, \bb} \aeq \GCC{\Tho}{\trou}{\identity}{\vao}$,
  $\Transltopid{\bv} \aeq \Topbv{\Thbv}{\subbv}{\vabv}$, and
  $\TUpe{y}{\is}{\fcont}{\bb}{\transl{\e}} \aeq
  \GCC{\Thi}{\Cont}{\identity}{\identity}$ (they have these shapes by
  definition).  We have $\sub \meq (\vao \rond \vabv \rond \subbv)$,
  $\Th \aeq (\hfu{\Tho}{\hfu{\Thbv}{\Thi}})$.  By
  Lemma~\ref{lemma-translation-of-access}, we obtain $\Thv$ and $\loc$
  (because $\val \notin \Vars$) such that $\Transl{\val} \aeq
  \config{\Thv}{\loc}$, $\Thv \subseteq \Thbv$, and $(\vabv \rond
  \subbv)(x) \meq \loc$.  Here, this immediately gives $\Thv \subseteq
  \Thbv \subseteq \Th$ and $\sub(x) \meq \loc$.
\end{pf}

\begin{Lemma}
  \label{lemma-updateabs}
  For all $\e$ and $\e'$, if $\e \xrightarrow{\textsc{\scriptsize
      Update${}'_{\circ}$}} \e'$, then $\CoUpdate{\e}
  > \CoUpdate{\e'}$ and $\TT{\e} \eqalloc \TT{\e'}$.
\end{Lemma}

\begin{pf}
  Obviously, $\CoUpdate{\e} > \CoUpdate{\e'}$. Furthermore, we have
  $\e \aeq (\letrecin{(\bv, y \ex{n} x, \bb)}{\e_0})$. Let $\dcont
  \aeq (\letrecin{(\bv, y \ex{n} \trou, \bb)}{\e_0})$.  Since
  $\dcontext{x} \xrightarrow{\textsc{\scriptsize Update${}'_{\circ}$}}
  \e'$, we have some non-variable value $\val$ such that
  $\scrape{\bv}{x} \aeq \val$ and $\Sizeab{\val} \meq n$.  By
  Lemma~\ref{lemma-translation-of-scraping}, and letting
  $\TTranslid{\dcont} \teq \CC{\Th}{\Cont}{\sub}$, we have
  $\Transl{\val} \teq \config{\Thv}{\loc}$ such that $\Thv \subset
  \Th$ and $\sub(x) \meq \loc$.  Then, let $\BBdb \qet \TDum{\bb}$ and
  $\BBdy \aeq \CC{\heap{\loc' \heql \alloc~n}}{\trou}{\vay} \qet
  \TDum{y \ex{n} \trou}$ for some location $\loc'$, with $\vay \meq
  \sbst{y \repl \loc'}$.  Let $\BBd \aeq \BBdy \rond \BBdb$, and
  $\Ttop{\bv} \teq \CC{\Th_1}{\trou}{\sub_1},$ such that $\Thv
  \subseteq \Th_1 \subseteq \Th$ and $\sub \meq \Substof{\BBd} \rond
  \sub_1 \meq \Substof{\BBdb} \rond \vay \rond \sub_1$.

We have
$$\TT{\dcontext{x}} \teq \Appcontext{(\BBd \rond \Ttop{\bv})}{\letin{\wc \eql \update~y~x, \Up{\bb}}{\transl{\e_0}}}.$$

But $y \notin \supp{\sub_1}$, so
$$\TT{\dcontext{x}} \teq \Appcontext{(\BBd \rond \Ttop{\bv})}{
\letseq~{\wc \eql \update~\loc'~\loc, \Up{\bb}}
\inlet~{\transl{\e_0}}}.$$
Furthermore, $\Sizeab{\val} \meq \Sizealloc{\Thv(\loc)} \meq n$, by
Hypothesis~\ref{hypothesis-size}, and moreover by construction of the
translation, $\Thv$ only contains one binding.  So, in fact, the
update copies $\Transl{\val}$ entirely, and the previous configuration
reduces by \RuleUpdatealloc and \RuleLetalloc to
$$\Appcontext{(\BBdb \rond \Ttop{\bv, y \ex{n}
    \val})}{\letin{\Up{\bb}}{\transl{\e_0}}}.$$
Finally, by Proposition~\ref{proposition-update-tupdate}
and Lemma~\ref{lemma-finish-translation-of-bindings-bis}, 
this is $\eqalloc$-equivalent to $\TT{\dcontext{\val}}$, which is
exactly $\TT{\e'}$. 

\end{pf}

We obtain that $\lambdaabsem$ simulates $\lambdaabs$, and hence 
also simulates $\lambdaab$. 

\begin{Lemma}\label{lemma-abs-absem}
  For all $\e$, if $\e$ reduces to an answer, loops, or 
  is faulty in $\lambdaabs$, then so it does in $\lambdaabsem$.
\end{Lemma}

\begin{pf}
  The only non-trivial point is non termination.  By
  Lemmas~\ref{lemma-external-merging-letrec}
  and~\ref{lemma-updateabs}, and since \RuleUpdateabs preserves
  $\Nbletrecfun$, the lexicographic order
  $(\Nbletrecfun,\CoUpdatefun)$ stricly decreases by \RuleEMab and
  \RuleUpdateabs. Thus, there is no infinite reduction sequence in
  $\lambdaabs$ involving only these rules.
\end{pf}

\subsection{\Correctness}

We now have the tools to prove the expected correctness theorem. 
We first notice the following useful property of $\mergecont{\cont}{\bv}{\e}$. 

\begin{Proposition}
  For all $\cont,\bv,\e,$ if $\mergecont{\cont}{\bv}{\e}$ is
  defined, then $\TT{\mergecont{\cont}{\bv}{\e}} \teq
  \Appcontext{\TT{\cont} \rond \TT{\bv}}{\Transl{\e}}.$
\end{Proposition}

\begin{pf}
  By commutation of $\Ttop{\bv}$ with $\MkGen{\Transl{\fcont}}$, where
  $\fcont$ is the nested lift context part of $\cont$. 
\end{pf}

\begin{Lemma}[Correctness]
\label{lemma-correctness}
  For all $\e$ and $\e'$, if $\e \reductsem \e'$, then $\TT{\e}
  \reducta^+ \TT{\e'}$. 
\end{Lemma}

\begin{pf}
We proceed by case analysis on the rule used. 

\begin{description}
\item \RuleBetaabs~~There exist $\cont, \val_0,$ and $\val$ such that
  $\e \aeq \context{\val_0~\val}$, $\scrapeval{\cont}{\val_0} \aeq \fun x . 
  \g$, and $\e' \aeq \mergecont{\cont}{x \uu \val}{\g}$. Let
  $\Transl{\val} \teq \config{\Thv}{\V}$ and $\Transl{\fun x . \g} \teq
  \config{\Th_1}{\loc}$ (with $\Th_1 \aeq \heap{\loc \heql \fun x . 
    \transl{\g}}$).  Let $\TT{\cont} \teq \CC{\Th}{\Cont}{\sub}.$
  \begin{itemize}
  \item If $\val_0 \aeq \fun x . \g$, then 
    $$\begin{array}[t]{lcl}
      \TT{\e} & \eqalloc & \Appcontext{\TT{\cont}}{\Transl{\val_0~\val}} \\
      & \eqalloc & \Appcontext{\TT{\cont}}{\config{\hfu{\Th_1}{\Thv}}{\loc~\V}} \\
      & \reducta & \Appcontext{\TT{\cont}}{\config{\hfu{\Th_1}{\Thv}}{\sbst{x \repl \V} (\transl{\g})}} \\
      & \eqalloc & \Appcontext{\TT{\cont}}{\config{\Thv}{\sbst{x \repl \V}(\transl{\g})}} \\
      & \eqalloc & \Appcontext{\TT{\cont} \rond \TT{x \uu \val}}{\config{\emptyheap}{\transl{\g}}} \\
      & \eqalloc & \Appcontext{\TT{\cont} \rond \TT{x \uu \val}}{\Transl{\g}} \\
      & \eqalloc & \TT{\e'}. 
     \end{array}$$
   \item If $\val_0 \aeq y$ with $\scrapecont{\cont}{y} \aeq \fun x . \g$,
     then by Lemma~\ref{lemma-translation-of-scraping} we have
     a location $\loc \meq \sub(y)$ such that $\Th(\loc) \aeq \fun x . 
     \transl{\g}.$ So we have 
    $$\begin{array}[t]{lcl}
      \TT{\e} & \eqalloc & \Appcontext{\TT{\cont}}{\Transl{y~\val}} \\
      & \eqalloc & \Appcontext{\TT{\cont}}{\config{\Thv}{y~\V}} \\
      & \eqalloc & \Appcontext{\TT{\cont}}{\config{\Thv}{\loc~\V}} \\
      & \reducta & \Appcontext{\TT{\cont}}{\config{\Thv}{\sbst{x \repl \V}(\transl{\g})}} \\
      & \eqalloc & \TT{\e'} \mbox{ (as above)}. 
     \end{array}$$
  \end{itemize}

\item \RuleSelectabs~~There exist $\cont, \val_0$, and $\X$ such that
  $\e \aeq \context{\val_0.\X}$, and $\scrapeval{\cont}{\val_0} \aeq
  \record{\sv}$ with $\sv(\X) \aeq z$.  Let $\Transl{\record{\sv}} \teq
  \config{\Th_1}{\loc}$ (with $\Th_1 \aeq \heap{\loc \heql
    \record{\sv}}$).  The whole expression reduces
  to $\context{z}$.  Let $\TT{\cont} \teq \CC{\Th}{\Cont}{\sub}.$
  \begin{itemize}
  \item If $\val_0 \aeq \record{\sv}$, then 
    $$\begin{array}[t]{lcl}
      \TT{\e} & \eqalloc & \Appcontext{\TT{\cont}}{\Transl{\val_0.\X}} \\
      & \eqalloc & \Appcontext{\TT{\cont}}{\config{\Th_1}{\loc.\X}} \\
      & \reducta & \Appcontext{\TT{\cont}}{\config{\Th_1}{z}} \\
      & \eqalloc & \Appcontext{\TT{\cont}}{\config{\emptyheap}{z}} \\
      & \eqalloc & \TT{\context{z}}. 
     \end{array}$$
   \item If $\val_0 \aeq y$ with $\scrapecont{\cont}{y} \aeq \record{\sv}$,
     then by Lemma~\ref{lemma-translation-of-scraping} we have
     a location $\loc \meq \sub(y)$ such that $\Th(\loc) \aeq \record{\sv}.$
     So we have $$\begin{array}[t]{lcl}
       \TT{\e} & \eqalloc & \Appcontext{\TT{\cont}}{\Transl{y.\X}} \\
       & \eqalloc & \Appcontext{\TT{\cont}}{\config{\emptyheap}{\loc.\X}} \\
       & \reducta & \Appcontext{\TT{\cont}}{\config{\emptyheap}{z}} \\
       & \eqalloc & \TT{\context{z}}. 
     \end{array}$$
  \end{itemize}

\item \RuleContextab with \RuleLiftab~~We have $\e \aeq \context{\e_1}$,
  with $\e_1 \aeq \lcontext{\letrecin{\bb}{\e_0}}$, and $\e' \aeq \context{\e_2}$, 
  with $\e_2 \aeq \letrecin{\bb}{\lcontext{\e_0}}$.  Let $\Gcont \aeq \CC{\Th}{\Cont}{\sub} \qet
  \TTranslid{\cont}$ and $\Transl{\lcont} \teq
  \CC{\Thl}{\Lcontl}{\identity}$.  We have 
  $$\begin{array}[t]{lcl}
    \TTranslid{\context{\e_1}} & \eqalloc &
  \Gcontext{\config{\Thl}{\appcontext{\Lcontl}{\letin{\Dummy{\bb},\Update{\bb}}{\transl{\e_0}}}}} \\  
  & \reducta & \Gcontext{\config{\Thl}{\letin{\Dummy{\bb},\Update{\bb}}{\appcontext{\Lcontl}{\transl{\e_0}}}}} \\
  & \eqalloc & \Gcontext{\config{\emptyheap}{\transl{\e_2}}} \\
  & \eqalloc & \transl{\e'}.
  \end{array}$$

\item \RuleIMab~~We have $\e \aeq \letrecin{\bb}{\e_0}$ and $\e' \aeq
  \letrecin{\bb'}{\e_0}$, with $\bb \aeq (\bv, x \is
  (\letrecin{\bb_1}{\e_1}), \bb_2)$, and $\bb' \aeq
  (\bv, \bb_1, x \is \e_1, \bb_2)$. 

  Let $\BBbv \qet \Transltopid{\bv}$, $\bb_0 \aeq (x \is
  (\letrecin{\bb_1}{\e_1}), \bb_2)$ and $\bb'_0 \aeq (x \is \e_1,
  \bb_2)$. 

  By definition of the translation, we have
  $\TTranslid{\letrecin{\bb}{\e_0}} \teq \Appcontext{\TDum{\bb_0} \rond
    \BBbv \rond \TUp{\bb_0}}{ \config{\emptyheap}{\transl{\e_0}}}$. 

Let now
$(\bd, \Fcont) \aeq \left \{
\begin{array}{l}
(x, \trou) \mbox{ if ${\is} \meq {\uu}$} \\
(\wc, \update~ x~\trou) \mbox{ otherwise.} 
\end{array} \right .$

We have 
$$
\begin{array}{rcl}
\Transl{\letrecin{\bb_1}{\e_1}} & \teq & (\config{\emptyheap}{\transl{\letrecin{\bb_1}{\e_1}}}) \\
& \aeq & 
(\config{\emptyheap}{\letin{\Dummy{\bb_1}, \Up{\bb_1}}{\transl{\e_1}}}) \\
& \aeq & \config{\emptyheap}{\E_1}.
\end{array}
$$
So $\TTranslid{\letrecin{\bb}{\e_0}} \teq
\Appcontext{\TDum{\bb_0} \rond \BBbv}{
\config{\emptyheap}{\letin{\bd \eql \Appcontext{\Fcont}{\E_1}, \Up{\bb_2}}{\transl{\e_0}}}}.$

But this reduces by (maybe rule \RuleLiftalloc and) rule \RuleIMalloc
to $$\TDum{\bb_0} \rond \BBbv [ \config{\emptyheap}{
\begin{array}[t]{l}
\letseq
\begin{array}[t]{l}
\Dummy{\bb_1}, \Up{\bb_1}, \\
\bd \eql \Appcontext{\Fcont}{\transl{\e_1}}, \Up{\bb_2} 
\end{array} \\
\inlet~
\transl{\e_0}}]. 
\end{array}
$$

But we recognize $\Up{\bb_1, \bb'_0}$, so the obtained configuration is equal to
$$\Conf \aeq \Appcontext{\TDum{\bb_0} \rond \BBbv}{
\config{\emptyheap}{\letin{\Dummy{\bb_1}, \Up{\bb_1, \bb'_0}}{
\transl{\e_0}}}}.$$

Then, by Propositions~\ref{proposition-weak-context-reduction} and~\ref{proposition-dummy},
we obtain
$$\Conf \eqalloc \Appcontext{\TDum{\bb_0} \rond \BBbv \rond \TDum{\bb_1}}{
\config{\emptyheap}{\letin{\Up{\bb_1, \bb'_0}}{
\transl{\e_0}}}}.$$

But as $\dom{\bb_1} \orth \dom{\bv} \cup \FV{\bv}$, this is equal to 
$$\Appcontext{\TDum{\bb_1,\bb'_0} \rond \BBbv}{
\config{\emptyheap}{\letin{\Up{\bb_1, \bb'_0}}{
\transl{\e_0}}}},$$

which by Proposition~\ref{proposition-update-tupdate} and
Lemma~\ref{lemma-finish-translation-of-bindings-bis} is
$\eqalloc$-equivalent to $\TTranslid{\letrecin{\bb'}{\e_0}}$, which
concludes the proof.
\end{description}

\end{pf}

This yields:
\begin{Corollary}\label{corollary-absem-allocs}
  For all $\e$, if $\e$ reduces to an answer, loops, or is faulty in
  $\lambdaabsem$, then so does $\TT{\e}$ in $\lambdaallocs$.
\end{Corollary}
\begin{pf}
  Since $\TTfun$ maps answers to answers
  (Proposition~\ref{prop-answers}), if $\e$ reduces to an answer, then
  so does $\TT{\e}$. Moreover, because Lemma~\ref{lemma-correctness}
  uses $\reducta^+$, if $\e$ loops, so does $\TT{\e}$. Finally, if
  $\e$ is faulty, then it reduces to a term $\e_0$ in normal form of
  one of the shapes in Proposition~\ref{prop-answers}, hence $\TT{\e}$
  reduces to $\TT{\e_0}$, which is faulty by
  Proposition~\ref{prop-faulty}. Hence $\TT{\e}$ is faulty.
\end{pf}

We finally have:
\begin{pf}[of Theorem~\ref{theorem-correctness}]
  By composing Lemmas~\ref{lemma-encoding-lambdaab-lambdaabs},
  \ref{lemma-abs-absem}, Corollary~\ref{corollary-absem-allocs}, and
  Lemma~\ref{lemma-alloc-allocs}, we obtain the result for $\TT{\e}$.
  But $\TT{\e} \eqalloc \transl{\e}$ in $\lambdaallocs$, hence they
  behave the same by Lemma~\ref{lemma-alloc-allocs}.
\end{pf}

\section{Related work}
\label{section-related-work}

Ariola and Blom \cite{Ariola98} study $\lambda$-calculi with $\Letrec$, in relation
with the graphs they represent. The $\lambdaab$ language presented
here is mostly a deterministic variant of their call-by-value
calculus. The main difference lies in our size indications, which
specialize the language for efficient compilation.


Lang et al \cite{Lang99} study $\lambda$-calculi with sharing and
recursion, resulting in the notion of
\intro{Addressed Term Rewriting Systems}. Unlike in
$\lambdaab$, cyclic data structures are represented using addresses:
each node of a term is given an address, which can be referred to by a
\intro{back pointer}. Addresses can be shared among instances of the
same term. Moreover, addresses are not bound in the considered term,
whereas in $\lambdaab$, $\letrec$ does bind variables. Thus, addressed
terms must satisfy a number of coherence conditions, which appear to
be far from trivial. This explains our choice of Ariola et al.'s
approach. 

Erk{\"o}k and Launchbury \cite{Erkok00} consider the interaction of recursion with
side effects. In the setting of monadic meta-languages,
Moggi and Sabry \cite{Moggi04} devise an operator named
$\kwd{Mfix}$, with an operational semantics, which unifies different
language constructs for recursion. This very interesting work is more
abstract than ours, in the sense that it unifies several recursion
constructs from both eager and lazy languages, whereas our work is
specific to call-by-value. Also, we are not specifically interested in
the interaction between recursion and side effects, although we treat
it with care. Moreover, Erk\"ok and Launchbury and Moggi and Sabry
are not concerned with compilation. 

Another work on recursion, already discussed in
Section~\ref{subsection-backpatching-ipu}, is Boudol's calculus
\cite{Boudol04}. From the standpoint of expressive power, this calculus is
incomparable with $\lambdaab$.  On the one hand, the semantics of
$\lambdaab$, based on Ariola et al.'s work, allows to represent cyclic
data structures such as $\Letrec~x \eql \texttt{cons}~1~x$, while such
a definition loops in Boudol's calculus.  On the other hand, the
unrestricted $\Letrec$ of Boudol's calculus avoids the difficult guess
of correct size indications.

From the standpoint of compilation, Boudol and Zimmer \cite{Boudol02}
use a backpatching approach, thus increasing the number of run-time
tests and indirections.  A similar backpatching approach is used in
Russo's extension of ML with recursive modules \cite{Russo01},
implemented in Moscow ML, and in Dreyer's work on typing of extended
recursion \cite{Dreyer04}.

Syme \cite{Syme05} extends the F\# language with generalized recursive
definitions where the right-hand sides are arbitrary computations.
Haskell-style lazy evaluation is used to evaluate these recursive
definitions: a strong, forward reference to a recursively-defined
variable $x$ is not an error, but causes the definition of $x$ to be
evaluated at this point.  In the application scenario considered by
Syme, namely interfacing with libraries written in object-oriented
languages, no compile-time information is available on dependencies
and object sizes, rendering our approach inapplicable and essentially
forcing the use of lazy evaluation.  However, lazy evaluation has some
additional run-time costs and makes evaluation order hard to guess in advance.

Nordlander, Carlsson and Gill \cite{Nordlander08} describe an original
variant of the in-place update scheme where the sizes of the
recursively-defined values need not be known at compile-time.
Consider a recursive definition $\kwd{rec}\, x = e$.  The variable $x$
is first bound to a unique marker; then, $e$ is evaluated to a value
$v$; finally, the memory representation of $v$ is recursively
traversed, replacing all occurrences of the unique marker with a
pointer to $v$.  This recursive traversal can be much more costly than
the updating of dummy blocks performed by the in-place update scheme:
a naive implementation runs in time $O(N)$ where $N$ is the size of
the value $v$.  (This size can be arbitrarily large even if the
evaluation of $e$ is trivial: consider $\kwd{rec}\, x =
\kwd{Cons}~l~x$ where $l$ is a $10^6$ element list previously
computed.)  Assuming linear allocation and a copying garbage
collector, the traversal can be restricted to blocks allocated during
the evaluation of $e$, resulting in a reasonable complexity $O(\min(N,
M))$ where $M$ is the number of allocations performed by the
evaluation of $e$.  However, this improvement seems impossible for
memory managers that perform non-linear allocation like those of OCaml
and F\#.

Mutually-recursive definitions of functions (syntactic
$\lambda$-abstractions) is a frequently-occurring special case that
admits a very efficient implementation \cite{Kranz86,Appel92}.
Instead of allocating one closure block for each function, containing
pointers to the other closure blocks, it is possible to share a single
memory block between the closures, and use pointer arithmetic to
recover pointers to the other closures from any given closure.  No
in-place update is needed to build loops between the closures.  We
believe that this trick could be combined with a more general in-place
update scheme to efficiently compile recursive definitions that
combine syntactic $\lambda$-abstractions and more general
computations.  However, significant extensions to $\lambdaalloc$ would
be needed to account for this approach.

\section{Conclusions and future work}
\label{section-future-work}

In this article, we have developed the first formal semantic account
of the in-place update scheme, and proved its ability to implement
faithfully an extended call-by-value recursion construct, as
characterized by our source language $\lambdaab$. 

At this point, one may wonder whether $\lambdaab$ embodies the most
powerful call-by-value recursion construct that can be compiled via
in-place update.  The answer is no, because of the requirement that
the sizes (of definitions that are forward-referenced) be known
exactly at compile-time.  In a context of separate compilation and
higher-order functions, often the only thing that the compiler knows
about definitions is their static types.  With some data
representation strategies, the sizes are functions of the static
types, but not with other strategies.  For example, the closures that
represent function values can either follow a ``two-block'' strategy
(a closure is a pair of a code pointer and a pointer to a
separately-allocated block holding the values of free variables) or a
``one-block'' strategy (the code pointer and the values of the free
variables are in the same block).  With the two-block strategy, all
definitions of function type $\tau_1 \rightarrow \tau_2$ have known
size~2; but with the one-block strategy, the size is $1 + n$ where $n$
(the number of free variables) is not reflected in the function type
and is therefore difficult to guess at compile-time.

\begin{figure}
\begin{framed}
$$\begin{array}{l@{\quad}rcll}
\mbox{Variable:} &x & \in  & \Vars \\
\mbox{Name:}     &\X & \in & \Namesfun \\[8pt]
\mbox{Expression:}
&\e \in \expr 
& \bnf & x \alt \fun x.\e \alt \e_1~\e_2 & \textrm{$\lambda$-calculus}\\
&& \alt & \record{\s} \alt \e \rsel \X & \textrm{Record operations} \\
&& \alt & \letrecin{\bb}{\e} & \textrm{Recursive definitions} \\[8pt]
\mbox{Record row:} &\s & \bnf & \emptysequence \alt \X \eql x, \s \\
\mbox{Binding:}    &\bb & \bnf & \emptysequence \alt x \is \e, \bb \\[8pt]
\mbox{Size indication:}
&\is & \bnf & \ex{\e} \alt \uu & \vspace{8pt}
\end{array}$$
\end{framed}
\caption{Syntax of generalized $\lambdaab$}
\label{figure-syntax-lambdaab-gen}
\end{figure}

There are several ways to relax the size requirement and therefore
increase the usability of $\lambdaab$ as an intermediate language.
First, one could permit values of size smaller than expected to fill
the pre-allocated blocks.  In this case, updating a pre-allocated
block changes not only its contents but also its size, an operation
that most memory managers support well.  All we now need to determine
statically is a conservative upper bound on the actual size.  For
example, if the type of a definition is a datatype (sum type), we can
take the maximum of the sizes of its constructors.  In the case of
one-block closures, we can allocate dummy blocks with a fixed size,
say 10 words, and instruct the compiler to never generate closures
larger than this, switching to a two-block representation for
closures with more than 9 free variables (such closures are uncommon).
This simple extension can be formalized with minimal changes to
$\lambdaab$, $\lambdaalloc$ and the proofs presented in this paper.

Another way to relax the size requirement is to notice that
the sizes of pre-allocated blocks do not need to be compile-time
constants: the in-place update scheme works just as well if these
sizes are determined by run-time computations that take place before
the recursive definition is evaluated.  For example, in the encoding
of mixins outlined in Section~\ref{subsubsection-mixin-modules},
each component of a mixin could be represented not just as a generator
function $f$, but as a pair $(n, f)$ where $n$ is the size of the
result of $f$.  The recursive definition implementing the
\verb/close/ operation could, then, extract these sizes $n$ from
the run-time representation of the mixin and use them to pre-allocate
dummy blocks. 

In preparation for future work, we now sketch an extension of
$\lambdaab$ where the size indications over bindings are no longer
compile-time constants but arbitrary expressions.  
Figure~\ref{figure-syntax-lambdaab-gen} gives the syntax of this
extended language.  In bindings, the size indications are all
evaluated before the evaluation of definitions begins, and cannot
refer to the recursively defined variables.

From the standpoint of compilation, we believe that in-place update
applies straightforwardly. However, a serious issue with this
extension is how to ensure statically that the predicted sizes are
correct: given a definition $x \ex{\e_1} \e_2$, we would like to
guarantee that $\e_2$ will evaluate to a value of size the value of
$\e_1$.  If $e_1$ is an arbitrary expression, a type system or another
static analysis can not try and evaluate $\e_1$ because this would make
it undecidable.  Instead, we have to find static means of ensuring the
validity of definitions in the useful cases. 

For this, we plan to start from Hirschowitz's type system for
$\lambdaab$~\cite{Hirscho-PhD} and extend it with dependent product
types and a special sized type $\sized{\val}{\abt}$, denoting the set
of values of type $\abt$ and of size $\val$. Given $n$ and $\e$ of
size $n$, one could give a dependent product type to the pair $(n,
\e)$, namely $\langle x : \id{int}, \sized{x}{\abt} \rangle$. 
Conversely, take for example a dependent pair $\e$ of type $\langle x
: \id{int}, \abt_1 \to \sized{x}{\abt_2} \rangle$, the expression
$(\snd{\e}~\e')$ has size $\fst{\e}$, and this can be checked
statically.  This guarantees that the definition $x \ex{\fst{\e}}
(\snd{\e}~\e')$ is correct \wrt~sizes.

\begin{acknowledgements}
  The authors warmly thank the anonymous referees for their
  detailed comments and helpful suggestions for improving the presentation.
\end{acknowledgements}

\let\french=\undefined
\bibliographystyle{spmpsci}
\bibliography{./biblio}

\end{document}